%% file: ms.tex
\documentclass[iop,apj,numberappendix,appendixfloats]{emulateapj}
\usepackage{graphicx}
\pdfoutput=1

\LongTables
\def \lya  {Ly$\alpha$}

\def \lyb  {Ly$\beta$}

\def \ly5  {Ly-5}
\def \ly6  {Ly-6}
\def \ly7  {Ly-7}
\def \nhi  {$N_{\rm HI}$}
\def \nsiiv  {$N_{\rm SiIV}$}
\def \novi  {$N_{\rm OVI}$}
\def \nhii  {$N_{\rm HII}$}
\def \mnhi  {N_{\rm HI}}
\def \mnhii  {N_{\rm HII}}
\def \lnhi {$\log N_{\rm HI}$}

\def \mlnovi  {\log N_{\rm OVI}}
\def \lnovi {$\log N_{\rm OVI}$}

\def \mlnhi {\log N_{\rm HI}}

\newcommand{\km}{${\rm km\,s}^{-1}$}
\newcommand{\kms}{${\rm km\,s}^{-1}$}

\def\lesssim{\mathrel{\hbox{\rlap{\hbox{%
 \lower4pt\hbox{$\sim$}}}\hbox{$<$}}}}
\def\gtrsim{\mathrel{\hbox{\rlap{\hbox{%
 \lower4pt\hbox{$\sim$}}}\hbox{$>$}}}}
\let\la=\lesssim                
\let\ga=\gtrsim

\newcommand{\hi}{H$\;${\small\rm I}\relax}
\newcommand{\hii}{H$\;${\small\rm II}\relax}
\newcommand{\heii}{He$\;${\small\rm II}\relax}
\newcommand{\hei}{He$\;${\small\rm I}\relax}
\newcommand{\alii}{Al$\;${\small\rm II}\relax}
\newcommand{\aliii}{Al$\;${\small\rm III}\relax}
\newcommand{\di}{D$\;${\small\rm I}\relax}

\newcommand{\cii}{C$\;${\small\rm II}\relax}

\newcommand{\civ}{C$\;${\small\rm IV}\relax}

\newcommand{\nv}{N$\;${\small\rm V}\relax}
\newcommand{\oi}{O$\;${\small\rm I}\relax}

\newcommand{\ovi}{O$\;${\small\rm VI}\relax}
\newcommand{\ovii}{O$\;${\small\rm VII}\relax}
\newcommand{\oviii}{O$\;${\small\rm VIII}\relax}

\newcommand{\sii}{S$\;${\small\rm II}\relax}
\newcommand{\siii}{Si$\;${\small\rm II}\relax}

\newcommand{\siiv}{Si$\;${\small\rm IV}\relax}

\newcommand{\hit}{H$\;${\scriptsize\rm I}\relax}
\newcommand{\hiit}{H$\;${\scriptsize\rm II}\relax}

\newcommand{\ciit}{C$\;${\scriptsize\rm II}\relax}

\newcommand{\civt}{C$\;${\scriptsize\rm IV}\relax}

\newcommand{\nvt}{N$\;${\scriptsize\rm V}\relax}
\newcommand{\oit}{O$\;${\scriptsize\rm I}\relax}

\newcommand{\ovit}{O$\;${\scriptsize\rm VI}\relax}
\newcommand{\oviit}{O$\;${\scriptsize\rm VII}\relax}

\newcommand{\siit}{S$\;${\scriptsize\rm II}\relax}
\newcommand{\siiit}{Si$\;${\scriptsize\rm II}\relax}

\newcommand{\siiiit}{Si$\;${\scriptsize\rm III}\relax}

\newcommand{\aliiit}{Al$\;${\scriptsize\rm III}\relax}
\newcommand{\aliit}{Al$\;${\scriptsize\rm II}\relax}

\newcommand{\siivt}{Si$\;${\scriptsize\rm IV}\relax}

\submitted{Accepted Version by the Astrophysical Journal}
\shortauthors{Lehner et al.}
\shorttitle{KODIAQ I: Galactic and Circumgalactic \ovi\ at $2<\lowercase{z}\la 3.5$}

\begin{document}

\title{Galactic and Circumgalactic \ovi\ and Its Impact on the \\ Cosmological Metal and Baryon Budgets at $2<\lowercase{z}\la 3.5$\altaffilmark{1} }

\author{
N. Lehner\altaffilmark{2},
J.M. O'Meara\altaffilmark{3},
A.J. Fox\altaffilmark{4},
J.C. Howk\altaffilmark{2},
J.X. Prochaska\altaffilmark{5},
V. Burns\altaffilmark{2}, \&
A.A. Armstrong\altaffilmark{3}
}
\altaffiltext{1}{We dedicate this paper and the KODIAQ project to the memory and families of Wal Sargent and Arthur M. Wolfe.  Without the vision and terrific efforts of these two scientists, this survey would not exist.  Their careers have greatly inspired and influenced our own, and we hope that their work continues to flourish with this archival dataset.
}
\altaffiltext{2}{Department of Physics, University of Notre Dame, 225 Nieuwland Science Hall, Notre Dame, IN 46556}
\altaffiltext{3}{Department of Chemistry and Physics, Saint Michael's College, One Winooski Park, Colchester, VT 05439}
\altaffiltext{4}{Space Telescope Science Institute, Baltimore, MD 21218}
\altaffiltext{5}{University of California/Lick Observatory, Santa Cruz, 1156 High Street, CA 95064}

\begin{abstract}
We present the first results from our NASA Keck Observatory Database of Ionized Absorbers toward Quasars (KODIAQ) survey which aims to characterize the properties of the highly ionized gas of high redshift galaxies and their circumgalactic medium (CGM) at $2<z<4$. We select absorbers optically thick at the Lyman limit ($\tau_{\rm LL}>1$, $\mlnhi > 17.3$) as probes of these galaxies and their CGM where both transitions of the \ovi\ doublet have little contamination from the Ly\,$\alpha,\beta$ forests. We found 20 absorbers that satisfy these rules:  7 Lyman limit systems (LLSs), 8 super-LLSs (SLLSs) and 5 damped \lya\ (DLAs). The \ovi\ detection rate is  100\% for the DLAs, 71\% for the LLSs, and 63\% for the SLLSs.  When \ovi\ is detected, $\log \langle N_{\rm OVI}  \rangle = 14.9 \pm 0.3$, an average \ovi\ column density substantially larger and with a smaller dispersion than found in blind \ovi\ surveys at similar redshifts. Strong \ovi\ absorption is therefore nearly ubiquitous in the CGM of $z \sim 2$--$3$  galaxies. The total velocity widths of the \ovi\ profiles are also large ($200 \le \Delta v_{\rm OVI}  \le 400$ \km). These properties are quite similar to those seen for \ovi\ in low $z$ star-forming galaxies, and therefore we hypothesize that these strong CGM \ovi\ absorbers (with  $\tau_{\rm LL}>1$)  at $2<z\la 3.5$ also probe outflows of star-forming galaxies. The LLSs and SLLSs with no \ovi\ absorption have properties consistent with those seen in cosmological simulations tracing cold streams feeding  galaxies.  When the highly ionized (\siiv\ and \ovi) gas is taken into account, we determine that the $\tau_{\rm LL}>1$ absorbers could contain as much as 3--14\% of the cosmic baryon budget at $z\sim 2$--3, only second to the \lya\ forest. We conservatively show that $5$--$20\%$ of the metals ever produced at $z\sim 2$--3 are in form of highly ionized metals ejected in the CGM of galaxies. 
\end{abstract}


\section{Introduction}\label{s-intro}

The redshift $2\la z \la 4$ interval corresponds to a turning point in the formation and evolution of galaxies as observations of the star-formation-rate density, QSO density, stellar mass density all  peak at these redshifts \citep[e.g.,][]{madau96,shapley01,dickinson03,reddy08}. It is also at these redshifts that there has been a difficulty in accounting for all the metals \citep[e.g.,][]{pagel02,pettini04,pettini06,bouche06}, with an apparent discrepancy of  $\sim 40\%$ observed between the mass in metals derived from integrating the star formation history of the universe compared to the metal budget at $z\sim 2$.   Based on the accounting by, e.g., \citet{bouche06} and \citet{pettini06} at $z\sim 2$,  a large fraction of metals is found in the environments probed by QSO absorbers, the intergalactic medium (IGM), circumgalactic medium (CGM), and near or in galaxies. About $\la15\%$ of the metals are found  in the diffuse gas of the \lya\ forest \citep[e.g.,][]{schaye03,simcoe04,bergeron05}, 5\% in the {\it neutral} gas of the damped \lya\ absorbers (DLAs) \citep[e.g.,][]{herbert-fort06}, and between 2\%  and 17\% in the neutral and photoionized super Lyman Limit systems (SLLSs, a.k.a., sub-DLAs) \citep{peroux06,prochaska06,kulkarni07}.\footnote{We use in this paper the standard \hit\ column density separation between the different types of $\tau_{\rm LL}>1$ QSO absorbers, i.e., the DLAs have  $\log N_{\rm HI} \ge 20.3$  \citep[e.g.,][]{wolfe05}, the SLLSs (or sub-DLAs) have with $19 \le \log N_{\rm HI} < 20.3$,  \citep[e.g,][]{peroux03,omeara07}, and the LLSs with with $17 \le \log N_{\rm HI} < 19$  \citep{tytler82,sargent89,ribaudo11}.} However, missing from the cosmic metal census is the contribution from the highly ionized gas probed by \ovi\ associated with absorbers optically thick at the Lyman limit (i.e., absorbers with $\tau_{\rm LL}>1$ or $\mlnhi > 17.3$). \ovi\ has often been set aside at high $z$ because the \ovi\ doublet lies deep in the dense \lya\ and \lyb\ forests and is often (but not always) blended with \hi\ and other lines. 

 \citet{fox07a} first led this search for \ovit\ in DLAs at $z>2$, showing in particular DLAs have a substantial amount of highly ionized gas in form of \ovi.  This and other works on \ovi\ associated with  $\tau_{\rm LL}>1$ absorbers have demonstrated the potential of the \ovi\ to constrain the physics of young galactic disks and halos as well as filling up the reservoirs of baryons and metals \citep{lu93,kirkman97,kirkman99,fox07c,lehner08}. However, most of the current conclusions regarding the census of baryons and metals in $\tau_{\rm LL}>1$ absorbers are tentative owing to a lack of knowledge on the detailed \ovi\ properties and the absence of sizable samples of \ovi\ associated with absorbers over the entire \hi\ column density range of the optically thick absorbers ($17.3 \le \mlnhi \la 22$). For example, it is unclear if all the high $z$ $\tau_{\rm LL}>1$ absorbers have large quantities of \ovi, or what is the ionization fraction in these absorbers and how it varies with \nhi\ or metallicity. 

We have therefore undertaken a program with the Keck observatory archive (KOA) to study the highly ionized gas of the galaxies and their CGM and assess their contribution to the cosmological metal and baryon budgets at $2<z<4$. We use absorbers with $\mlnhi > 17.3$ as tracers of these environments because we know that these absorbers probe galaxies and the densest regions of the CGM according to observational results  \citep[e.g.,][]{morris91,bergeron94,lanzetta95,tripp98,penton02,bowen02,chen05,morris06,prochaska11,rudie12} and cosmological simulations \citep[e.g.,][]{springel03,kobayashi04, oppenheimer10,smith11,voort12,sijacki12,rahmati13,bird13}. The rate of incidence of these absorbers is proportional to the product of the comoving number density of galaxies giving rise to the LLSs, SLLSs, or DLAs and the average physical cross-section of the galaxies \citep[e.g.,][]{tytler87,ribaudo11}, and therefore their selection is independent of the galaxy luminosity, possibly probing other types of galaxies than the rest-frame UV selected Lyman break galaxies \citep[e.g.,][]{steidel10}. 

We focus  on the highly ionized gas traced by the \ovi\ doublet at 1031.926 and 1037.617 \AA\ in these $\tau_{\rm LL}>1$ absorbers because the \ovi\ has been shown to be an excellent tracer of galaxy feedback with typically large enhancement in the \ovi\ column toward actively star forming galaxies relative to more passive galaxies  \citep[][and see also \citealt{lehner09,grimes09,heckman02}]{tumlinson11}.  Among the highly ionized  absorber metals (the so-called ``high ions'', \siiv, \civ, \nv, \ovi), the \ovi\ doublet is in fact unique as a tracer of galaxy feedback and collisionally ionized processes: the ionization potentials involved in the creation and destruction of \ovi\  (113.9--138.1 eV) are significantly higher than all the other rest-frame UV ions, ensuring that in dense environments of the CGM or galaxies it is a good indicator of collisionally ionized processes rather photoionization by galaxies or QSOs \citep[e.g.,][and see also discussion in this paper]{simcoe02,fox07a,lehner08,lehner09}. The \civ\ and \siiv\ ions have ionization potential ranges 33.5--45.1 eV and 47.9--64.5 eV, respectively, below or including the \heii\ ionization potential at 54.4 eV, and hence they can be more readily produced by photoionization of hot stars or the extragalactic UV background (EUVB). So even though at $z>2$ the \ovi\ doublet is often blended with \hi\ and other lines, the \ovi\ gas-phase is critical to study (and as we will show in this paper there is no other high ions that can be used as a proxy for \ovi). On the positive side, a major advantage of studying \ovi\ at $z>2$ is that \ovi\ can be observed at both high signal-to-noise (S/N) and high spectral resolution, which is not the case for most studies of the low redshift universe, including our own Milky Way. 

Previous blind high redshift \ovi\ surveys have assembled samples selected directly using \ovi\ and hence typically targeting lower \hi\ column density absorbers \citep{simcoe02,muzahid12} or associated with \lya\ forest lines \citep{carswell02}. Our hunt for \ovi\ absorbers differs from these studies in that we only search for \ovi\ associated with $\tau_{\rm LL}>1$ absorbers, i.e., the \ovi\ absorption overlaps in velocity with more weakly ionized and neutral gas with $\mlnhi \ge 17.3$. Our first KODIAQ\footnote{Keck Observatory Database of Ionized Absorbers toward Quasars.} survey spans a factor $>1000$ in \nhi\ from $\mlnhi = 17.75$ to $\mlnhi = 20.8$, spanning a critical regime in \nhi\ where the gas transitions from being predominantly ionized to predominantly neutral. In this first KODIAQ survey, we adopt stringent criteria in our selection of the \ovi\ absorbers, requiring that the \ovi\ doublet has little contamination in both the weak and strong transitions (or in the case of absence of \ovi\ absorption, no contamination over the velocities where other metal lines are detected). We also require that we can estimate \nhi\ and metallicity of the gas, which  allows us to characterize the properties of the \ovi\ absorption (as well as \civ, \siiv, and \nv) associated with $\tau_{\rm LL}>1$ absorbers as a function of metallicity and \hi\ column density. Twenty absorbers satisfy our selection criteria: 7 LLSs, 8 SLLSs, and 5 DLAs.  For each absorber, we systematically estimate the total column densities of the \hi\ and metal ions, and the metallicity. When \ovi\ absorption is observed, we fitted the individual components of the high ions (\ovi, \civ, \siiv) to determine the temperature and kinematics of the highly ionized gas and the relationship between the high ions. 

This paper is structured as follows. In \S\ref{s-pres} we describe briefly the KODIAQ survey, its data reduction and its future release. In \S\ref{s-anal}, we present the analysis of the data, including the estimates of the total column densities and the profile fitting method, and additional details for each absorber are provided in the appendix. Our main results are presented in \S\ref{s-prop}, while the implications, including for the cosmic metal and baryon budgets, are discussed in \S\ref{s-disc}. Finally, in \S\ref{s-sum} we summarize our main results.  Throughout the manuscript, we adopt a $\Lambda$CDM cosmology with $\Omega_\Lambda = 0.7$, $\Omega_{\rm m} = 0.3$ and $H_0 = 70$ \km\,Mpc$^{-1}$.

\section{Presentation of KODIAQ}\label{s-pres}
\subsection{KODIAQ Database and Data Reduction}\label{s-redux}
All data in this sample were acquired with the HIgh Resolution Echelle Spectrometer (HIRES) \citep{vogt94} on the Keck\,I telescope on Mauna Kea. These data were obtained by different PIs from different institutions with Keck access, and hundreds of spectra of QSOs at $0<z<6$, most being at $z\sim 2$--$4$, were collected. This provides one of the richest assortments of high-$z$ QSO spectra at high spectral resolution (6--8 \km) and high signal-to-noise (many with S/N\,$>20$--$50$). A large fraction of these data is now publicly available in a raw form from the Keck Observatory Archive (KOA).\footnote{Available at http://www2.keck.hawaii.edu/koa/} However, the spectra are not readily useable without intensive processing.  We were awarded a NASA Astrophysics Data Analysis Program (ADAP) grant (PI Lehner) to perform the data reduction and coaddition of the individual exposures of the entire KOA QSO database to study in detail the highly ionized plasma associated with $\tau_{\rm LL}>1$ absorbers at $2<z<4$. We plan to release the entire KODIAQ database to the scientific community in  2015. 

The data presented here represent the first data products from our science program including some of the best \ovi\ absorbers associated with $\tau_{\rm LL}>1$ \hi\ absorber at high $z$ (see next section). The spectra were uniformly reduced, coadded, and continuum normalized.  The specifics of the data reduction pipeline and associated tools will be presented in a later paper (J. O'Meara et al. 2014, in preparation), but we discuss the major points now. As \ovi\ requires sensitivity in the blue, most of the exposures were obtained with HIRES configured in its current form, namely with the three CCDs installed in August, 2004; only a few spectra are from the previous single CCD configuration, in particular where red coverage was lacking.   The 2D HIRES images were reduced with the XIDL HIRES redux pipeline,\footnote{Available at http://www.ucolick.org/$\sim$xavier/HIRedux/} which extracts and coadds the data.  Continuum fitting was performed by a procedure within the pipeline that assigns a continuum level to each spectral order to be coadded and the continuum normalized spectra were then combined into a single 1D spectrum. Each spectrum contains spectral gaps due to the gaps between the 3 CCDs, along with occasional additional gaps due to the size of the CCD not completely covering the full spectral range of an echelle order. 

\subsection{KODIAQ I: This Survey}\label{s-kodiaq}
From the sample of 108 reduced and normalized QSO spectra presently in KODIAQ, we first searched for \ovi\ absorbers associated with $\tau_{\rm LL}>1$ absorbers, i.e., we searched for the presence of strong ($\mlnhi \ga 17.3$) \hi\ absorption, revealed either by the presence of a break in  the flux at the absorber rest wavelengths $\lambda_{r} \sim 912$ \AA\ or by the presence of damping features in the \lya\ line.  For each $\tau_{\rm LL}>1$ absorber, we generated continuum-normalized absorption profiles and apparent column density profiles ($N_a(v)$, see below) as a function of velocity to visually inspect the spectra for the presence and contamination of the \ovi\ absorption. In the great majority of cases ($\sim$70\%), the detection of \ovi\ is highly uncertain due to the presence of \lya\ or \lyb\ forest absorption in the region of the expected \ovi\ absorption. However, for 20 absorbers, our selection criteria are met: 
\begin{enumerate}
\item the absorbers have $\mlnhi \ga 17.3$ and \nhi\ can be estimated with an error less than $\pm 0.3$ dex;
\item the contamination in {\it both}\ transitions of the \ovi\ doublet within about $\pm 200$ \km\ of the redshift-frame of the absorbers is small, i.e., there is a good match between the apparent column density profiles of the weak and strong transitions of \ovi\ over most of the observed absorption;
\end{enumerate}
Seven of these absorbers are LLSs, eight are SLLSs, and five are DLAs;  one SLLS overlaps with a DLA classification within 1$\sigma$ error on \nhi. In our analysis, we also consider some of the \ovi\ absorbers associated with DLAs from the sample assembled by \citet{fox07a}, but only those that satisfied our second criterion, reducing their sample of 12 DLAs to 6. We hereafter refer to this DLA sample as the F07 sample.\footnote{Specifically the DLAs observed in the spectra of the following QSOs are part of the F07 sample: Q0027-186, Q0112+306, Q0450-131, Q2138-444, Q2243-605, Q0841+129. }

Parallel to this effort, we have undertaken a search for strong \lya\ absorbers ($\mlnhi \ga 16$)  where \ovi\ can be estimated. That database is much larger (about 50 \ovi\ absorbers) and will be presented in a future KODIAQ paper. The database of \ovi\ absorbers associated with strong \hi\ absorbers will increase in the near future, but we may not be able to determine accurately \nhi\ and the metallicity of the cool gas as in this paper. 

In Table~\ref{t-data}, we summarize the redshifts and coordinates of the QSOs, redshifts of the absorbers, spectral resolution $R$ of the data and signal-to-noise S/N in the continuum near the \ovi\ absorption, and the original PI who acquired the data. In the last column of this table, we also list the velocity separation between the QSO and absorber redshifts (for the QSO redshift, we adopted those listed in Simbad and NED). For $\delta v \le 3000$ \km, we define the absorber as ``proximate'', while for $\delta v > 3000$ \km, the absorber is defined as ``intervening''. In the remainder of the paper, we will differentiate the sample that include proximate absorbers from the sample that does not. In Figs.~\ref{f-q1009} to \ref{f-j1211} of the appendix, we show the normalized and  apparent column density profiles of the metal lines for this sample of absorbers. A few of these have already appeared in the literature:  the absorber toward J1211+0422 \citep{lehner08}, Q1009+2956 \citep{simcoe02}, and Q1442+2931 \citep{crighton13} (all of these were observed with Keck HIRES), and Q2206-199 \citep{fox07a} (UVES observations). None of the other \ovi\ absorbers were previously reported in the literature (although for some, the \hi\ and other metal lines may have been analyzed previously as we discuss below). In all cases, the high-ion absorption profiles and metallicities are newly analyzed from the data that we reduced; only for some, estimates of \nhi\ were adopted from previous works (see \S\ref{s-nhi} and the appendix).

\section{Analysis}\label{s-anal}
\subsection{Metal line Absorption}\label{s-ion}
We employed the apparent optical depth (AOD) method described by \citet{savage91} to estimate the total column density of the metal ions and the degree of contamination of the line. The absorption profiles are converted into apparent column densities per unit velocity, $N_a(v) = 3.768\times 10^{14} \ln[F_c(v)/F_{\rm obs}(v)]/(f\lambda)$ cm$^{-2}$\,(\km)$^{-1}$, where $F_c(v)$ and $F_{\rm obs}(v)$ are the modeled continuum and observed fluxes as a function of velocity, respectively, $f$ is the oscillator strength of the transition and $\lambda$ is the wavelength in \AA. The atomic parameters are from \citet{morton03}. In Figs.~\ref{f-q1009} to \ref{f-j1211}, the bottom panels show the $N_a(v)$ profiles of \ovi\ $\lambda$$\lambda$1031, 1037 and \civ\ $\lambda$$\lambda$1548, 1550. Although not shown in the paper, we have also systematically compared the $N_a(v)$ profiles of the strong and weak transitions of \siiv\ and \nv\ to check for either saturation or contamination (as discussed in \S\ref{s-fit}, we also used a profile fitting method, which helped to identify contamination in the individual components).

As the wavelengths of \siiv\ and \civ\ ions are sufficiently offset from \lya, they generally suffer little  contamination from the \lya\ and \lyb\ forests and other lines. However, these ions  have sometimes relatively narrow and strong components, and saturation can be more an issue than for \ovi\ or \nv. On the other hand, while \ovi\ and \nv\ do not have generally strong narrow components, the \lya\ and \lyb\ forests can contaminate these features to a much larger extent. As the sample of \ovi\ was chosen so that there is little contamination in the \ovi\ profile, blending with the \lya\ forest (or other lines) is small (although often not completely absent). However, this is not the case for \nv, where several absorbers have both \nv\ transitions contaminated. In the appendix, we provide more information regarding possible contamination for each absorber, and we refer the reader to Figs.~\ref{f-q1009} to \ref{f-j1211} for the visualization of the AOD profiles and Figs.~\ref{f-fitq1009} to \ref{f-fitj1211} where we show the profile fitting results to \civ, \siiv, and \ovi\ (see \S\ref{s-fit} for more details). For the singly ionized species and \oi, we follow the same procedure, i.e., we always employ all the available transitions (and in particular some of the \oi\ transitions in the far-UV rest-frame of the absorber) to estimate the column density of a given ion or atom to be confident that saturation or contamination does not affect our results. 

The total column density for a specific ion was obtained by integrating over the absorption profile  $N_a = \int N_a(v)dv$. When no absorption is observed for a given species, we measure the equivalent width (and 1$\sigma$ error) over the same velocity range found from a species that is detected (preferably \siiv, but also a singly ionized species). The 3$\sigma$ upper limit on the equivalent width is defined as the 1$\sigma$ error times 3. The 3$\sigma$ upper limit on the column density is then derived assuming the absorption line lies on the linear part of the curve of growth. In Tables~\ref{t-ions} and \ref{t-met}, we summarize our apparent column density estimates of the metals. 

\subsection{\hi\ Absorption}\label{s-nhi}
The determination of the \hi\ column density was made using Voigt profile modeling of the \hi\ Lyman series lines. The exact methodology was dependent on the \nhi\ values.  As a general rule, for the SLLSs and DLAs, the \nhi\ is large enough to create damping wings and extended zero-flux regions in the absorption profiles. The \lya\ transition was, however, generally not used since the damping wings of the SLLSs and DLAs often cover multiple echelle orders in the HIRES data, making an accurate continuum level determination, and thus a constraint on \nhi\ from the extended damping wings, very difficult. The \hi\ \lyb\ line was used to best bound the total \nhi\ for the absorption system, along with additional information from neutral and low ions such as \oi.  The non-zero flux portions of the core region ($\pm 500$ to 1000 \kms\ of line-center, depending on \nhi) of \lya\ in the DLAs and SLLSs were also often used to add additional lower limit constraints modulo blending from other lines from the \lya\ forest. For the LLSs, we used \lya\ along with the higher order Lyman series lines  to constrain \nhi. In the appendix, we discuss for each absorber how \nhi\ is determined or provide the references if \nhi\ was adopted from previously published works. The results are summarized in  Table~\ref{t-ions} and subsequent tables. 

\subsection{Metallicity}\label{s-metal}
The metallicity is critical for determining the chemical enrichment level and the ionized fraction of the gas (since the ionized gas is only observed via the metal lines). In order to estimate the metallicity of the cool/warm gas traced predominantly by the singly ionized and neutral species, we compare their total column densities with \nhi. For the velocity profiles with very few components and a small velocity-width extent, this approximation is likely to hold. However, for more complicated profiles that spread over $\Delta v > 200$ \km, this may not always hold as shown, e.g., for the absorber at $z=3.54996$ toward Q2000-330  where the metallicity varies from $[{\rm X/H}] \equiv \log N_{X}/N_{\rm H} - \log ({\rm X/H})_\sun =-2.1$ dex in the LLS at $+140$ \km\ to $-0.61$ at 0 \km\ \citep[][as we discuss later, we will adopt the results from this study to estimate the \nhii]{prochter10}. In this work, we assume the solar heavy element abundances from \citet{asplund09}. 

For the DLAs, the singly ionized species are mostly associated with the neutral gas, and we can estimate the metallicity by using \sii\ or \siii\ with no ionization correction \citep{prochaska96}. For the SLLSs and LLSs, we systematically use \oi\ to estimate the metallicity as \oi\ is an excellent tracer of \hi\ since its ionization potential and charge exchange reactions with hydrogen ensure that the ionizations of \hi\ and \oi\ are strongly coupled as long as the ionization spectrum is not too hard and/or the densities too low \citep{jenkins00,lehner03}. We did test if ionization corrections were needed using Cloudy photoionization models \citep[version c13.02][]{ferland13}. In these models, we assume the gas is photoionized, modeling it as a uniform slab in thermal and ionization equilibrium. We illuminate the slab with the  Haardt-Madau EUVB radiation field from quasars and galaxies (HM05, as implemented within Cloudy). For each absorber, we a priori set the metallicity to that estimated from [\oi/\hi] and vary the ionization parameter, $U =n_\gamma/n_{\rm H}=$\,H-ionizing photon density/total hydrogen number density (neutral + ionized) to search for models that are consistent with the constraints set by the total column densities determined from the observations (in particular, the column densities of \oi, \siii, \siiv, \alii\ and other available ions for a specific absorber). In Table~\ref{t-met}, we summarized the metallicity without and with ionization correction, and except for the LLS toward Q1009+2956, the ionization corrections are small. 

Our adopted metallicities are summarized in the last column of Table~\ref{t-met}. The main uncertainties are from the \nhi\ determination and the assumption of constant metallicity throughout the velocity profiles (the former is included in the error estimates, but not the latter).  The spread in metallicity is quite large, with  metallicities ranging from $<0.3\%$ solar to about solar metallicity. This spread is observed in the LLSs and SLLSs; for the DLAs, the spread is smaller and consistent with the DLA metallicity distribution over a similar redshift interval derived from a much larger sample \citep[e.g.,][]{rafelski12}. 

\subsection{Profile Fitting}\label{s-fit}
Assessing the temperature of the gas probed by the high ions is critical for understanding the physical conditions that govern this gas. To obtain that information, we need to fit the individual components observed in the absorption profiles, which gives the Doppler broadening of the individual components. The Doppler parameter $b$ is a direct estimate of the temperature since $b^2=b^2_{\rm th} + b^2_{\rm nt}= 2 k T/(A m_p)+b^2_{\rm nt}$, where $A$ is the atomic weight and $m_p$ the proton mass, $k$ is Boltzmann's constant, $T$ is the temperature,  and $b_{\rm nt}$ is the non-thermal contribution to the broadening. The Keck observations  have enough resolution (4 to 8 \km\ FWHM, see Table~\ref{t-data}) to study the profiles of the high ions in detail. This contrasts from our Milky Way where \civ, \siiv, and \nv\ have been observed at high resolution, but \ovi\ observed only at 20 \km\ resolution \citep{bowen08,lehner11}.  

In order to separate an absorption profile into individual components, we used the method of component fitting which models the absorption profile as individual Voigt components. We used a modified version of the software described in \citet{fitzpatrick97}.  The best-fit values describing the gas were determined by comparing the model profiles convolved with an instrumental line-spread function (LSF) with the data.  The LSF were modeled as a Gaussian with a FWHM derived from the $R$-value listed in Table~\ref{t-data}.  The three parameters $N_i$, $b_i$, and $v_i$ for each component, $i$, are input as initial guesses and were subsequently varied to minimize $\chi^2$.  The fitting process enables us to find the best fit of the component structure using the data from one or more transitions of the same ionic species simultaneously.

We applied this component-fitting procedure to the \civ, \siiv, and \ovi\ doublets. All the ions were fitted independently  (i.e., we did not assume a common component structure for all the ions), but both lines of the doublet are simultaneously fitted.  We always started each fit with the smallest number of components that reasonably modeled the profiles, and added more components as necessary.  In this manner, if the addition of a particular component did not improve  substantially the $\chi^2$ goodness of fit parameter, we adopted the solution without this additional component.  We allowed the software to determine the components freely (i.e., we did not fix any of the input parameters). However, we did not allow the software to fit unrealistically broad component (typically a component with $b_i\ga 90$ \km\ was not allowed); in the cases where the software fitted broad components, we checked systematically our continuum placement and/or added more components.

The results for the component fitting of \siiv, \civ, and \ovi\ are given in Table~\ref{t-fit}, and the component models can be seen in Figs.~\ref{f-fitq1009} to \ref{f-fitj1211} of the appendix where both the individual components and global fits are shown. We stress that the errors summarized in this table reflect only those derived by the profile fitting procedure for the adopted solution. In complicated profiles, where several components are blended, the fit results may not be unique, and in particular some broad components with $b\ga 40$ \km\ may sometimes compensate for our inability to decipher the true velocity structure. For example, when the profiles are complex, a very broad component may be fitted principally to reduce the $\chi^2$  while several narrower components could be more adequate. In order to test the robustness and uniqueness of the results, two of us (Burns and Lehner) independently profile fitted all the profiles shown in Figs.~\ref{f-fitq1009} to \ref{f-fitj1211} using the same program and methodology. Using these independent results, in the last column of Table~\ref{t-fit}, we flag the results either as reliable (flag `0', similar results in the independent fits) if there is an agreement or as less reliable (flag `$>0$') if there is a disagreement or there is saturation in one of the components  (see the footnote of Table~\ref{t-fit} for the detailed flag description). 

For non-saturated profiles, there is a  good agreement for the total column densities between the AOD and profile fitting methods. In saturated regions of the profiles (where the flux reaches zero), it is impossible to accurately model these regions because there is no way to determine the correct number of components. However, in these cases, $b$ cannot be larger in the saturated components, and therefore the output $b$-value is still a robust upper limit (but as $N$ directly depends on $b$ in the saturated regime,  $N$ cannot be considered a lower limit; in this case the AOD method gives a more robust lower limit).

 We note that while each species was fitted independently, some of the components seen at similar velocities in the various species may trace similar gas. Therefore, a posteriori, we can compare the results between the high ions to further support (or not) the results from the profile fits. In the cases where there is some doubt, we did not revisit the fits because in order to do so we would need to alter our rule of minimum number of fitted components with no statistical improvement in the $\chi^2$ (and we would actually have little constraints to define any additional components). In a future KODIAQ paper with a larger sample of absorbers, we will revisit some of these issues. 

\section{Characteristics of the \ovi\ Gas in High-\lowercase{z} Galaxies and their CGM}\label{s-prop}

Among the questions we want to address for the \ovi\ associated with $\tau_{\rm LL}>1$ \hi\ absorbers are: 1) are they the same population of \ovi\ as observed in \lya\ forest absorbers?; 2) do they have analogs in the low redshift universe?; 3) do they trace all the same phenomena or a mixture of phenomena (e.g., shocked gas in outflows from galaxies, shocked infall on large-scale structure)?; 4) are they collisionally ionized or photoionized? 5) are they hot or cool?; and 6) how many baryons and metals are hidden in the \ovi\ plasma? 

 In order to address these questions we need to review the integrated properties (total column density, profile width) as well as the properties of the individual components determined from the Voigt profile fitting of the absorption profiles. Constraining the properties of the high ions is not only critical for understanding their origin(s), but also for estimating the cosmic baryon and metal budgets at these redshifts as their cosmic densities are directly inversely proportional to the ionization fraction.

\begin{figure}[tbp]
\epsscale{1.2} 
\plotone{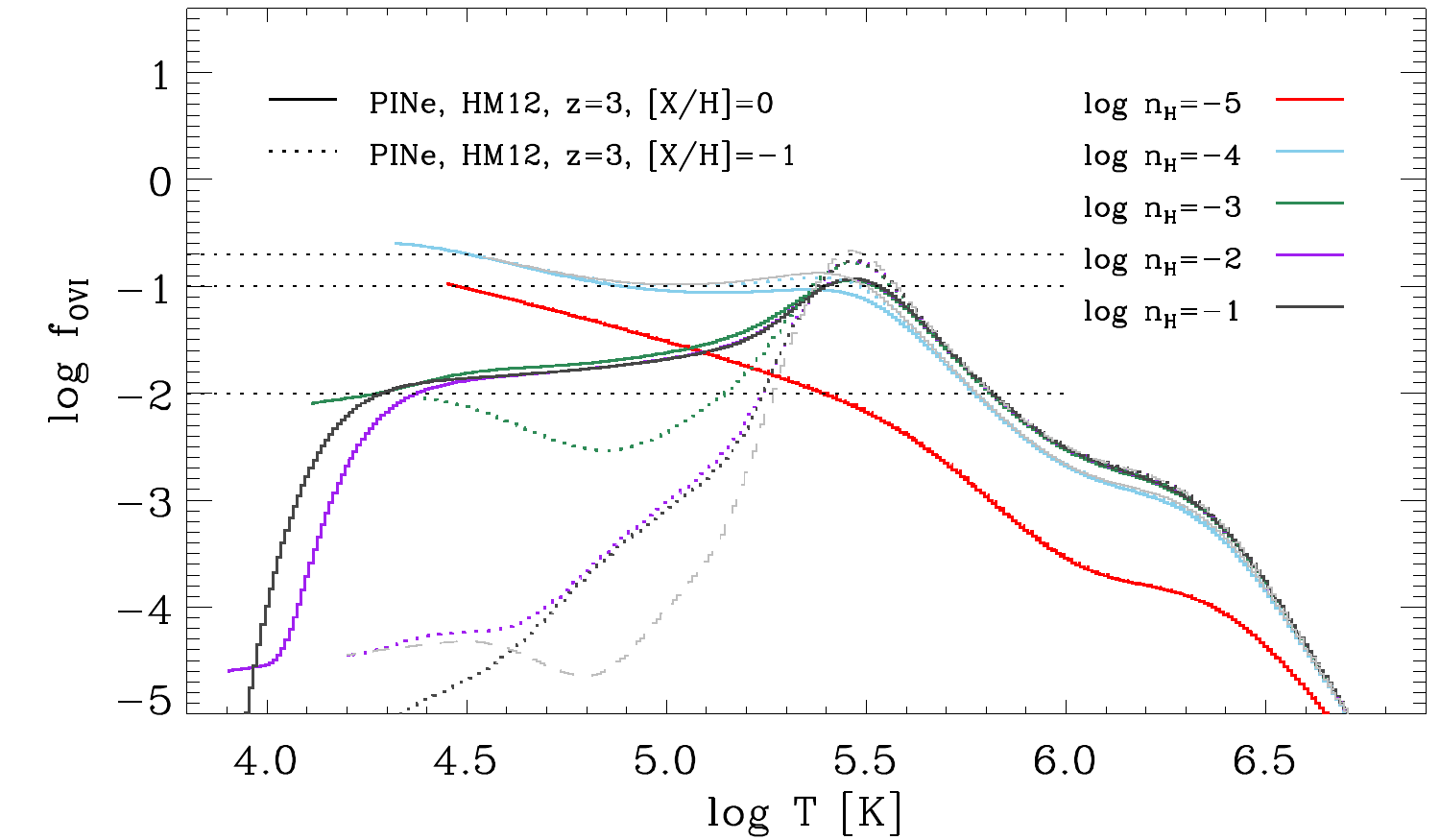}  
  \caption{Ionization fraction of \ovit\ as a function of the temperature in an isochoric, radiatively cooling diffuse gas in the presence of a \citet{haardt12} extragalactic background at $z=3$ for different densities and two metallicities according to the models of \citet{oppenheimer13}. The solid and dashed gray lines are CIE calculations in the presence of the same extragalactic background at $z=3$ for $\log n_{\rm H} =-4$ and $-2$, respectively (neither photoionization nor CIE is dependent on the metallicity). The horizontal dotted lines highlight some specific ionization fractions. 
 \label{f-model}}
\end{figure}

\subsection{Ionization Constraints on the Fraction of \ovi}\label{s-ovi}

One of the challenges in constraining the ionization fraction of \ovi\ and other high ions is that they  probe low-density gas with temperatures where their abundances peak that are significantly less stable than at higher or lower temperatures. Gas at $0.5 \times 10^5 <T<3\times 10^5$ K with near solar metallicities cools very rapidly since free thermal electrons are able to excite the valence electrons into the upper states of the strong resonance transitions of these same high ions. The subsequent rapid removal of energy from the gas through spontaneous photon emission cools the gas much more rapidly than it can typically recombine. Thus, low-density plasmas at these temperatures are not likely to be in a state of collisional ionization equilibrium (CIE), especially if the gas has high metallicity. 

To illustrate this for the \ovi, we show in Fig.~\ref{f-model} predictions for the ionization fraction of \ovi, $f_{\rm OVI}$, made from recent CIE and isochoric radiative cooling models by \citet{oppenheimer13}  \citep[and see also][]{gnat07,vasiliev11},  where the gas is also exposed to photoionization by the EUVB radiation calculated for different gas densities with a solar and tenth solar metallicity. The CIE and non-equilibrium ionization (NEI) fraction of \ovi\ peaks around $f_{\rm OVI}\simeq 0.2$ and $0.1$, respectively, at $T_p \simeq 3 \times 10^5$ K (for \nv, \civ, and \siiv, the peak abundance temperatures are $2\times 10^5$, $10^5$, $0.7\times 10^5$ K, respectively). This value provides a strict upper limit on $f_{\rm OVI}$. At very low densities, $\log n_{\rm H}\la -5$, the \ovi\ production is dominated by the extragalactic background, but as the density increases, collisional ionization dominates the production of \ovi\ at $T\ga 2\times 10^5$ K (which corresponds to $b_{\rm OVI}\ga 15$ \km\ for purely thermal broadening); at $T>10^6$ K, $f_{\rm OVI}< 0.003$ or about 50--60 times less than its peak ionization. At $T<2\times 10^5$ K, photoionization by the EUVB radiation is only efficient for $\log n_{\rm H}\la -3.5$; for higher densities ($\log n_{\rm H}< -3$ and $T\ga 2\times 10^5$ K), the gas must be solar metallicity or above in order to have $0.01<f_{\rm OVI}< 0.1$, otherwise $f_{\rm OVI}\ll 0.001$. The ionization of the \ovi\ (and similarly for the other high ions) can therefore be well below from their peak abundances, especially if the gas is relatively high metallicity or hot.

\begin{figure}[tbp]
\epsscale{1.2} 
\plotone{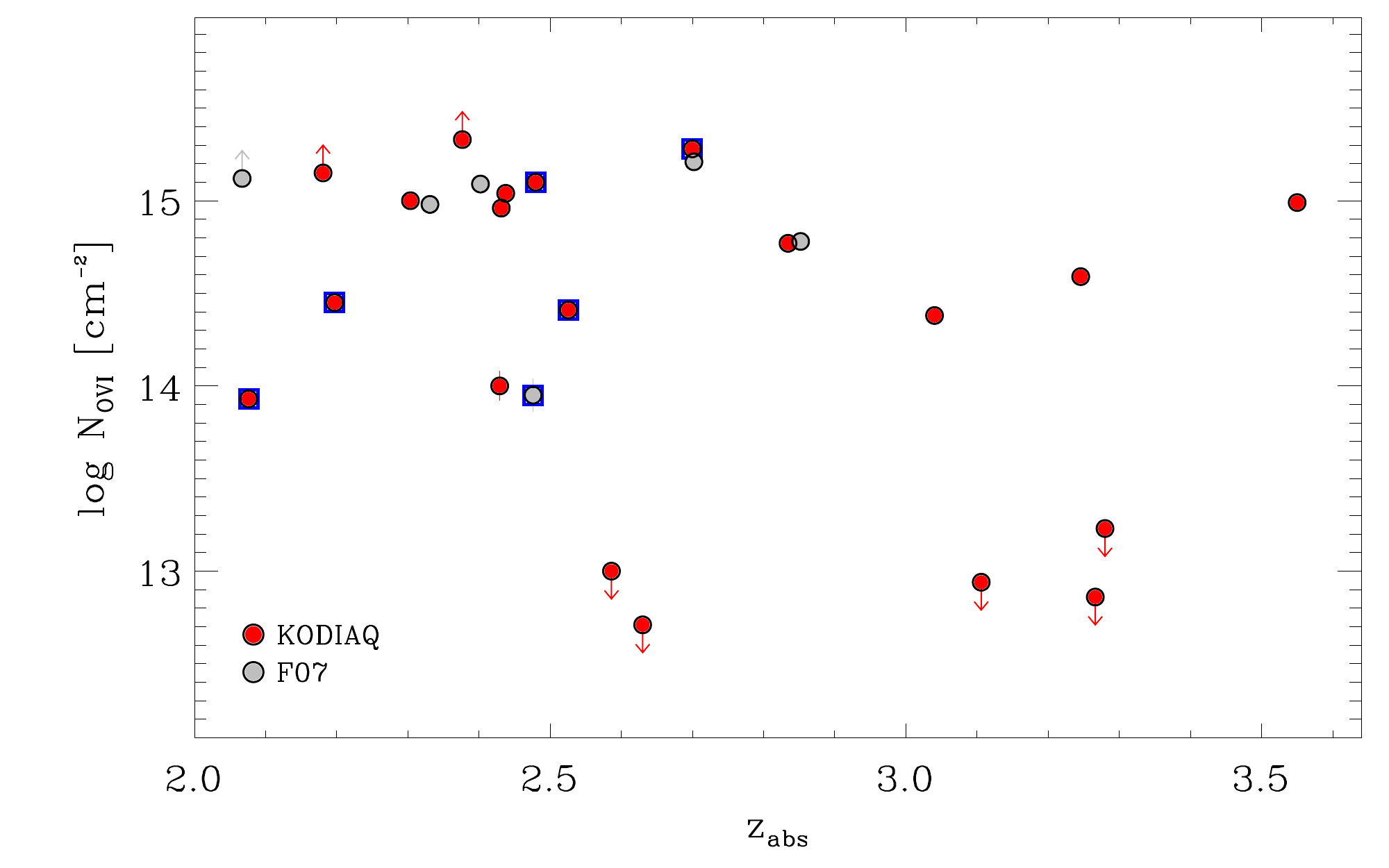}  
  \caption{Total column density of \ovit\ associated with optical thick absorbers as a function of the redshift of the absorber. Circles with overplotted blue squares highlight the proximate systems.  When the $1\sigma$ error is not apparent, it is less than the size of the circle.   \label{f-zno6}}
\end{figure}

\begin{figure*}[tbp]
\epsscale{1} 
\plottwo{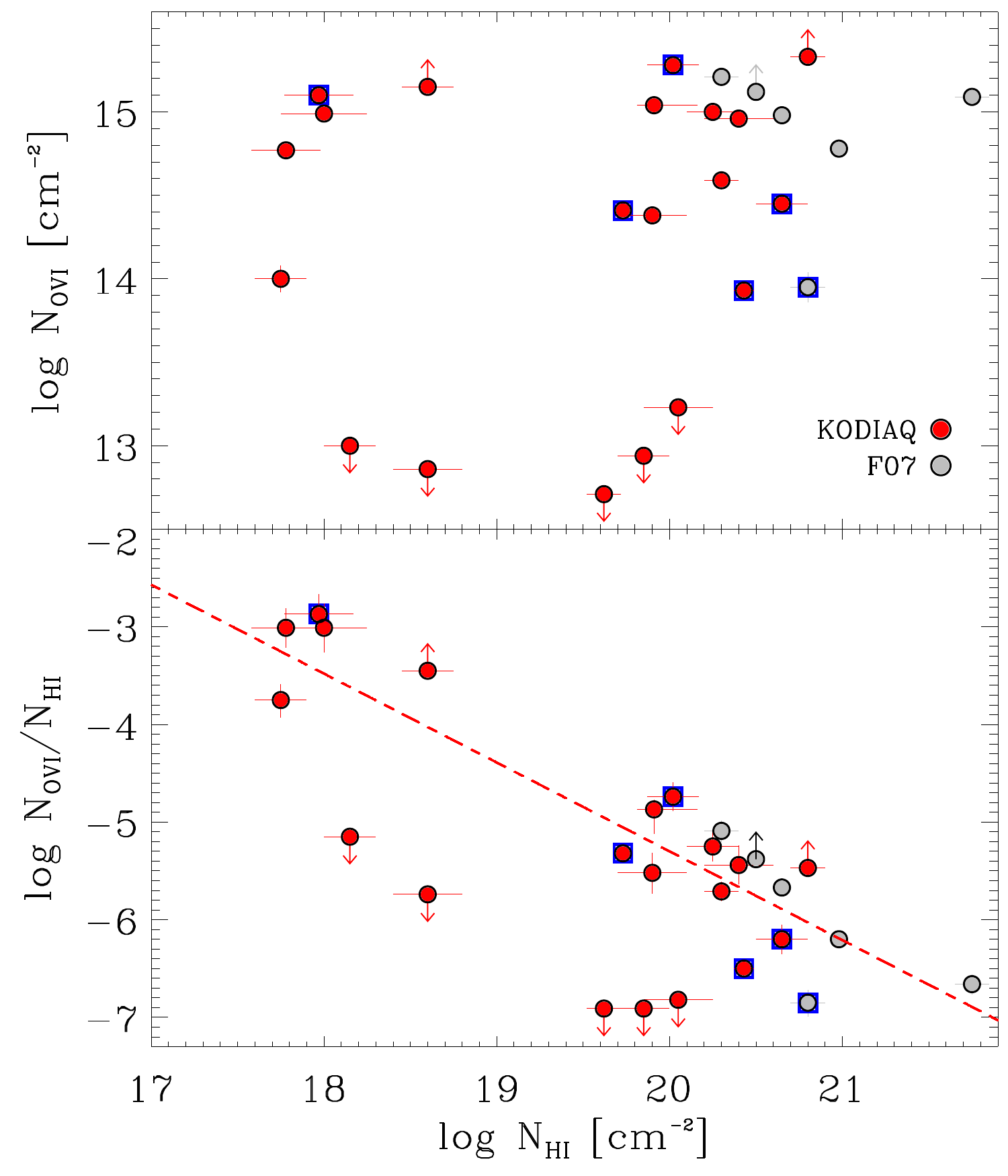}{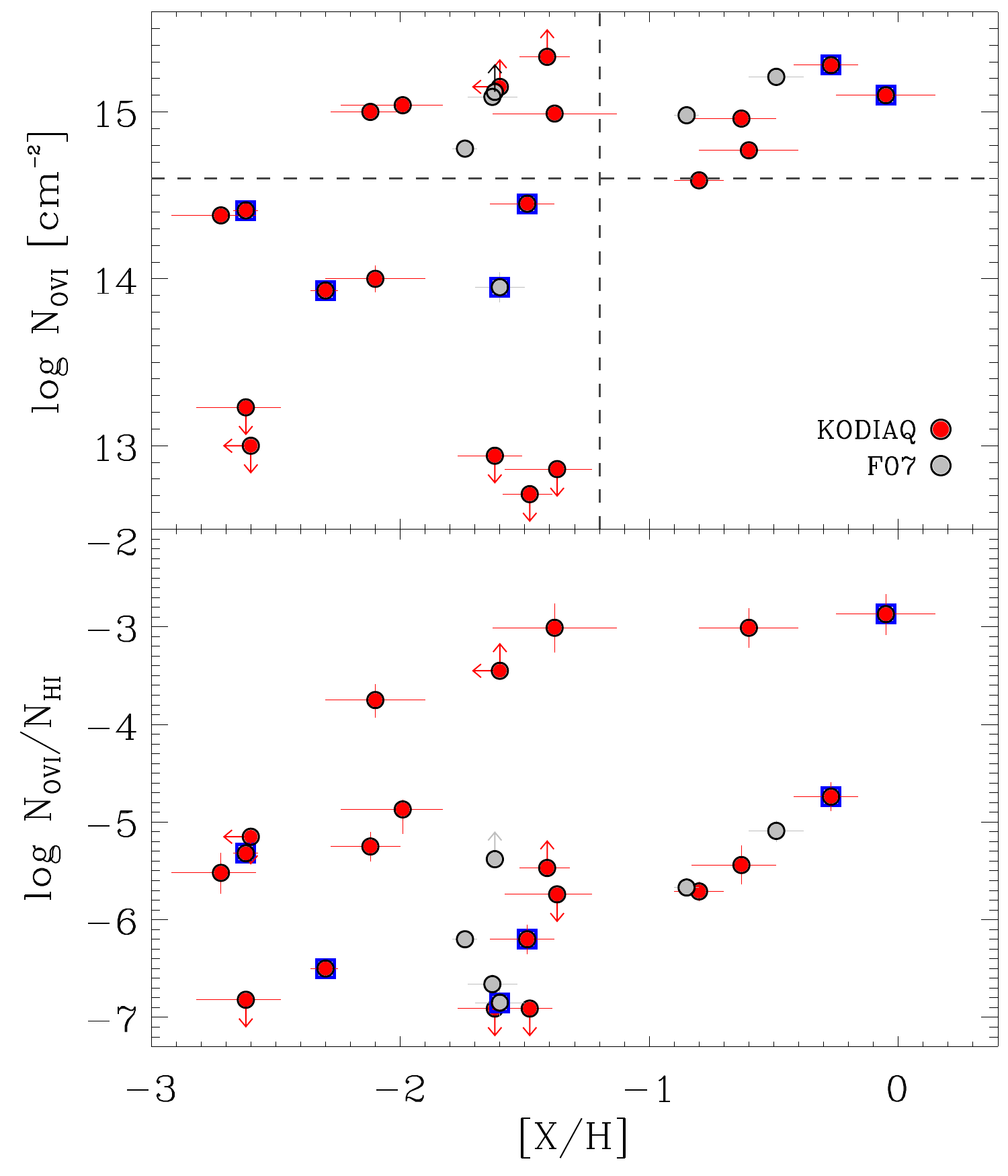}  
  \caption{{\it Left:}\  Total column density of \ovit\ and \novi/\nhi\ vs. \nhi. The dashed line in the bottom panel is the fit derived for \ovit\ absorbers associated with \lya\ forest absorbers ($\mlnhi \la 16$) at similar redshifts  from \citet{muzahid12}. {\it Right:}\ Total column density of \ovit\ and \novi/\nhi\ vs. the metallicity of the cool gas. Circles with overplotted blue squares highlight the proximate systems.  When the $1\sigma$ error is not apparent, it is less than the size of the circle. 
 \label{f-no6met}}
\end{figure*}

\subsection{General Properties from the Integrated Profiles}\label{s-int}

The detection rate of the \ovi\ absorption in the KODIAQ sample is $15/20$ ($64$--$83$\% assuming a Wilson score with a 68\% confidence interval). For the LLSs, the \ovi\ detection rate is 5/7 ($53$--$85$\%), 5/8 for the SLLSs ($45$--$77$\%), and 5/5 ($83$--$100\%$) for the DLAs; combining the KODIAQ and F07 samples yields a 11 detections of \ovi\ in 11 DLAs,  i.e., a detection rate $>92\%$ at the 68\% confidence level.  All the upper limits on \novi\ are well below the detection threshold, typically $\la 13$ dex, which is at the level or less than that observed in the local interstellar medium \citep{savage06}. When \ovi\ is not detected, 1) \civ\ is not detected either (for the absorbers where there is coverage of \civ), 2) \siiv\ can be present (there are 2 non-detections and 2 detections of \siiv\ when \ovi\ is not detected), and 3) the velocity structure in \oi\ and singly ionized species is simple (1 or 2 components). However, a simple velocity structure does not necessarily imply the absence of high-ion absorption (see, e.g., Fig.~\ref{f-q1243} or \ref{f-j0831}). 

 When \ovi\ is detected, \lnovi\ varies from 13.93 to $>15.33$ (see Figs.~\ref{f-zno6} and \ref{f-no6met}), i.e., the \ovi\ absorption is quite strong  even though weak \ovi\ absorbers are quite typical in \ovi\ absorbers  associated with $\mlnhi < 17$ absorbers  \citep{simcoe02,muzahid12}. In fact, in the KODIAQ+F07 sample, when \ovi\ is detected, \novi\ is remarkably large: 86\% (18/21) of the absorbers have $\mlnovi \ga 14.4$, 65\% (13/21) have $\mlnovi \ga 14.8$, and on average $\langle N_{\rm OVI}  \rangle = 10^{14.9 \pm 0.3}$  cm$^{-2}$. There is no apparent difference in the total column density of \ovi\ between the proximate and intervening absorber samples as displayed in Fig.~\ref{f-zno6}. 

\begin{figure}[tbp]
\epsscale{1.2} 
\plotone{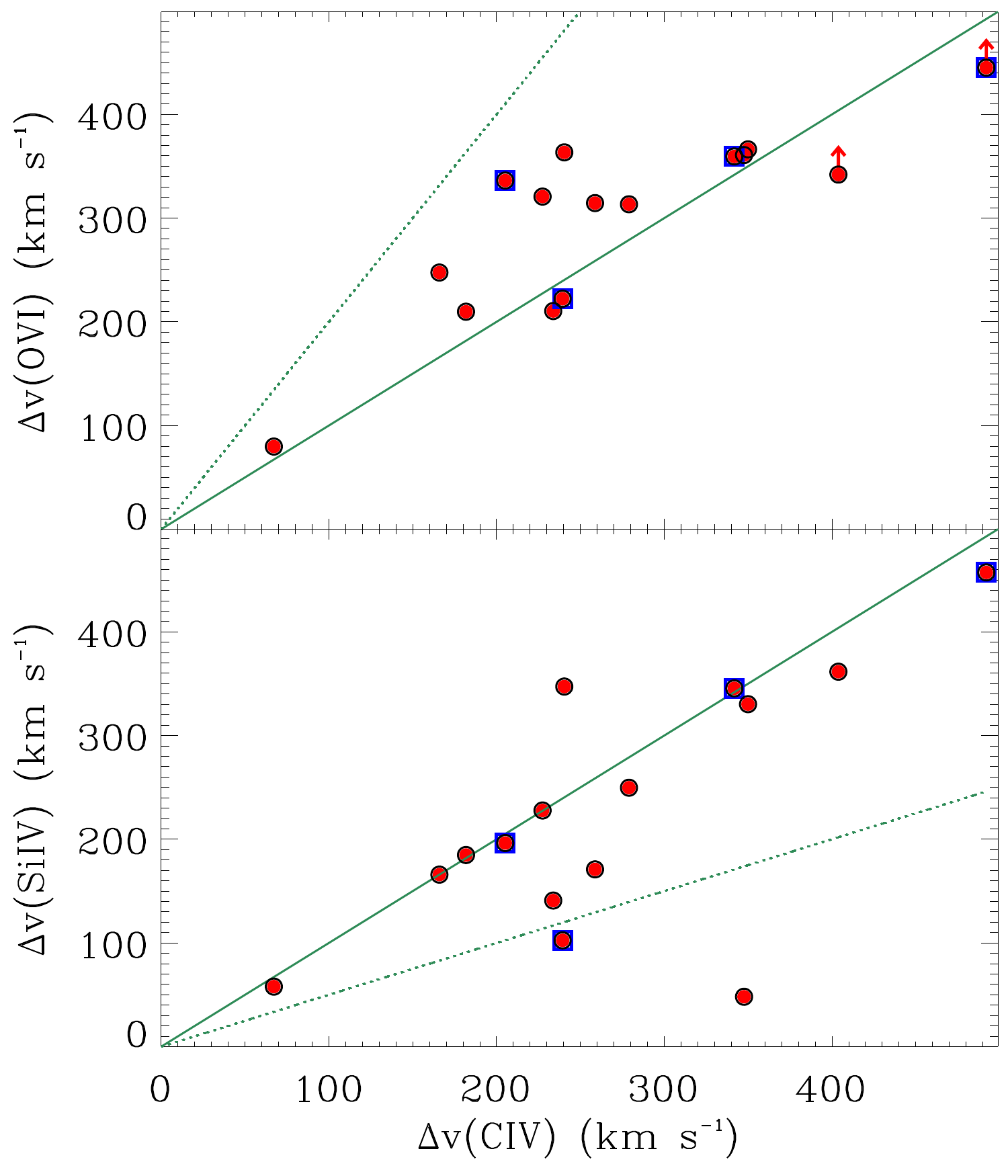}   
  \caption{Full-width comparison of the high ions. The solid green line shows a 1:1 relationship, while the dotted line is a 2:1, 1:2 relationship in the top and bottom panels, respectively. A data point with an up arrow highlight an absorber where the \ovit\ profile is contaminated at one of the edges where \civt\ absorption is observed. The circles with overplotted blue squares highlight the proximate systems. 
 \label{f-fw}}
\end{figure}

In the KODIAQ sample, the reason for the large total \ovi\ column densities is that the \ovi\ profiles typically extend over $200$ \km: we compare the full-widths, $\Delta v$, of the \ovi, \siiv, and \civ\ profiles in Fig.~\ref{f-fw}.\footnote{The full-width is estimated from the profile fitting results such as $\Delta v =  \max(v_i + 2 \sqrt{\ln 2}\, b_i ) - \min(v_j - 2 \sqrt{\ln 2}\,  b_j)$, where the first term corresponds to the highest velocity and the second term the lowest velocity,  or  $\Delta v =  4 \sqrt{\ln 2} \, b$ if only a single component was fitted.} The average and dispersion of $\Delta v$ for \ovi, \civ, and \siiv\ are $299 \pm 90$, $269 \pm 104$, and $ 225 \pm 121$ \km, respectively. Except for two data points (excluding the two where \ovi\ is partially contaminated), $\Delta v_{\rm OVI}> \Delta v_{\rm CIV}$, while for \civ\ and \siiv, about half the sample has $\Delta v_{\rm SiIV} \simeq \Delta v_{\rm CIV}$ and the other half $\Delta v_{\rm SiIV} < \Delta v_{\rm CIV}$.\footnote{The data with  $\Delta v_{\rm SiIV} \gg \Delta v_{\rm CIV}$ corresponds to the $z=2.43749$ absorber toward Q1442+2931. The most negative velocity component of \siivt\ is not observed in \civt, but where strong absorption of \oit\ and singly ionized species is observed (see Figs.~\ref{f-q1442} and \ref{f-fiths0757} and Table~\ref{t-fit}).}  This is again quite different from the results of \ovi\ absorbers not selected for strong \hi\ where the line-spreads of \ovi\ are smaller with  a median of 66 \km\ (with typically $\Delta v \la 100$--$150$ \km) and $\Delta v_{\rm OVI} \sim \Delta v_{\rm CIV}$ \citep{muzahid12}. However, the \ovi\ properties in the KODIAQ sample are quite similar to those derived from the stacked DLA SDSS spectrum at $z\sim 2.9$ \citep[ $ N_{\rm OVI} \simeq 10^{14.6}$ cm$^{-2}$ and  $\Delta v_{\rm OVI}> \Delta v_{\rm CIV} > \Delta v_{\rm SiIV}$, see][and also \citealt{fumagalli13} who show strong \ovi\ absorption in the composite spectrum of 38 $\tau\ge 2$ LLSs]{rahmani10}.

In the local universe, \ovi\ absorption with  $\mlnovi \ga 14.4$ is only observed in the SMC or starburst galaxies, and with $\mlnovi \ga 14.8$ only in starburst galaxies \citep{heckman02,grimes09}. The strength of the \ovi\ absorption in the halos of galaxies and its relation with the level of star formation in the galaxies is demonstrated by  \citet{tumlinson11} who found at $z<0.5$ that the dichotomy between star-forming and passive galaxies strongly affect the \ovi\ column density in their CGM within at least 150 kpc: star-forming galaxies have typically  $\mlnovi \ga 14.5$ while more passive galaxies have $\mlnovi \ll 14.5$ in their CGM. The total velocity interval where high-ion absorption is observed is also similar to the breadths of the \ovi\ profiles observed in local starburst galaxies \citep{heckman01,grimes09} despite the different geometries. The large \ovi\ velocity breadths and column densities therefore strongly suggest that the \ovi\ absorption associated with $\tau_{\rm LL}>1$ absorbers at $z>2$ is strongly linked to star-forming galaxies.

There is no trend between the amount of \ovi\ and the redshift or \nhi\ as displayed in Figs.~\ref{f-zno6} and \ref{f-no6met}. The strong anti-correlation between \novi/\nhi\  and \nhi\ confirms that \novi\ and \nhi\ are not correlated. This  anti-correlation extends with a similar slope to the \lya\ forest at similar redshifts \citep[see][and see dashed line in the bottom left panel of Fig.~\ref{f-no6met}]{muzahid12}.  

On the right-hand side  of Fig.~\ref{f-no6met}, we show \novi\ and \novi/\nhi\ as a function of the metallicity [X/H] of the cool gas. There is no evidence for a relation between  \novi/\nhi\ and [X/H]. There is a moderate correlation between \novi\ and [X/H] according to the rank Spearman test ($r=0.5$) significant at the 97\% level for the KODIAQ+F07 excluding the upper limits. However, what seems more remarkable is the apparent division at $[{\rm X/H}] = -1.2$: For  $[{\rm X/H}] \la -1.2$,  a large scatter in \novi\ is observed (spanning a factor $>100$ from $\mlnovi < 13 $ and $\mlnovi >15.4$), while  for $[{\rm X/H}] \ga -1.2$, there are only strong \ovi\ absorbers with  $14.7 \la \mlnovi \la 15.3$. If our interpretation above is correct,  high-$z$ metal-enriched galaxies are actively forming massive stars, enriching their CGM with large quantity of \ovi\ (and hence metals).

\subsection{General Properties from the Velocity Profiles}\label{s-prof}

The typically large velocity breadth and column density of \ovi\ suggests that the \ovi\ absorbers associated with $\tau_{\rm LL}>1$ absorbers at $z>2$ are tracing large-scale feedback of active star-forming galaxies. We want now to determine if there is any relationship between \ovi\ with the other high ions and singly ionized species and how the kinematic profiles compare between the high ions. As we selected our sample based on the small contamination level of the \ovi\ absorption by the \lya\ and \lyb\ forests  and the spectral resolution is the same for all the ions for a given absorber, we can directly compare the \siiv, \civ, and \ovi\ profiles to assess for broad similarities or dissimilarities.\footnote{We do not systematically use \nvt\ in our comparison because it is detected for only 25\% of the sample because the lines are either contaminated or too weak owing to the N nucleosynthesis history.} 

We first consider four absorbers that highlight some of the overall similarities and differences between the high-ion profiles: 

 {\it - Absorber $z=2.42903$ toward Q1009+2956 (see Figs.~\ref{f-q1009} and \ref{f-fitq1009}):}\  \civ\ and \siiv\ have both narrow and broad components and very similar component structures. A similar velocity structure is observed in the singly ionized species except for the component at $-167$ \km, which is only observed in \civ\ and \siiv. On the other hand, the \ovi\ profiles differ entirely from all the other ions: the overall \ovi\ profile is smoother, in particular where the main absorption is observed in \civ\ and \siiv\ between $-80$ and $+80$ \km; the \ovi\ profile is more extended at positive velocities; and even where the narrow \ovi\ absorption is observed at $-71$ \km, there is surprisingly no corresponding absorption in \civ\ and \siiv. Hence the bulk of \civ\ and \siiv\ as well as the singly ionized species trace gas that is kinematically related. Our Cloudy simulations (see \S\ref{s-metal}) show that \siiv\ and \siii\ can be produced by photoionization by the EUVB radiation, but falls short to produce enough \civ\ by a factor 4. Some other ionization sources may include a harder ionizing source \citep[e.g., from the interface of a hot gas; see][]{knauth03,borkowski90}, or photoionization of a very low density gas \citep{oppenheimer13}, or possibly turbulent mixing layers \citep{slavin93,kwak10}.  \ovi\ must, however, trace a much different type of gas that is hotter ($T \la 2\times (10^5$--$10^6)$ K implied from the individual $b$-values but one --- the $-71$ \km\ component has $T<3\times 10^4$ K) and/or much more turbulent. No \nv\ is observed in this absorber. 

 {\it -Absorber $z=2.83437$ toward J1343+5721 (see Figs.~\ref{f-j1343} and \ref{f-fitq1009}):}\  For this absorber, the situation is quite different with no evidence of narrow components in the high ions that align with the cooler gas. The \civ\ and \siiv\ profiles have again matching components, with the bulk of the absorption shifted by $-131$ and $-70$ \km\ from the strongest absorbing component in the singly ionized species. However, while the velocities of  the \ovi\ and \civ\ profiles at $v<-50$ \km\ overlap, the bulk of the \ovi\ absorption is at $-28$ \km. In this absorber, the high ions trace gas that is hot and/or kinematically disrupted gas, but there is a large variation in the ionic ratios of \ovi\ to \civ\ (or \siiv), implying changes in the physical conditions of the gas. In this case \nv\ $\lambda$1238 is detected and free of contamination and its profile structure follows that of \ovi.  

  {\it - Absorber $z=3.04026$ toward HS0757+5218 (see Figs.~\ref{f-hs0757} and \ref{f-fiths0757}):} \ The bulk of the absorption for the low-ions and \oi\ is between $-30$ and $+30$ \km, while for the high ions, it is between $-150$ and $-30$ \km, so again the components of the high and low ions do not align. There is again some similarity between \civ\ and \siiv\ and the profiles of these ions show a larger number of velocity components than observed in the \ovi\ profile. However, the AOD comparison also shows that the broad \ovi\ component is also visible  in the \civ\ profile. So in this case, the peak optical depth of the absorption in the high ions is largely shifted from that of the low ions, but \ovi\ and some of the \civ\  trace hot and/or kinematically disrupted gas.  In this case, there is no detection of \nv.

    {\it - Absorber $z=2.52569$ toward Q1243+3047 (see Figs.~\ref{f-q1243} and \ref{f-fitq1217}):} \ All the absorbers described above are intervening, while this SLLS is a proximate absorber. The low ions probes relatively cool and very low metallicity ($0.2\%$ solar) gas \citep[see also][]{kirkman03}. The bulk of the observed absorption in the high-ion profiles is again displaced by about 40 to 120 \km\ relative to low-ion absorption.  There is, however, an excellent match between the \civ\ and \ovi\ (and \siiv\ profiles) even though there is more structure visible in the \civ\ and \siiv\ profiles.  So in this case, all the high ions trace gas governed by similar physical conditions. This better match between \civ\ and \ovi\ is also observed in 3/5 proximate absorbers in the KODIAQ sample.  There is also an intervening absorber ($z=2.18076$ toward Q1217+499) where all the ion profiles follow each other quite well, but in this case, all the high-ion profiles (including \nv) are saturated at some level (see Fig.~\ref{f-q1217}).
  
From this description and considering the entire sample, we can draw additional properties of the highly ionized gas probed by \siiv, \civ, \ovi\ associated with strong \hi\ absorbers at $2<z<4$: 

$\bullet$ While the high-ion absorption profiles often overlap with those of the low ions or \oi, the high-ion profiles have typically a larger velocity breadth with the bulk of their absorption often offset from the low ions by several tens or hundreds of kilometers per second. This is similar to the observations of local starbursts where the kinematics of the hot gas as traced by the \ovi\ absorption are different from that seen in the cooler gas \citep{grimes09}, and therefore consistent with the idea the high ions probe  outflowing gas or are strongly influenced by large-scale stellar feedback. 
\\
$\bullet$  While the velocity breadths are smaller for \civ\ and \siiv\ than \ovi, a larger number of individual components is found in \civ\ and \siiv, or stated differently, the \ovi\ profiles are generally smoother than those of \civ\ and \siiv.   From the profile fitting results, we estimate that on average the number of components in the \civ\ and \siiv\ profiles are about the same, but is about 1.8 times the number of \ovi\ components.\footnote{For the intervening systems, on average the number of \civt\ components is twice the number of \ovit\ components compared with 1.3 for the proximate systems; in both intervening and proximate systems, the numbers of \civt\ and \siivt\ components are the same.} A direct consequence of this is that on average, the $b$-values of the individual components increases with the ionization potential of the high ion (see also \S\ref{s-bval}).

\begin{figure}[tbp]
\epsscale{1.2} 
\plotone{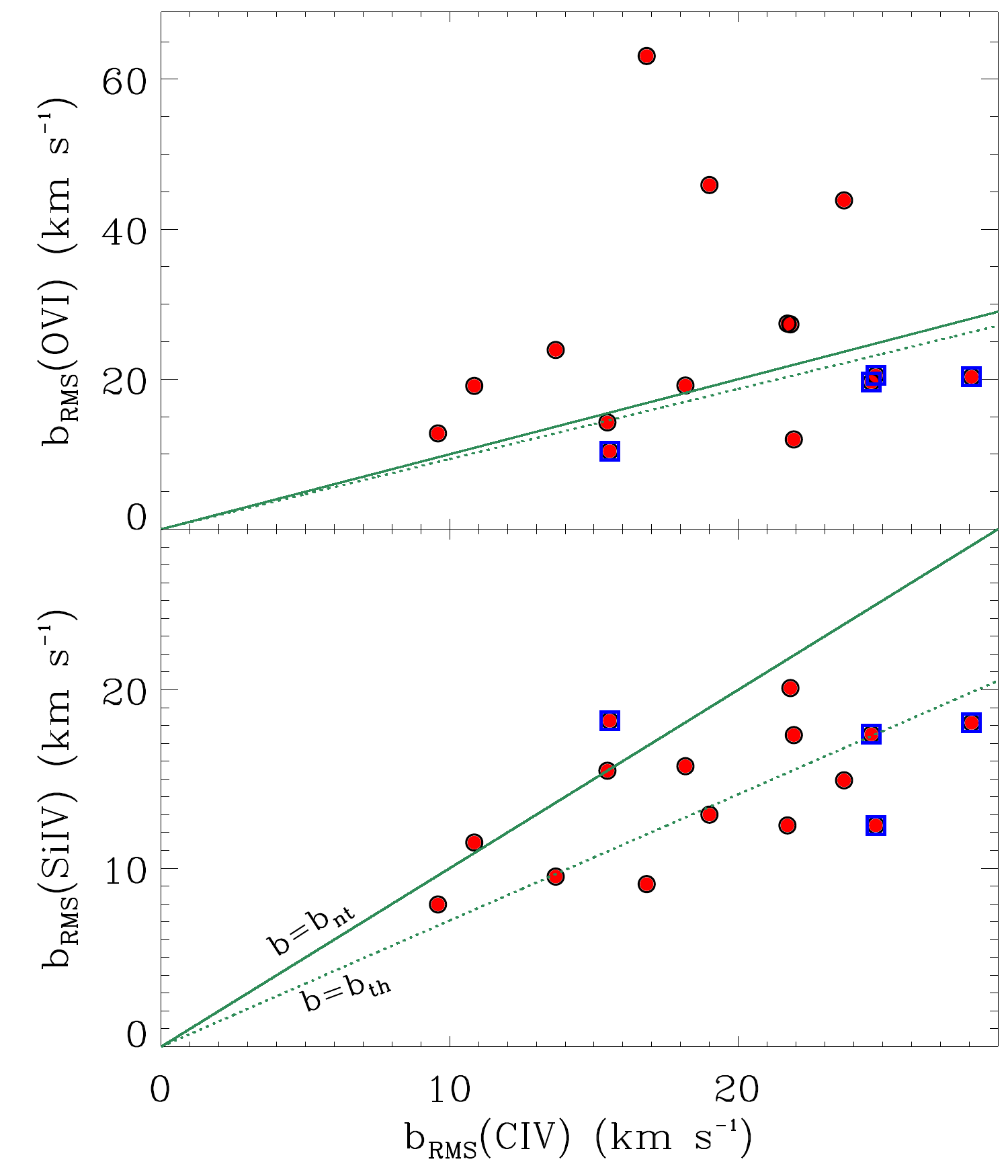}  
  \caption{Comparison of the RMS $b$-values of the high ions. $b_{\rm rms}$ is determined from the profile fit results for a given ion of a given absorber.  The solid line shows a 1:1 relationship, which corresponds to the case where non-thermal motions dominate the profile broadening. The dotted line corresponds to the broadening dominated by thermal motions. The circles with overplotted blue squares highlight the proximate systems. 
\label{f-fwhm}}
\end{figure}

In Fig.~\ref{f-fwhm}, we compare the RMS of individual $b$-values of the high ions for each absorber: for the intervening systems, all but one absorber have $b_{\rm OVI}^{\rm rms} > b_{\rm CIV}^{\rm rms}$, while for the proximate systems $b_{\rm OVI}^{\rm rms} \le b_{\rm CIV}^{\rm rms}$. For \civ\ and \siiv, there is no difference between the proximate and intervening absorbers, and for all but one absorber $ b_{\rm SiIV}^{\rm rms} \la b_{\rm CIV}^{\rm rms}$; in this case, the larger difference in the atomic weight between Si and C better separates the non-thermal and thermal broadening contributions, and since many components of \civ\ and \siiv\ are aligned it is expected that most data lie roughly between the pure-thermal and non-thermal solutions as displayed in Fig.~\ref{f-fwhm} (the detailed comparison of the profiles is beyond the scope of this paper and will be made in a future KODIAQ paper). 

These properties imply that the \ovi\ absorption traces gas that is hotter and/or is more kinematically disturbed for the intervening absorbers than the gas probed by \siiv\ and the bulk of \civ. So even though all these ions trace highly ionized gas, \ovi\  often traces a different gas-phase than \civ\ and \siiv. The strong correlation observed here between \civ\ and \siiv\ for all the $\tau_{\rm LL}>1$ absorbers was already noted for the DLAs by \citet{wolfe00}. The absence of correlation between the velocity components between \ovi\ and \civ\ (or \siiv) and the fact that the average $b$ for \ovi\  is quite large, typically larger than for \civ, almost certainly rule out that photoionization is a major contribution in the production of the observed \ovi\ (see \S\ref{s-bval} for more details). On the other hand, for the proximate systems,  there is often a much better match between the \civ\ and \ovi\ profiles, implying that they may probe gas with the same physical conditions.  It could also be possible that photoionization  plays a major in the production of the high ions in the case that the Hubble broadening dominates the line broadening; the absorbers would then probe $\la 60$--100 kpc path length structure \citep[see ionization models in][]{simcoe02}, however, as we discuss below this would be contrary to the large differences observed in the broadenings of the individual components between \ovi\ and \civ.

\begin{figure}[tbp]
\epsscale{1.2} 
\plotone{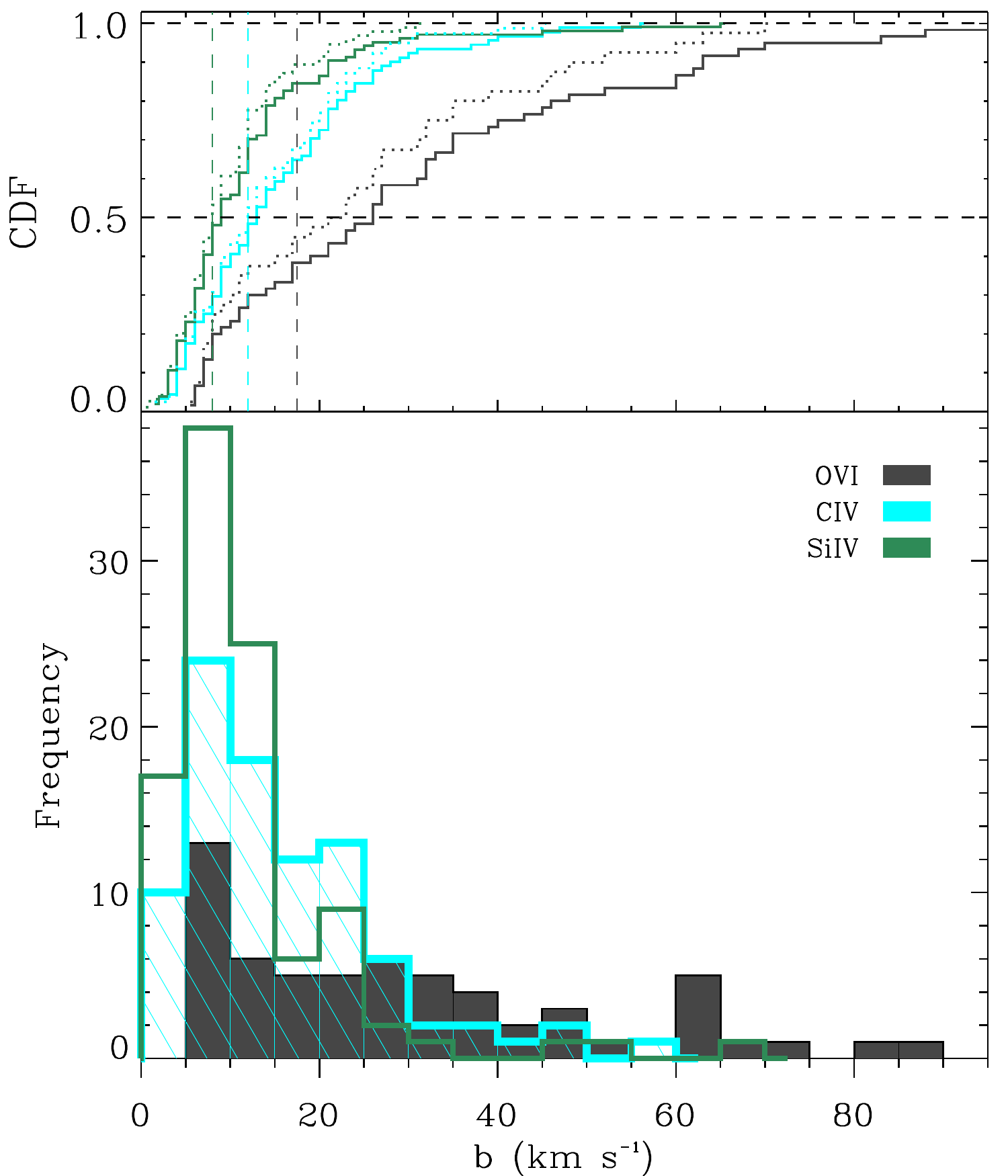}  
  \caption{Frequency and cumulative  distribution functions of the $b$-value of the individual components for \ovit, \civt, and \siivt\   (using the results from Table~\ref{t-fit}). In the bottom panel, the entire sample is used. In the top panel, the solid lines show the results for the entire sample while the dotted lines show the results for the robust sample. The vertical dashed lines in the top panel show the $b$-value corresponding at the peak-abundance temperatures for \siivt, \civt, \ovit\  in collisional ionization if thermal motions dominate the broadening.
 \label{f-bval}}
\end{figure}

\subsection{Doppler Parameter}\label{s-bval}

We can make a step further in characterizing the highly ionized gas by studying in more detail the individual $b$-values, which gives an upper limit on the temperature. For our analysis, it is useful to consider a cutoff on the observed Doppler parameter. Following \citet{lehner11}, we define a $b$ cutoff,  $b_c$, between broad and narrow components where the narrow components probe gas temperature of $7 \times 10^4$ K in the absence of non-thermal broadening ($5 \times10^4$ K if we assume roughly equal thermal and non-thermal contributions to the line width). The $b_c$-value varies for each species such that  $b_c= 6.5$ \km\ for \siiv, 10 \km\ for \civ, and 9 \km\ for \ovi. In Table~\ref{t-bval}, we summarize the median, mean, dispersion of $b$, and fraction of components with $b>b_c$ for the various samples. In Fig.~\ref{f-bval}, we show the $b$-value distribution and cumulative distribution for \ovi, \civ, and \siiv\ for the entire sample of absorbers. Except for a slight decrease in the mean, median, dispersion of $b$ for \ovi\ between the robust and entire samples, there is no major difference between the various samples, and in particular between the samples that include and exclude proximate absorbers. The top panel of Fig.~\ref{f-bval} shows the cumulative distributions for entire (solid line) and robust (dotted line) samples, demonstrating only small differences between the two samples. We therefore consider hereafter in this section the entire sample for our discussion of the Doppler parameters.

The $b$ distributions for \siiv\ and \civ\ have a prominent peak around their mean, at about 10 and 15 \km\ ($T<2\times 10^5$ K), respectively, and a shallow extended wing at higher $b$. In contrast, the distribution of \ovi\ is much flatter with about the same frequency of $b$ in each bin. Fig.~\ref{f-bval} and Table~\ref{t-bval} show that the mean and median of $b$ increase with the ionization energy $E_i$ of the ions. A similar trend is observed in the Milky Way \citep{lehner11}, although $\langle b \rangle$ for \ovi\ is much larger ($\sim$40 \km) for the Milky Way (likely because the instrumental resolution of the Milky Way observations is much cruder, about 20 \km\ compared to 6--8 \km\ for the Keck HIRES spectra). Comparing our results with the high redshift study of \civ\ and \ovi\ associated with the \lya\ forest \citep{muzahid12}, we find a small increase in the difference of 3 \km\ in the median $b$-values for \civ\ in our sample, but a significant one for \ovi\ of 12 \km, implying that the \ovi\ associated with $\tau_{\rm LL}>1$ absorbers is not produced in the same conditions or is more turbulent than in the more diffuse gas. 

As it can be seen from the $b$ distributions in Fig.~\ref{f-bval}, both narrow and broad absorption components are observed in the \siiv, \civ, and \ovi\ profiles, where the narrow components imply gas with $T<7\times 10^4$ K. About 20\% of the \ovi\ individual components and 30\% of the \siiv\ and \civ\ components are narrow. For the \ovi, this is direct evidence that the highly ionized plasma has radiatively cooled or is photoionized by the EUV and soft X-ray radiation from, e.g., cooling hot gas, or  EUVB radiation.  If NEI dominates the ionization of the \ovi\ production, then the gas  must be solar or higher metallicity as otherwise no significant amount of \ovi\ would be observed at these low temperatures (see Fig.~\ref{f-model}). We show below the $b$--$N$ correlation for \ovi\ strongly suggests that \ovi\ is more likely produced by collisions rather than in a gas irradiated by EUV and soft X-ray photons. We also find that 30\%--40\% of components of the high ions are so broad (i.e., for \ovi, \civ, and \siiv, $b\ge 31,21,10$ \km, respectively) that they would imply ionization fraction for a given high ion $<0.003$ for any types of ionization \citep[][and see Fig.~\ref{f-model}]{oppenheimer13} if thermal motions dominated the broadening mechanisms or if thermal and non-thermal motions contribute similar amounts of broadening. We note a large bulk flow driven by the Hubble expansion is unlikely to be a dominant broadening mechanism because there is so little correspondence between the high ion profiles; a  better match between the centroid velocities and $b$-values would be expected if the Hubble flow was important.

\begin{figure}[tbp]
\epsscale{1.2} 
\plotone{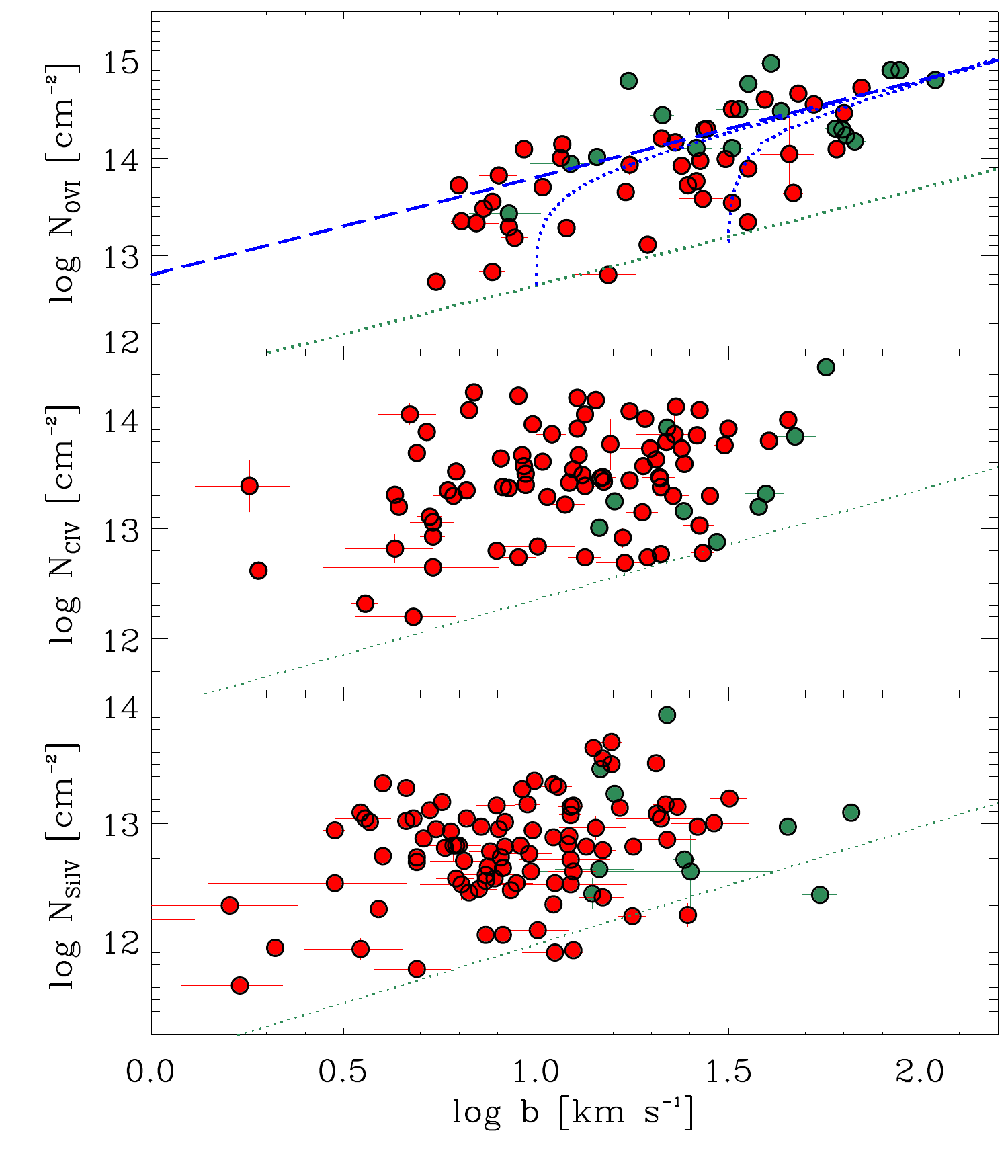}  
  \caption{Column density against the $b$-value for each individual component of \ovit, \civt, and \siivt\ for the KODIAQ sample (using the results from Table~\ref{t-fit}). Red data are more robust estimates than green data (flag `0' and `$>0$' in Table~\ref{t-fit}, respectively).   The dotted line in each panel shows $N$ vs. $b$ for a Gaussian line shape with 10\% central optical depth, which represents the typical detection limit. A Spearman ranking test indicates a strong correlation between $b$ and $N$ for \ovit,  but not for \civt\ or \siivt. In the top panel, we reproduce results from \citet{heckman02} for reference: the curved dotted lines indicate the predictions for radiatively cooling gas for  temperatures of $T_{\rm OVI} = 10^5, 10^6$ K assuming that the cooling flow velocity is equivalent to the $b$-value; the dashed lines corresponds to the effect of blending of multiple narrow components with a velocity separation   $\Delta v = 0$ \km\ (if $\Delta v \ge 10$ \km, then the effect would resemble to that of the radiatively cooling lines, see Heckman et al.)
      \label{f-bn} }
\end{figure}

In the physics of radiatively cooling or conductively heated gas, a relationship between the observed Doppler parameter and the column density is expected \citep{heckman02,edgar86}. While \citet{heckman02} present a heterogeneous sample of \ovi\ measurements,  several works afterwards study this relationship in more controlled environments. In the IGM at both high and low $z$, the correlation is marginal or not present \citep{lehner06,tripp08,muzahid12}. However, in the Milky Way disk, while \citet[and see also \citealt{bowen08}]{lehner11} show no relationship between $N$ and $b$ for \civ\ and \siiv, they found a strong correlation for \nv\ and \ovi. A strong correlation between $b$ and $N$ is also observed in starburst galaxies \citep{heckman02,grimes09}, which led \citeauthor{grimes09} to conclude that the bulk of the \ovi\ absorption is produced in a radiatively cooling gas produced in the interaction between the hot outflows seen in X-rays and much cooler gas seen in H$\alpha$ emission. The $b$--$N$ plot for the high ions in  our sample is shown in Fig.~\ref{f-bn}.  For any $b$-values within $0.5 \le \log b \le 1.5$ (where 90\% of the data lie for \civ\ and \siiv), $N$ for \civ\ and \siiv\ can have any values between about the maximum and minimum observed values of $N$.  As noted above, for \ovi, there is an overall increase in $b$ as 90\% of the \ovi\ components have $0.8 \le \log b \le 1.9$, but over this $b$ interval, there is also correlation between $N$ and $b$. This holds for any subsamples of \ovi, i.e., the entire sample, the intervening robust or entire sample. From Fig.~\ref{f-bn}, there is visually an apparent correlation for \ovi, but not \civ\ or \siiv. This is confirmed using a ranking Spearman test:  this test shows a strong correlation for the \ovi\ intervening systems with $r = 0.70$  (with a statistical significance $p \ll 0.1\%$) and for the robust intervening sample $r = 0.65$ ($p \ll 0.1\%$), statistically supporting a strong correlation between $b$ and $N$ for \ovi. The same result is reached if one adopts various cut offs in $N$ over all $b$-values (to account for the potential that there are different sensitivity limits as a function of $b$). In contrast, for \civ\ and \siiv,  only a very weak correlation is observed with  $r = 0.2$, and at the $3\sigma$ level this relationship could happen just by chance. 

The absence of strong correlation between $N$ and $b$ for \civ\ and \siiv\ and strong correlation for \ovi\ on one hand, and the significant difference in the $b$ distribution of \ovi\ and \civ\ (or \siiv) on the other hand imply that not only the gas in the LLSs, SLLSs, and DLAs is multiphase, but the bulk of the \ovi\ may trace altogether a different type of gas, i.e., the physics that govern the \ovi\ and the other high ions are different. The correlation between $b$ and $N$ for \ovi\ is naturally explained by a radiatively cooling gas produced between a shock-heated outflowing gas driven possibly by massive starbursts and a cooler gas traced by singly ionized species \citep[see Fig.~\ref{f-bn} where we reproduced the models from][]{heckman02}.\footnote{Although a heating mechanism can take place, e.g., in the evaporative phase of a conductive interface, this process is transient \citep[$<2$\,Myr, see][]{borkowski90} and hence unlikely to be seen often in absorption.} The scatter can be explained by various temperatures and velocities in the cooled/heated gas as well as to some photoionization by the hot plasma \citep{heckman02,knauth03}.  The large scatter for \civ\ and \siiv\ suggests that photoionization by the EUVB radiation or more localized sources (stars, hot plasmas) could play an important role, as a relationship between $N$ and $b$ is not expected in that case; this is consistent with the fact that \civ\ and \siiv\ are more easily photoionized than \ovi\ owing to their significantly smaller ionizing energy values. In \S\ref{s-star}, we discuss in more details the implications for the origins of the gas probed by the high ions.

\begin{figure}[tbp]
\epsscale{1.2} 
\plotone{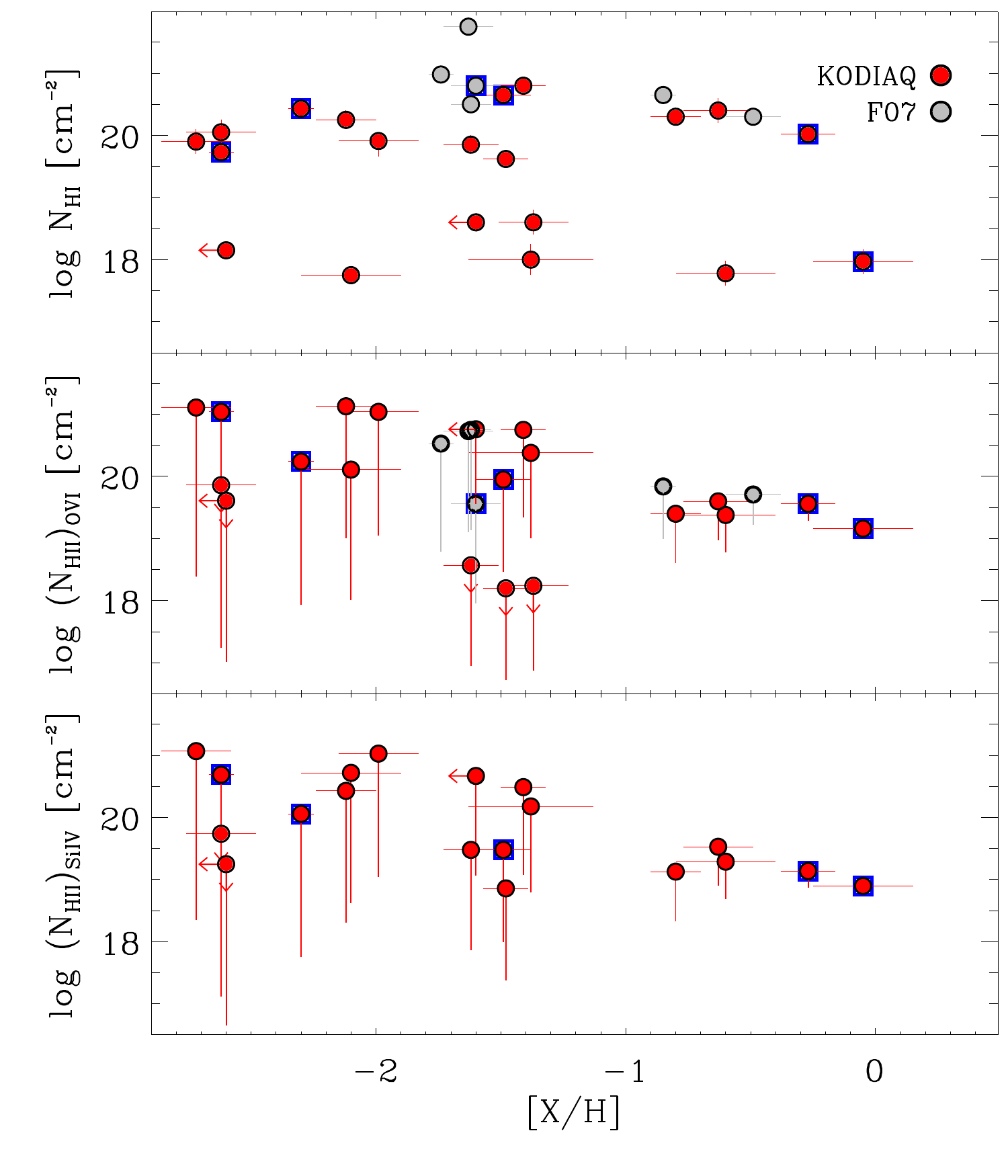}  
  \caption{The \hit\ and \hiit\ column densities against the metallicity of the cool gas. The \hiit\ column densities are derived from \ovit\ and \siivt\ assuming the ionization fractions of 0.2 and 0.4, respectively, and the metallicity of the cool photoionized gas (see  \S\ref{s-nh} for more details).  There is no relation between \nhi\ and the metallicity, but there is a strong anti-correlation between  \nhii\ and the metallicity according to the rank Spearman test, which is expected since \nhii\ derived from a metal line is inversely proportional to the metallicity. That anti-correlation disappears when a solar metallicity in the \ovit\ and \siivt\ phases is assumed and shown with the downward error bars. Circles with overplotted blue squares highlight the proximate systems.
      \label{f-nhmet}}
\end{figure}

\begin{figure}[tbp]
\epsscale{1.2} 
\plotone{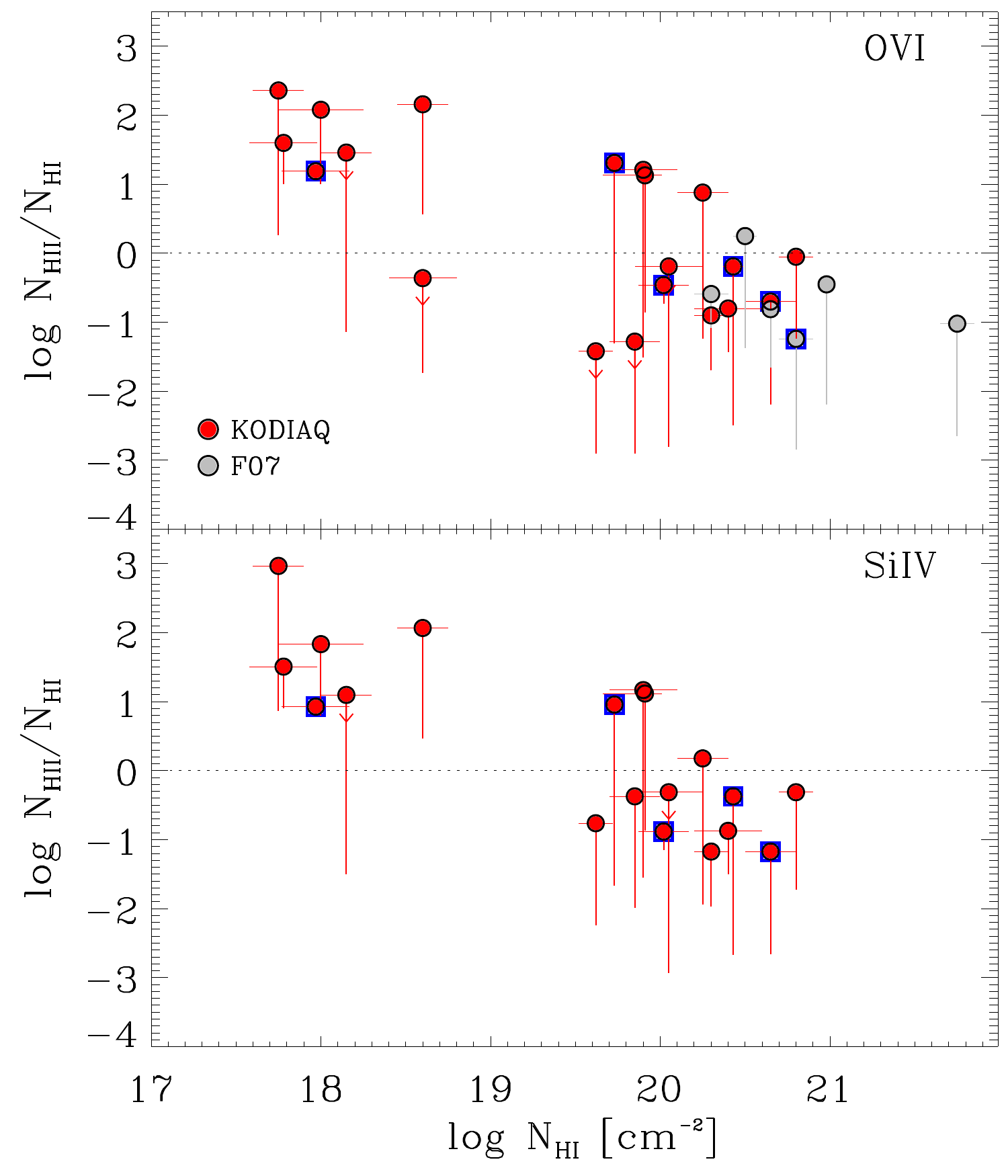}  
  \caption{Ionization-to-neutral ratio as a function of the  \hit\ column density. The filled circles are data where \nhii\ is derived from \ovit\ and \siivt\ assuming the ionization fractions of 0.2 and 0.4, respectively, and the metallicity of the cool photoionized gas (see  \S\ref{s-nh} for more details). The downward error bars indicate the effect of using a solar metallicity in the \ovit\ and \siivt\ phases. As the highly ionized gas is not in CIE, these ionization fraction values are  upper limits, so when \ovit\ or \siivt\ is detected, \nhii\ must increase for a given metallicity. Circles with overplotted blue squares highlight the proximate systems.
 \label{f-nh2nh1}}
\end{figure}

\subsection{Total H Column Density}\label{s-nh}
In the previous sections, we have advanced several arguments that \siiv\ and \ovi\ do not probe generally the same ionized gas, while \civ\ can be in both types of gas at various levels. To gauge the importance of the highly ionized gas relative to the neutral gas, we use \siiv\ and \ovi\ to estimate the amount of \hii\ associated with the warm highly ionized gas and hotter or more turbulent highly ionized gas, respectively. We estimate the amount of hydrogen in the ionized phase probed by \ovi\ from $ N^{\rm OVI}_{\rm HII}= N($\ovi$)/(f_{\rm OVI} \, ({\rm O/H}))$, where $f_{\rm OVI} = N($\ovi$)/N({\rm O})$ is the ionization fraction of \ovi. As we discussed in \S\ref{s-ovi} (and see Fig.~\ref{f-model}), the most conservative value for the ionization fraction of \ovi\ is  $f_{\rm OVI}= 0.2$  \citep[see also, e.g.,][]{sutherland93,gnat07}.  We can similarly estimate the amount of ionized hydrogen in the \siiv-bearing gas by replacing O by Si and \ovi\ by \siiv\ in the previous equation. The ionization fraction of \siiv\ is  $f_{\rm SiIV}\le 0.4$, and for many ionization conditions and densities, $f_{\rm SiIV} \sim 0.1$ \citep[e.g.,][]{oppenheimer13}. As for \ovi, we  first adopt $f_{\rm SiIV}= 0.4$ to have a lower limit on $ N^{\rm SiIV}_{\rm HII} \, {\rm (Si/H)}$. In Table~\ref{t-nh}, we summarize the results from these calculations.  

\begin{figure*}[tbp]
\epsscale{1.3} 
\plotone{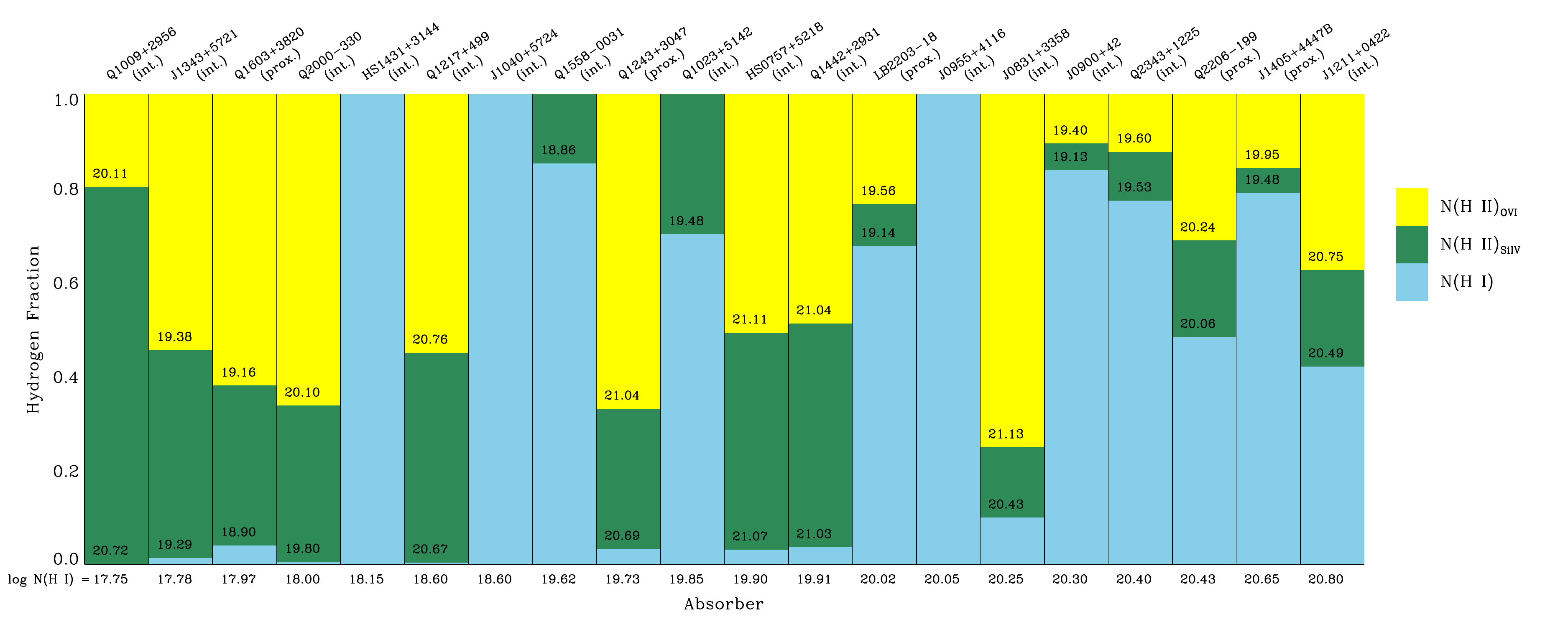}
  \caption{Summary of the level of  the hydrogen ionization and neutral breakdown in each absorber in the KODIAQ sample as a function of increasing \hit\ column density.  \nhi\ is directly estimated from the data (see \S\ref{s-nhi}), while \nhii\ is derived from \ovit\ and \siivt\ assuming the ionization fractions of 0.2 and 0.4, respectively, and the metallicity of the cool photoionized gas (see  \S\ref{s-nh} for more details). When no \ovit\ or \siivt\ absorption is detected, no ionized contribution is shown.  Absorbers with $18\la \mlnhi\  \la 20.1$ can be either predominantly ionized or neutral. For lower \nhi\ values, the gas is dominantly ionized, while for higher values predominantly neutral. When \ovit\ is detected, its contribution to the ionized and often total baryon fraction is never negligible. We emphasize that neither weakly ionized gas (e.g., \aliiit) or hotter highly gas (e.g., \oviit) is accounted for in this figure. 
  \label{f-barplot}}
\end{figure*}

In both cases, the major assumptions that we have to make are that the metallicity is the same in the ionized  and cool neutral gas-phases, and the metallicity is the same in each component. There is plenty of evidence in our Milky Way halo that the metallicity can change by a factor 3--10 or more between components separated by 100--200 \km, e.g., high-velocity clouds or tidal debris metallicity relative to the Milky Way disk metallicity \citep[e.g.,][]{wakker01,lehner08a,zech08,fox13a}. \citet{prochter10} show for one of the absorbers in the KODIAQ sample (the $z=2.43749$ absorber toward Q2000-330) a spread in metallicity between near solar to 1/100 solar, and for that sightline we take this into account. \citet{crighton13} also noted a factor 30 between two components of the many components observed in the KODIAQ absorber at $z=2.4379$ toward Q1442+2931. Interestingly in both cases, the strongest absorption of \ovi\ overlaps with the velocity components with the highest metallicity estimates, while the strongest low-ion absorption tends to be in components with low metallicities, i.e., the metallicity could be systematically underestimated in the \ovi-bearing gas when the metallicity of the cool gas is low. At low redshift, \citet{savage11a,savage11b} directly estimated the metallicity in the cool and highly ionized gas, and found in one case (a LLS) that the highly ionized gas had more metals, and in the other case (a \lya\ forest absorber) tentatively the opposite; in both cases, the difference in metallicity is a factor $\sim$3. Although it is too soon to draw any firm conclusion from these sparse studies, they suggest that the metallicity of the \ovi-bearing gas associated with strong \hi\ absorbers could be higher than that of the cool gas. We argued in \S\ref{s-bval} that some of the properties of the \ovi\ absorbers in our sample are consistent with the idea that near solar or higher metallicities are quite likely for the \ovi-bearing gas (e.g, the low $b$-values of \ovi, which implies gas at temperatures far from the peak-temperature in CIE). Therefore, we also estimate \nhii\ assuming a solar metallicity and display both estimates in Fig.~\ref{f-nhmet}, which shows \nhi\ and \nhii\ as a function of the metallicity. Unsurprisingly there is no trend between \nhi\ and the metallicity. There is, however, an anti-correlation between \nhii\ estimated assuming $[{\rm X/H}]_{\rm cool}$ and the metallicity, which is not surprising either as $\mnhii \propto 1/({\rm X/H})$. That anti-correlation disappears when solar metallicity is assumed. Similar conclusions were found at low $z$ in the \ovi\ absorbers associated with weak LLSs \citep{fox13}.

At $2<z<4$, the average metallicity of the gas probed by the DLAs is around 3--12\% with a very large scatter at any given $z$ \citep[e.g.,][]{prochaska03,rafelski12}, but with most values being sub-solar. So adopting a solar metallicity and the peak ionization fraction in CIE provides the most conservative estimates of \nhii. In this case, the highly ionized component would only be important compared with the neutral gas column for the LLSs as displayed in Fig.~\ref{f-nh2nh1}. However, the properties of the \ovi\ imply that the gas is in NEI (see above sections), and in this case $0.015 \le f_{\rm OVI}\le 0.10$ for a wide range of models and temperatures, which would correspond to the high values displayed in Figs.~\ref{f-nhmet} and \ref{f-nh2nh1}  and listed in Table~\ref{t-nh} if the solar metallicity is assumed. Similar conclusions also apply to \siiv. With that in mind, we show the magnitude of the baryon reservoir in the various forms of hydrogen as a function of increasing \nhi\ in Fig.~\ref{f-barplot} as reported in Table~\ref{t-nh}. The strength of the green and yellow bars relative to the blue bars clearly indicates that \hi\ is a trace constituent for most (71\%) LLSs, i.e., \hii\  dominates their baryon content. In the cases where high ions are detected, the total $N_{\rm H}\simeq \mnhii$ in LLSs can be as large as in the DLAs.  The DLAs are on the other hand pre-dominantly neutral, but with a non-negligible amount of warm and hot and/or turbulent highly ionized gas. For the SLLSs, both naming conventions (SLLSs and sub-DLAS) can apply for these absorbers as about 60\% of the SLLSs have  $\mnhii \gg  \mnhi$  and the other 40\%   $\mnhii \le  \mnhi$. Contrary to earlier assumptions, the ionization fraction can be important for any SLLSs, with low or high \nhi. We also emphasize that these estimates only take into account the highly ionized gas, not the weakly warm ionized medium (WIM) probed by, e.g., \aliii. \citet{wolfe00} show indeed that \aliii\ and singly-ionized metal line profiles are strongly correlated, but those of \aliii\ and \civ\ (or \siiv) are not correlated. In the Milky Way, \citet{lehner11} show as well that \aliii\ is mostly produced in the WIM, implying that \siiv\ and \civ\ have different sites of origin than does \aliii. The reason for this is that most of the observed \siiv\ and \civ\ (even the narrow components) do not arise in the WIM because the WIM is strongly affected by the presence of the \hei\ absorption edge at 24.6 eV in stellar atmosphere \citep{lehner11}. Therefore a photoionized phase is present and not accounted for in this paper. In \S\S\ref{s-barbud}, \ref{s-metbud}, we discuss the implications for the baryonic and metal cosmic budgets.

\section{Discussion}\label{s-disc}
\subsection{Origin(s) of the \ovi\ in Intervening $\tau_{\rm LL}>1$ Absorbers: Outflows, Hot Halos, Accretion}\label{s-star}
Before discussing the implications of our findings, we remind the reader the \ovi\ absorbers were not selected randomly. We searched specifically for LLSs, SLLSs, or DLAs for which the expected \ovi\ absorption was not too contaminated near the velocities where \hi, \oi, or singly ionized species are observed. This selection ensures that the observed \ovi\ is associated with galaxies and their CGM rather than the more diffuse IGM, and indeed the properties of the \ovi\ in the KODIAQ sample are different from those derived in blind surveys of \ovi\ at similar redshifts \citep{muzahid12,simcoe02}: 1) when \ovi\ is detected, the total column and line spread of \ovi\ are typically much larger in the KODIAQ sample than in blind \ovi\ surveys (see \S\ref{s-int}), 2) there is a strong correlation between $N$ and $b$ of the individual components for \ovi\ in the KODIAQ sample, which is not observed in blind \ovi\ surveys at similar redshifts, and is neither observed for \civ\ or \siiv\ in any surveys (see \S\ref{s-bval}). Although the conditions could be different at high $z$ than at low $z$, the \ovi\ properties in the KODIAQ sample are astoundingly similar to those observed in starburst galaxies at low redshift \citep{grimes09}, suggesting the origin of the \ovi\ associated with $\tau_{\rm LL}>1$ absorbers is the hot cooling gas of large-scale outflows from high redshift star-forming galaxies, possibly the Lyman break galaxies \citep[e.g.,][]{steidel10}. The near ubiquity of strong \ovi\ absorption in our sample is also completely consistent with the near universal galaxy-scale outflows at $z \sim 1$--$4$ \citep{shapley03,weiner09}.

While the similarity of the \ovi\  properties in the KODIAQ sample and local starbursts is striking, other processes may be at play in the production of \ovi. For example, any absorbers with $\mlnhi >17$ and ${\rm [X/H]}\la -1$ could probe cold streams according to cosmological simulations \citep[][and see also \citealt{ribaudo11b,lehner13} for low-$z$ observational results]{fumagalli11}. It is also quite possible that some of the \ovi\ traces remnants of outflows or recycled materials \citep[e.g.,][]{oppenheimer10}. \citet[hereafter S13]{shen13} recently used zoom-in simulations to determine the properties of the metals (including \ovi, \civ, \siiv, etc.) and \hi\ in the CGM of massive galaxy (with a stellar mass of $1.5\times 10^{10}$ M$_\sun$) at $z\sim 3$, and since the same diagnostics  are investigated over similar redshifts in this simulation and our study, we primarily use these theoretical results for our discussion. 

The simulated spectra along two lines of sight in S13 at impact parameters of 23 and 34 kpc show complicated structures spanning $\Delta v \sim 400$ \km. A qualitative comparison of the observed  and simulated spectra shows that the simulated spectra have similar properties as described in \S\ref{s-prop}: 1) \ovi\ has a smoother structure than \civ\ and \siiv, 2) the component structures of \civ\ and \siiv\ are quite similar, but in some cases some \civ\ components align with \ovi. In these simulations, low and high ions probe both outflows and inflows. For one of the lines of sight, some of the components also trace the CGM of a dwarf galaxy companion.  Referring to Fig.~4 in S13, where the projected column density maps for the outflows and inflows  are shown separately, only outflows have a large covering factor for absorbers with  $\mlnovi \ga 14.4$ at any impact parameters.  Outflows and inflows have, however, similar covering factors within a virial radius of the massive galaxy for gas with $13.5 \le \mlnovi < 14.4$.  From our profile fitting analysis, 1/3 of the individual \ovi\ components in the intervening absorbers have $\mlnovi \ge 14.4$ (and $\langle b \rangle = 50 \pm 31 $ \km), and according to S13 simulation, these are consistent with outflows; about 1/2 have $13.5 \le \mlnovi < 14.4$ (and $\langle b \rangle = 27 \pm 16 $ \km), where both outflows and inflows could give rise to the \ovi\ absorption. The remaining 20\% of the \ovi\ components have $ \mlnovi \le 13.5$ (and $\langle b \rangle = 14 \pm 10 $ \km), and in the S13 model, these low \ovi\ column densities would be produced mostly at 2--3 virial radii from a massive galaxy. According to the S13 simulations, only 1/3 of the total \ovi\ within 2 virial radii is produced in inflows, compared to 44--50\% for \civ\ and \siiv\ and 77\% for \hi\ (or singly ionized species). Therefore, similar to our conclusions above, S13 finds that \ovi\ is mostly produced by outflows and the metallicities of the \ovi\ gas and cool gas are not necessarily the same. 

\ovi\ absorption appears to be ubiquitous in DLAs based on the 100\% detection rate in the combined KODIAQ--F07 sample, but not in LLSs and SLLSs, although the detection rate of \ovi\  in these absorbers is still quite high, 60--70\%. For the 5 $\tau_{\rm LL}>1$ absorbers where there is no detection of \ovi, the metallicity ranges between $<-2.6$ to $-1.4$ dex, which according to the simulation results from \citet{fumagalli11} would be likely candidates for cold streams. As \ovi\ appears ubiquitous in the halo of $L*$ star-forming galaxies at $z \sim 0.5$  according to COS-Halos observations \citep{tumlinson11} and has a near unity covering factor in massive galaxies at $z\sim 3$ according the S13 simulations, these $\tau_{\rm LL}>1$ absorbers with no associated \ovi\ may trace cold gas inflows in the CGM of lower mass galaxies. 

\subsection{Cosmological Baryon Budget}\label{s-barbud}
At $z\sim 3$, most of the cosmic baryons ($\sim$90\%) are in the diffuse IGM \citep{fukugita98,rauch97,weinberg97}. Adopting $\Omega_b = 0.044$ the ratio of the total baryon density to the critical density \citep[e.g,][]{spergel03,omeara01}, the neutral gas of the DLAs and SLLSs contributes to about 1.8\%  and 0.4\% of the cosmic baryons, respectively \citep[e.g.,][and see below]{peroux03,peroux05,prochaska05}. \citet{fox07a} show that about 1\% of the baryons are in form of highly ionized gas probed by \ovi\ in DLAs. Our present survey allows us to extend the evaluation of the cosmic baryon budget in all the $\tau_{\rm LL}>1$ absorbers and in both their neutral and highly ionized gas-phases. 

As we argued in \S\ref{s-prof}, \siiv\ and \ovi\ probe largely different types of gas in the intervening absorbers, and therefore the mean gas density relative to the critical density can be obtained from the \hi\ density distribution function  \citep[$f(N_{\rm H I})$, e.g.,][]{tytler87,lehner07,omeara07,prochaska10,ribaudo11}, assuming that \hii\ follows the same $f(N_{\rm H I})$ distribution: 
\begin{eqnarray*}
{\Omega_{\rm g} } &= & \frac{\mu_{\rm H} m_{\rm H} H_0}{\rho_c\, c }\Big[\int N_{\rm HI} f(N_{\rm HI}) d N_{\rm HI}  \\ 
& & + \int    N_{\rm HII}^{\rm OVI}  f(N_{\rm HI}) d N_{\rm HI} \\
& & + \int    N_{\rm HII}^{\rm SiIV} f(N_{\rm HI}) d N_{\rm HI}\Big]
\end{eqnarray*}
where $m_{\rm H} = 1.673\times 10^{-24}$ g is the atomic mass of hydrogen, $\mu_{\rm H} = 1.3 $ corrects for the presence of helium, $\rho_c = 3 H^2_0/(8\pi G) = 9.20 \times 10^{-30}$ g\,cm$^{-3}$ is  the current critical density, and $f(N_{\rm H I})$ the column density distribution function \citep[e.g.,][]{tytler87,lehner07,omeara07,prochaska10,ribaudo11}. Here we use the most recent model of $f(N_{\rm HI})$ at $z=2.5$ from \citet{prochaska13}. The first integral corresponds to the neutral gas, the second to the highly ionized gas probed by \ovi, and the third to  the highly ionized gas probed by \siiv. As there is no trend in $N_{\rm HII}^{\rm OVI}$ or $N_{\rm HII}^{\rm SiIV}$ with \nhi\ (see Fig.~\ref{f-barplot}), we estimate from Table~\ref{t-nh} the average value of $\log \mnhii$ and its dispersion for each category of absorbers. We exclude non-detections from this average, but we applied a correcting factor of 0.7 and 0.6 for the LLS and SLLS estimates, respectively, to take into account that the non-detections of \ovi\ in these absorbers (for \siiv, the correcting factor is 0.9 for the LLLs and SLLSs). We summarize $\Omega_{\rm g} $ for each gas-phase and each $\tau_{\rm LL}>1$  absorber in Table~\ref{t-omegab}; the tabulated range of values corresponds to the dispersion in  $\log \mnhii$. The  values for the neutral gas in SLLSs and DLAs compare well with previous results over similar redshift intervals (see above). The contribution of the neutral gas in the LLSs to the cosmological baryon budget is negligible. 

The last column of Table~\ref{t-omegab} lists the contribution from these 3 gas-phases to the  cosmic baryon fraction for the LLSs, SLLSs, and DLAs. The upper limit for the SLLSs  suggests that the SLLSs could dominate the cosmic baryon budget for the $\tau_{\rm LL}>1$ regime. However, we note that while the average metallicities for the LLSs and DLAs are similar ( $\langle {\rm [X/H]}\rangle = -1.39$, $-1.33$, respectively ), the average SLLS metallicity is a factor 4 smaller in our KODIAQ sample  ( $\langle {\rm [X/H]}\rangle = -1.93$). The average metallicity for the SLLSs in the KODIAQ sample  is also about a factor 4 smaller than  the average SLLS metallicity at $z\sim 2$--3 derived by \citet{peroux07}. Hence although the very metal-poor SLLSs could hide a large amount of baryons, this result could also the subject of small number statistics. Assuming a similar average metallicity for all the  $\tau_{\rm LL}>1$  absorbers, then  the SLLS contribution is $0.2$--4.0\%, and $\Omega_{\rm g}/\Omega_{\rm b}= 3$--14\%  at $2<z\la 3.5$.  We conclude that the highly ionized gas in the $\tau_{\rm LL}>1$  absorbers is very likely the second largest contributor after the \lya\ forest to the baryon budget at $z\sim 2$--3. We note that if the metallicity of the \ovi\ or \siiv\ gas is solar or higher (see \S\ref{s-bval}),  the contribution of the highly ionized gas in  $\tau_{\rm LL}>1$  absorbers could be much smaller  (a factor 10$--$20 smaller), except if this is counterbalances by much smaller ionization fraction of \ovi\ and \siiv. 

\subsection{Cosmological Metal Budget}\label{s-metbud}
While we know where the majority of the baryons are at $z\sim  2$--3 (see
 \S\ref{s-barbud}), the census of metals is far more uncertain at any $z$ \citep[e.g.,][]{bouche07,peeples14}. A large fraction of the metals produced in galaxies is thought to have been expelled by $z\sim 2$  \citep{pettini04,bouche05,ferrara05} by large-scale galaxy outflows \citep[e.g.,][]{shapley03}, the same outflows that we discuss in \S\ref{s-star}. With our KODIAQ survey, we can estimate the metal census at $z\sim 2$--3 in the neutral {\it and}\ highly ionized gas-phases of a sizable sample of LLSs, SLLSs, and DLAs, and hence determine if indeed a large fraction of metals are present in the CGM of galaxies at these redshifts. 

The comoving metal-mass density of the $\tau_{\rm LL}>1$ absorbers could be be simply estimated using the baryon density via $\Omega_{\rm Z} =\Omega_g Z_{\rm abs}$ where $Z_{\rm abs} $ is the metallicity in mass units, but this approach requires we make assumptions on the ionization fraction and metallicity of the gas. A more direct strategy uses the fact that \novi\ in LLSs, SLLSs, and DLAs -- when \ovi\ is detected -- does not vary much with \nhi\ with an average of $\langle N_{\rm OVI}  \rangle = 10^{14.9 \pm 0.3}$  cm$^{-2}$. This method does not require any knowledge of the metallicity and hence unlike $\Omega_{\rm g}$, the calculation of $\Omega_{\rm Z}$ does not depend on the metallicity.  We can therefore directly estimate the metal budget in the $\tau_{\rm LL}>1$ absorbers in the \ovi-bearing gas via,
$$
\Omega_{\rm Z}^{\rm O\,VI} = \frac{\mu_{\rm O} m_{\rm O} H_0}{\rho_c\, c } \frac{\langle N_{\rm OVI} \rangle}{f_{\rm OVI}} \int f(N_{\rm HI}) d N_{\rm HI},
$$
where $\mu_{\rm O} =2.33$ corrects for the fact that O is 43\% of the total solar metal mass,  $m_{\rm O} = 16 m_{\rm H}$ the mass of oxygen, $f_{\rm OVI}$ the ionization fraction of \ovi.

Here we choose  $f_{\rm OVI}=0.1$ as a more representative of the \ovi\ ionization fraction since the \ovi\ gas is very likely in NEI (see \S\ref{s-prop}) and since the conditions near the peak abundance of \ovi\ are mostly transient \citep[][and below]{gnat07,oppenheimer13}. Taking into account that 71\% of the LLSs and 63\% of the SLLSs have a detection of \ovi, and integrating the previous equation over the various \nhi\ limits of the LLSs, SLLSs, DLAs, we can estimate  $\Omega_{\rm Z}^{\rm OVI}$ for the LLSs, SLLSs, and DLAs. We summarize the results in Table~\ref{t-omega}; the tabulated range of values corresponds to the dispersion ($\pm 0.3$ dex) in \novi. 

As discussed in \S\ref{s-prof}, there is little correspondence between the \ovi\ and \siiv\ profiles in the intervening absorbers, implying that \ovi\ and \siiv\ probe in general distinct gas-phases. Hence, using the same procedure for \siiv, we can estimate $\Omega_{\rm Z}^{\rm Si\,IV}$ for the LLSs, SLLSs, and DLAs, with $\langle N_{\rm SiIV}  \rangle = 10^{13.7 \pm 0.3} $ cm$^{-2}$, $f_{\rm SiIV}=0.1$ (see \citealt{oppenheimer13} and below), $28 m_{\rm H}$ for the mass of silicon, and 4.7\% for the total solar metal mass of silicon. We list the results in Table~\ref{t-omega}, along with the total metal-mass density, $\Omega_{\rm Z}^{\rm h}$, of the highly ionized plasma in LLSs, SLLSs, and DLAs. As for the cosmic baryon census, these estimates do not include the WIM traced by \aliii, which is not produced by the same mechanisms as the highly ionized gas \citep{wolfe00,lehner11}, or the hotter gas probed by \ovii\ and \oviii. 

With our approach, $\Omega_{\rm Z}$ depends only one variable parameter, the ionization fraction of \ovi\ and \siiv. \citet{oppenheimer13} show that the ionization fraction of \siiv\ is around 0.1 for a large range of temperatures and ionization conditions, but it can reach  $f_{\rm SiIV}\simeq 0.4$ if the gas is in CIE and near the peak temperature or photoionized, and hence $\Omega_{\rm Z}^{\rm Si\,IV}$ in Table~\ref{t-omega} could decrease by a factor 4. On the other hand, the ionization of \ovi\ of 0.1 is quite conservative since $0.01<f_{\rm OVI}\la 0.1$ for a wide range of temperatures when the temperature is higher or lower than the peak temperatures (see Fig.~\ref{f-model}). Hence the \ovi\ ionization fraction could be a factor 2--5 smaller, i.e., $\Omega_{\rm Z}^{\rm O\,VI}$ in Table~\ref{t-omega} would increase by a similar factor. 

For comparison, the metals in the neutral gas of DLAs and SLLSs yield   $\Omega_{\rm Z}^{\rm n} \simeq   1.5\times 10^{-6}$ and $0.5 \times 10^{-6}$, respectively \citep[][and see also \citealt{peroux03,peroux05,kulkarni05,kulkarni07, prochaska05}]{bouche07}. Hence at least 13--38\% of the metals in DLAs are in the highly ionized plasma \citep[see also][]{fox07a}.  The highly ionized gas probed by \siiv\ and \ovi\ in the dense CGM have at least as much and up to a factor 4 more metals than in galaxies probed by the neutral gas of DLAs.

In the more diffuse ionized gas probed by the \lya\ forest ($\mlnhi <17$),  $\langle \Omega_{\rm Z}^{\rm Ly\alpha} \rangle \simeq   4.2\times 10^{-6}$ \citep{bouche07,schaye03,simcoe04,aguirre04,bergeron05}. The highly ionized CGM associated with LLSs and SLLSs in the KODIAQ sample could have therefore as much metals as the more diffuse metal-enriched \lya\ forest. We note that the mean metallicity of the \lya\ forest in these surveys is a factor 3 larger than that of the DLAs at similar redshifts, and hence the metals in the \lya\ forest may also probe the CGM of galaxies, especially in the high end of the \lya\ forest ($\mlnhi >14.5$). At low $z$, several surveys have shown that metals  arises preferentially in the CGM of galaxies \citep[e.g.,][]{prochaska11}. 

As presented in the introduction, there seems to be a missing metals problem at $z\sim 2$--3 \citep{pettini06,bouche06}. However, as recently shown by \citet{peeples14}, the uncertainties in the nucleosynthesis yields (which answers the question how many metals there are) are large enough to dominate this missing metals issue, which may disappear with more accurate yields. So, we just review this issue using the fiducial value of the metal density of $\Omega_Z^{\rm total} \simeq 3 \times 10^{-5}$ at $z\sim 2$ obtained from integrating the star formation from $z = 10$ to $z = 2$ \citep[see][]{bouche07}. With this value, we find that about  $5$--$20\%$ of metals are in the dense CGM probed by the LLSs, SLLSs, and DLAs. If we include the metals in the \lya\ forest, $19$--$34\%$ are in the highly ionized CGM phase,  compared to about 5\% in the neutral gas of galaxies probed by the DLAs. While the uncertainties on the metal census are large (for the absorbers studied in this work as well as for any categories of absorbers or galaxies), our findings show that a substantial fraction of these metals is found in the highly ionized CGM  of galaxies at $z\sim 2$--3. 

\section{Summary}\label{s-sum}

With our first NASA KODIAQ survey, we have undertaken an in-depth analysis of the \ovi, \civ, \siiv, and \nv\ absorption associated with 7 LLSs, 8 SLLSs, and 5 DLAs.  The high quality (S/N and resolution) of the Keck spectra has allowed us to reveal a wealth of new information on the properties of the highly ionized gas in the dense regions of the CGM and galaxies at $2<z\la 3.5$. Our main results are summarized as follows.

\begin{enumerate} 
\item  For 20 absorbers with strong \hi\ absorption and a clean \ovi\ region, strong \ovi\ absorption is detected in 15 cases; The \ovi\ detection rate is  100\% for the DLAs, 71\% for the LLSs, and 63\% for the SLLSs. Hence there is nearly ubiquitous \ovi\ absorption in the CGM of $z\sim  2$--$3$ galaxies. 
\item Narrow and broad absorption components are seen in \siiv, \civ, and \ovi, where the narrow components imply gas temperatures with $T<7\times 10^4$ K. About 20\% of the \ovi\ individual components and 30\% of the \siiv\ and \civ\ components are narrow. For the \ovi, this is evidence that the highly ionized plasma has radiatively cooled or is photoionized by the EUV and soft X-ray radiation from, e.g.,  a cooling hot plasma. We also find that 30\%--40\% of components of the high ions are so broad that it would imply an ionization fraction $<0.003$ for any types of collisional ionization or photoionization models if thermal motions dominate the broadening of the profiles or if thermal and non-thermal motions contribute similar amounts of broadening.
\item For the \ovi\ (but not for \civ\ and \siiv), we find a correlation between $N$ and $b$ of the individual components, which is indicative that the bulk of the \ovi\ absorption is produced in a radiatively cooling gas rather than in a very diffuse, extended photoionized gas by the EUVB radiation. 
\item The above conclusions imply that a large fraction of the observed \ovi\ is produced via non-equilibrium ionization processes, which also implies that some of the observed \ovi-bearing gas could have a high metallicity, much higher than the cooler gas probed by \hi\ and low ions, according to NEI models. 
\item  When \ovi\ is detected, the dispersion in \novi\ is much smaller than in blind \ovi\ surveys and \novi\ is large with  $\langle N_{\rm OVI}  \rangle = 10^{14.9 \pm 0.3}$  cm$^{-2}$. We also find that the high ions have a large-velocity breadth ($200 \le \Delta v  \le 400$ \km\ for \ovi), where the bulk of the absorption is often displaced by several tens to a couple hundreds of kilometers per second from the cooler gas. The strength and breadth of these \ovi\ absorbers are quite similar to those seen in low redshift starburst galaxies, which strongly suggest that the strong \ovi\ absorbers associated with $\tau_{\rm LL}>1$ absorbers probe gas associated with the outflows of star-forming galaxies at $z\sim 2$--3.
\item The comparison with recent cosmological simulations (S13) supports the conclusion that the majority ($\sim$70\%) of the \ovi\ is produced by outflows of galaxies (independent of their mass) at $z\sim 3$. The bulk of the cooler gas (\hi, \cii, \siii) in these simulations traces inflows, and therefore the metallicity of the major part of the absorption of \ovi\ and singly ionized or atomic species may not be the same as we independently argue in this work. The properties of the absorbers with no detection of \ovi\ make them especially good candidates for cold stream inflows in dwarf galaxies; for the absorbers with \ovi, inflows may also occur, but are dominated by cooler gas traced by \hi\ and singly ionized species according to S13 simulations.
\item When the highly ionized gas is taken into account, we show that the $\tau_{\rm LL}>1$ absorbers could have as much as $3$--$14\%$ of the cosmic baryon budget at $z\sim 2$--3. Hence, almost all the baryons at $z\sim 2$--3 are found in ionized gas outside galaxies.
\item We conservatively show that $5$--$20\%$ of the metals ever produced at $2<z\la 3.5$ are in form of highly ionized metals in the circumgalactic gas of galaxies. Combined with estimates of the metal reservoir in the \lya\ forest at similar redshifts, $19$--$34\%$ of the metals are in form of ionized gas outside galaxies, and hence a substantial fraction of metals have already been ejected by galaxy-scale outflows in the CGM at  $z\sim 2$--3.  
 
\end{enumerate}

\acknowledgements
We thank the referee for insightful comments that help improving our manuscript. 
Support for this research was made by NASA through the Astrophysics Data Analysis Program (ADAP) grant NNX10AE84G. This research has made use of the Keck Observatory Archive (KOA), which is operated by the W. M. Keck Observatory and the NASA Exoplanet Science Institute (NExScI), under contract with the National Aeronautics and Space Administration.  We are grateful to all the PIs listed in Table~\ref{t-data} for voluntarily sharing their data in the KOA. The data presented herein were originally obtained at the W.M. Keck Observatory, which is operated as a scientific partnership among the California Institute of Technology, the University of California and the National Aeronautics and Space Administration. The Observatory was made possible by the generous financial support of the W.M. Keck Foundation. The authors wish to recognize and acknowledge the very significant cultural role and reverence that the summit of Mauna Kea has always had within the indigenous Hawaiian community.

\input{tab1.tex}
\input{tab2.tex}
\input{tab3.tex}

\input{tab4.tex}
\input{tab5.tex}
\input{tab6.tex}

\input{tab7.tex}

\begin{appendix}
\makeatletter 
\renewcommand{\thefigure}{A\@arabic\c@figure} 

\renewcommand{\thetable}{A\@arabic\c@table}

\section{Notes on the column densities of \hi\ and high ions}

\subsection{$z_{\rm abs} = 2.42903$ toward Q1009+2956}

- While there is an overall good correspondence between the \ovi\ profiles, the lines are contaminated at $v \la -110$ \km, so the column density of \ovi\ does not match the absorption seen in \civ\ and \siiv\ at $-167$ \km.  

- The strongest \hi\ feature in this LLS is line-black at the Lyman limit, but cannot exceed \lnhi\,=$17.8$ from the lack of damping features at \lya\ modulo the continuum level errors.  A second LLS at $z=2.4069$ contributes absorption to the Lyman limit region, but cannot account for the strong saturated nature of the LLS of interest.  We thus adopt a conservative estimate a total \hi\ column density of \lnhi\,=$17.75 \pm 0.15$ for this absorber.

\subsection{$z_{\rm abs} = 2.83437$ toward J1343+5721}

- While there is an overall good correspondence between the \ovi\ profiles, \ovi\ $\lambda$1037 is partially contaminated at $-220\la v \la -110$ \km. The reported column density is from \ovi\ $\lambda$1031. Similarly, there is an overall agreement between the \nv\ profiles, but \nv\ $\lambda$1242 is contaminated at $-220\la v \la -170$ \km.

- The absorption in this LLS is optically thick at the Lyman limit, and is limited from having a column density higher than $\mlnhi \sim 18.2$ by the lack of damping features at \lya\, modulo continuum level errors.  A fairly narrow velocity width of $b=15$ \km\ best fits the higher order Lyman lines. Additional absorption with \lnhi $\simeq 16.7$ is required to account for higher order Lyman series absorption (and is likely given the observed \cii\ lines), but does not dramatically effect the total \hi\ column density, which we set at $\mlnhi =17.78 \pm 0.20$ for the LLS, where the error is dominated by continuum uncertainty and model degeneracies.

\subsection{$z_{\rm abs} = 2.47958$ toward Q1603+3820}

- \nv\ $\lambda$1242 is contaminated, so we have to rely only on \nv\ $\lambda$1238 to estimate the \nv\ column density. As the \nv\ $\lambda$1238 and \ovi\ $\lambda$$\lambda$1031, 1037 profiles are similar, it gives us confidence that \nv\ $\lambda$1238 is not contaminated. 

- This complex LLS was fitted with multiple components to account for the absorption seen in the higher order Lyman series lines, subject to the additional constraint that no weak damping features can be allowed to over-absorb the \lya. As it can be seen from the \oi\ absorption (see Fig.~\ref{f-q1603}), this absorber is dominated by a strong single component at $z=2.47949$ where $\mlnhi \simeq 17.9$.  Seven additional components ranging in column density from $\mlnhi =14.05$ to \lnhi\,=$16.65$ are required to account for the strongest features in the higher order Lyman series, bringing the total \hi\ column to $\mlnhi =17.97 \pm 0.20$ where the error is dominated by continuum level at \hi\ \lya\ and model degeneracies in the higher order Lyman series.  Noteworthy is that the line width of this LLS is quite narrow with $b < 20$ \km.

\subsection{$z_{\rm abs} = 3.54996$ toward Q2000-330}

-  While there is an overall good correspondence between the \ovi\ profiles, \ovi\ $\lambda$1031 is partially contaminated at $v\ga +100$ \km.  \ovi\ $\lambda$1031 is also partially saturated between 0 and 50 \km. \civ\ $\lambda$1548 is also somewhat contaminated at various velocities (see Fig.~\ref{f-q2000}), although the impact on the total column density is small (+0.08 dex). 

- We adopted the \nhi\ for this LLS from \citet{prochter10}. 

\subsection{$z_{\rm abs} =2.58615$ toward HS1431+3144}

- The weak line of the NV doublet is contaminated, \nv\ $\lambda$1238 can be used to place a upper limit on the amount of \nv. \ \ovi\ $\lambda$1031 is contaminated, and $\lambda$1037  is partially contaminated at $v>+50 $ \km\, but there is no absorption at the $-40 \le v \le 40$ \km\ where absorption would be expected from the low-ion profiles. There is no absorption in \siiv\ and \civ. 

- The \hi\ column density for this kinematically simple LLS comes from two primary constraints.  The first is a weak damping wing feature on the blue side of the \lya\ line.  The second is the termination of the Lyman series at the Lyman limit, which is line black.  From these constraints, we obtain $\mlnhi = 18.15 \pm 0.15$.

\subsection{$z_{\rm abs} = 2.18076$ toward Q1217+499}

- All the high-ion profiles are partially saturated, but the overall similarity between the profiles for a given high ion gives us confidence that they are mostly uncontaminated (except for \siiv\ $\lambda$1402 and the regions identified in Fig.~\ref{f-q1217}).

- As  \lyb\ is contaminated, the primary constraint on this LLS comes from the damping wing on the blue side of \lya. If a one-component model is adopted for the \hi\ absorption, we find $\mlnhi =18.60 \pm 0.15$. However, we note that no \oi\ is observed and most of the absorption in the singly ionized species is observed at about $-30$ \km\ relative to $z=2.18076$  (see Fig.~\ref{f-q1217}). If another component is added, the total column density is consistent with the solution from the single component fit, but the \nhi\ values in the individual components are unreliable. 

\subsection{$z_{\rm abs} =3.26620$ toward J1040+5724}

- The strong line of the \nv\ doublet is contaminated, \ovi\ $\lambda$1031 is contaminated, and $\lambda$1037  is partially contaminated at $v>+50 $ \km\, but there is no absorption at the $-40 \le v \le 40$ \km\ where absorption would be expected from the low-ion profiles. There is no wavelength coverage of \siiv\ and \civ. 

- The \nhi\ for this system is poorly constrained from \lya\ due to forest line blending.  Nevertheless, it cannot exceed 18.75 dex, or it would over-absorb in the damping wings of the line on either side of line center.  Further constraints come from the Lyman limit, where the absorption is black.  The termination of the Lyman series also favors $\mlnhi > 18.5$.  We adopt $\mlnhi = 18.60 \pm 0.20$ based on these constraints.

\subsection{$z_{\rm abs} = 2.62999$ toward Q1558-0031}

- Both lines of the \nv\ doublet are contaminated and cannot be used. \ovi\ $\lambda$1037 is contaminated, but there is no absorption of \ovi\ $\lambda$1031 making the non-detection of \ovi\ secured. Only \siiv\ is detected among the high ions, and only \siiv\ $\lambda$1393 can be used, as $\lambda$1402 is contaminated. The absorption profiles of \siiv\ $\lambda$1393 and the singly ionized species are quite similar, giving us confidence that it is not contaminated by an unrelated absorber. 

- The system has prominent damping wings in \hi\ \lya. The outer damping wings of \lya\ suffer from large continuum uncertainties, but the core of the line provides a good constraint on the total \nhi. \lyb\ places an upper limit on the \nhi\ value. We adopt $\mlnhi =19.62 \pm 0.10$ for this absorber.  A fluxed calibrated MIKE spectrum is also available for this SLLS \citep[see][]{omeara06}, which allows for an accurate placement of the continuum placement at \lya. The adopted value from the HIRES data provides a good fit to the \lya\ absorption observed with MIKE. 

\subsection{$z_{\rm abs} = 2.52569$ toward Q1243+3047}

- The high ions are not contaminated. 

- We adopted the \nhi\ value for this SLLS from \citet{kirkman03}.

\subsection{$z_{\rm abs} =  3.10586$ toward Q1023+5142}

- \ovi\ $\lambda$1037 is contaminated at $v>-200$ \km, while $\lambda$1031 is contaminated at $v<-200$ \km.There is, however, no absorption of \ovi\ at the 3$\sigma$ level where \civ, \siiv, or low-ion absorption is observed. Note that the main absorption of \civ\ is blueshifted by over $200$ \km\ relative to the low ions and \siiv. 

- The primary constraint on this SLLS comes from the damping feature on the red side of \lyb\ with an additional constraint from the core of the \lya\ absorption.  We adopt $\mlnhi =19.85 \pm 0.15$ for this system, with the dominant error coming from continuum level placement at \lyb.

\subsection{$z_{\rm abs} = 3.04026$ toward HS0757+521}

- Although \ovi\ $\lambda$1031 is contaminated at $-200 \la v \la -70$ \km, there is excellent agreement between the two lines of the \ovi\ doublet in the red wing of the absorption, giving us confidence that \ovi\ $\lambda$1037 is not contaminated (except for an insignificant absorption feature between 5 and 20 \km). 

- In this case, there is only coverage of \lya.  We adopt $\mlnhi =19.90 \pm 0.20$, which is constrained by the core region and the blue side of the damping wing of \lya. This is consistent with \citet{misawa07} who reported a value of 19.82 dex for the system using other \hi\ transitions in addition to \lya. 

\subsection{$z_{\rm abs} =  2.43749$ toward Q1442+2931}

- Both lines of the \nv\ doublet are contaminated and cannot be used. \civ\ is saturated and only a lower limit can be determined from the AOD method. 

- This SLLS is complex with two strong absorption components as revealed by the \oi\ (see Fig.~\ref{f-q1442}; note that for this absorber the redshift was adopted to be the average of the two strong \oi\ absorption components). The \hi\ absorption profiles are too complex to model without using a priori the information from the metal lines. As \oi\ is an excellent tracer of \hi, we use it to constrain the total \hi\ column density: we model this absorber by placing the \hi\ absorption components at the velocities of the 4 strongest \oi\ components (see Fig.~\ref{f-q1442}) and  we assume that the metallicity in these four components does not vary. We can therefore assign \hi\ column density values to the four components, with relative strengths equal to the relative strengths of the \oi\ lines.  We then vary the \hi\ strongest component until the system provides a reasonable fit to the data at \lya\ in the core region, and no over-absorption in \lyb.  This process results in a  total \hi\ column of $\mlnhi =19.91^{+0.10}_{-0.25}$, where the upper limit comes from the constraint on over-absorption, and the increased uncertainty in the lower limit arising from model uncertainties and the likelihood that other, weaker, components add more uncertainty to the fit.  We do not assign an uncertainty to the assumption that [O/H] does not change component to component. We note that our adopted value overlaps within $1\sigma$ with the value derived by \citet{crighton13} using the \di/\hi\ ratio. 

\subsection{$z_{\rm abs} = 2.69985$ toward LB2203-18}

- Despite the \ovi\ absorption being extremely strong, there is an excellent agreement between the profiles of the \ovi\ doublet, with only some contamination in the blue wing of \ovi\ $\lambda$1037 at $-200 \la v \la -100$ \km. 

- This SLLS shows multiple low ionization metal lines.  As with Q1442+2931, we will proceed under the assumption that [O/H] does not vary in the absorber, and we scale \hi\ absorption to the relative strengths of the \oi\ lines.  For this analysis, we use only the two strongest \oi\ lines.  To best fit the core region of \lya\ and not over-absorb in \lyb\, we arrive at two components at $z=2.69815$ and $z=2.69901$ with $\mlnhi = 19.85$ and $19.54$ respectively, giving a total $\mlnhi =20.02 \pm 0.15$ for the SLLS.  The errors are dominated by the continuum level placement.  Additional absorption is certainly needed to completely characterize the system (and is hinted by the metal lines), but their relative strengths contribute to the total \nhi\ at levels below our error estimate.

\subsection{$z_{\rm abs} =  3.27987$ toward J0955+4116}

- No absorption is detected in the high ions at the $3\sigma$ level. 

- This strong SLLS shows a simple velocity structure in the low ionization metal lines with a single component. To best fit the core region of \lya\ and not over-absorb in \lyb\, we arrive at one-component fit with $\mlnhi =20.05 \pm 0.20$ for the SLLS.  The errors are dominated by the continuum level placement.

\subsection{$z=2.30361$ toward J0831+3358}

- \nv\ $\lambda$1238 is contaminated, so we have to rely only on \nv\ $\lambda$1242 to estimate an upper limit on the \nv\ column density. The \ovi\ $\lambda$$\lambda$1031, 1037 AOD profiles are consistent with each other, only revealing some contamination in the blue wing of \ovi\ $\lambda$1037 at $-400 \la v \la -340$ \km.

- The blue core region of \lya\ and the over-absorption from \lyb\ provide the best constraints on the \hi\ for this SLLS/DLA.  The absorption in \oi\ and singly ionized species is dominated by two components at about $-10$ and $+10$ \km. A somewhat strong (as high as \lnhi = 18.75)  additional absorption component is needed to model this system, and is implied by the \civ\ and \siiv\ absorption at about $-240$ \km, but is very poorly constrained, and does not affect the total \nhi\ estimate as it is very weak relative to the strongest \hi\ component.  We adopt a value of $\mlnhi =20.25 \pm 0.15$ for this absorber with continuum uncertainties dominating the error budget.

\subsection{$z = 3.24571$ toward J0900+42}

- There is no detection of \nv\ (both transitions are uncontaminated). \ovi\ $\lambda$1031 appears free of contamination, but $\lambda$1037 is contaminated at $v \la -130$ \km. \civ\ $\lambda$1550 is also partially contaminated at $v\ga 30$ \km. 

- We adopted the \nhi\ value from \citet{prochaska07}. 

\subsection{$z = 2.43125$ toward Q2343+1225}

- While there is an overall match in the \ovi\ profiles, both transitions show some signs of contamination at different velocities. The reported column density is the sum of the column densities of \ovi\ $\lambda$1037 ($-61\le v \le 76$ \km) and \ovit\ $\lambda$1031 ($76\le v \le 350$ \km), which is fully consistent with the total column density derived from the profile fitting method. 

- This DLA is very complex, showing absorption in \oi\ in at least 8 distinct components. The redshift $z=2.43125$ is taken from the strongest of the \oi\ components. For simplicity, we opt to use the core of \lya\ and the damping wing on the red side of \lyb\ to model a single DLA absorber with $\mlnhi = 20.40 \pm 0.20$, which is consistent a previous estimate by \citep[][this QSO is named Q2343+1232 in this paper]{lu98}. 

\subsection{$z = 2.07617$ toward Q2206-199}

- Both \nv\ profiles are contaminated below $-10$ \km, but neither shows any absorption at velocities seen in the other high ions. The \ovi\ $\lambda$$\lambda$1031, 1037 AOD profiles are consistent with each other, only revealing some contamination in the red wing of \ovi\ $\lambda$1031 at $v \ga +35$ \km.

- We adopted the \nhi\ value for this DLA derived by \citet{pettini01}. 

\subsection{$z=2.16707$ toward J1405+4447B}

- There is no contamination in the high ions in this absorber despite the profiles spanning a large velocity interval, except for \civ\ $\lambda$1550 at $-30 \la v \la 20$ \km. 

- The damping features in \lyb\ provide an excellent constraint on the \hi\ for this DLA, and we adopt $\mlnhi =20.65 \pm 0.15$ with the errors being dominated by continuum uncertainties in \lyb.

\subsection{$z= 2.37654$ toward J1211+0422}

- This absorber has been analyzed in details by \citet{lehner08}. The \nhi\ value was adopted from their work, but the other column density estimates are from this work using the newly reduced data. Our new and previous analyses are independent, but fully consistent.

\section{Profile Fit Results}

The results for the component fitting of \siiv, \civ, and \ovi\ are given in Table~\ref{t-fit}, and the component models can be seen in Figs.~\ref{f-fitq1009} to \ref{f-fitj1211} where both the individual components and global fits are shown. Discussion on the profile fit method and its limitations can be found in \S\ref{s-fit}.

\begin{figure*}[tbp]
\epsscale{1} 
\plotone{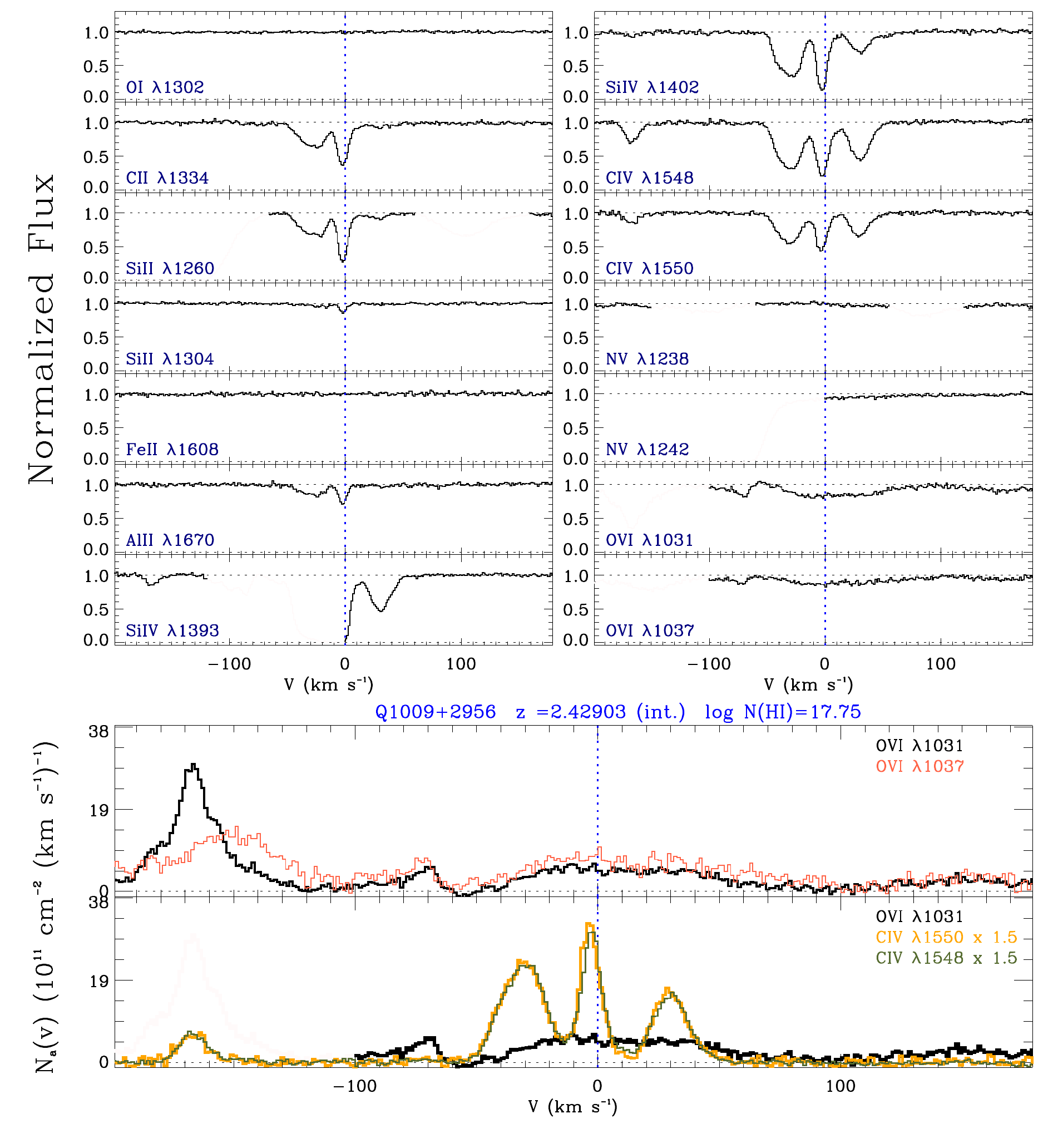}  
  \caption{{\it Top panels}: Normalized profiles as a function of the rest-frame velocity for the absorbers toward Q1009+2956. Grayed areas are contaminated regions. {\it Bottom panels}: Apparent column density profiles of the \ovit\ and \civt\ doublets. Note that \civt\ AOD  profiles have been scaled by the factor indicated in the figure. 
 \label{f-q1009}}
\end{figure*}

\begin{figure*}[tbp]
\epsscale{1} 
\plotone{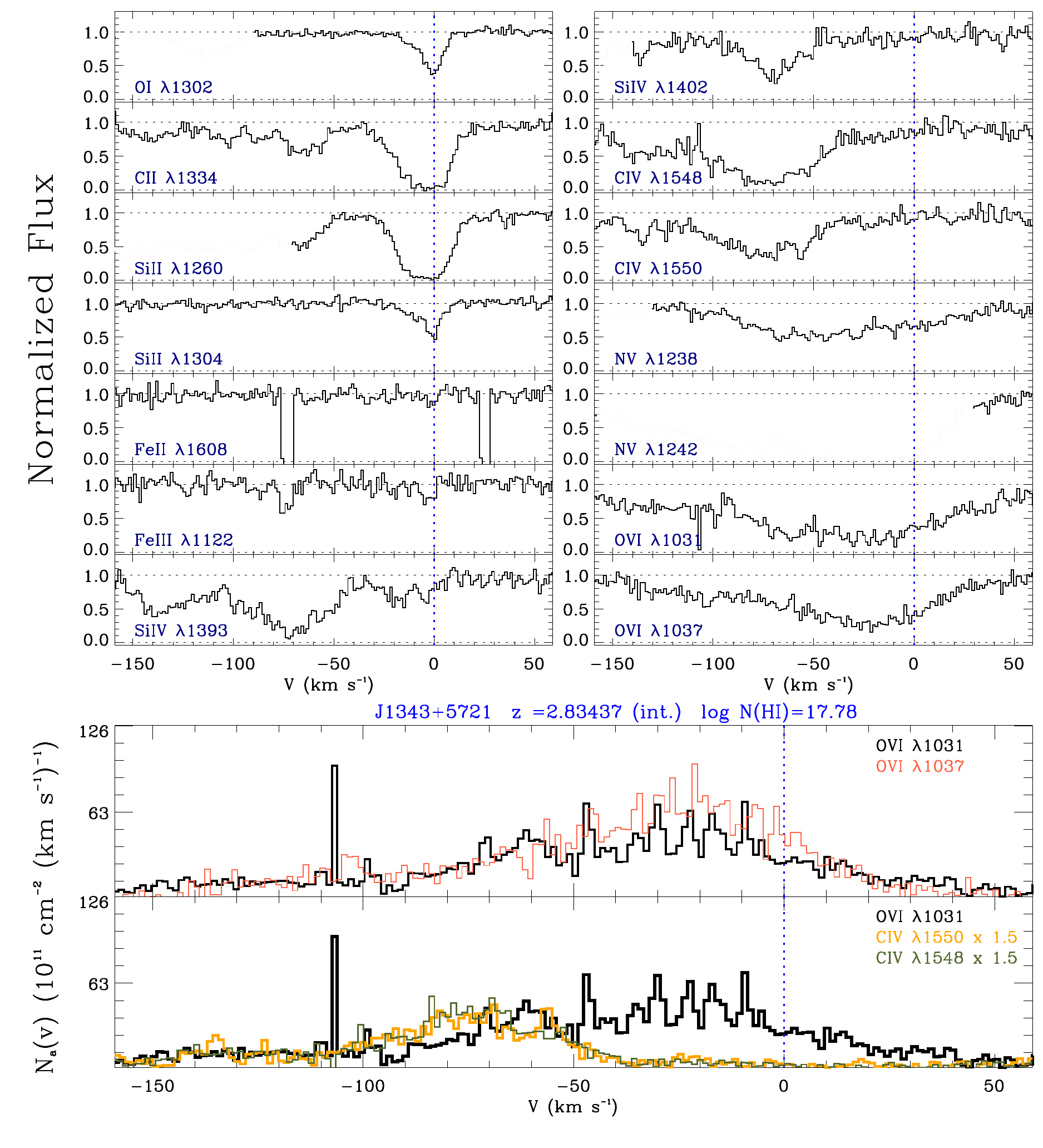}  
  \caption{Same as Fig.~\ref{f-q1009}, but for a different absorber.  \label{f-j1343}}
\end{figure*}

\begin{figure*}[tbp]
\epsscale{1} 
\plotone{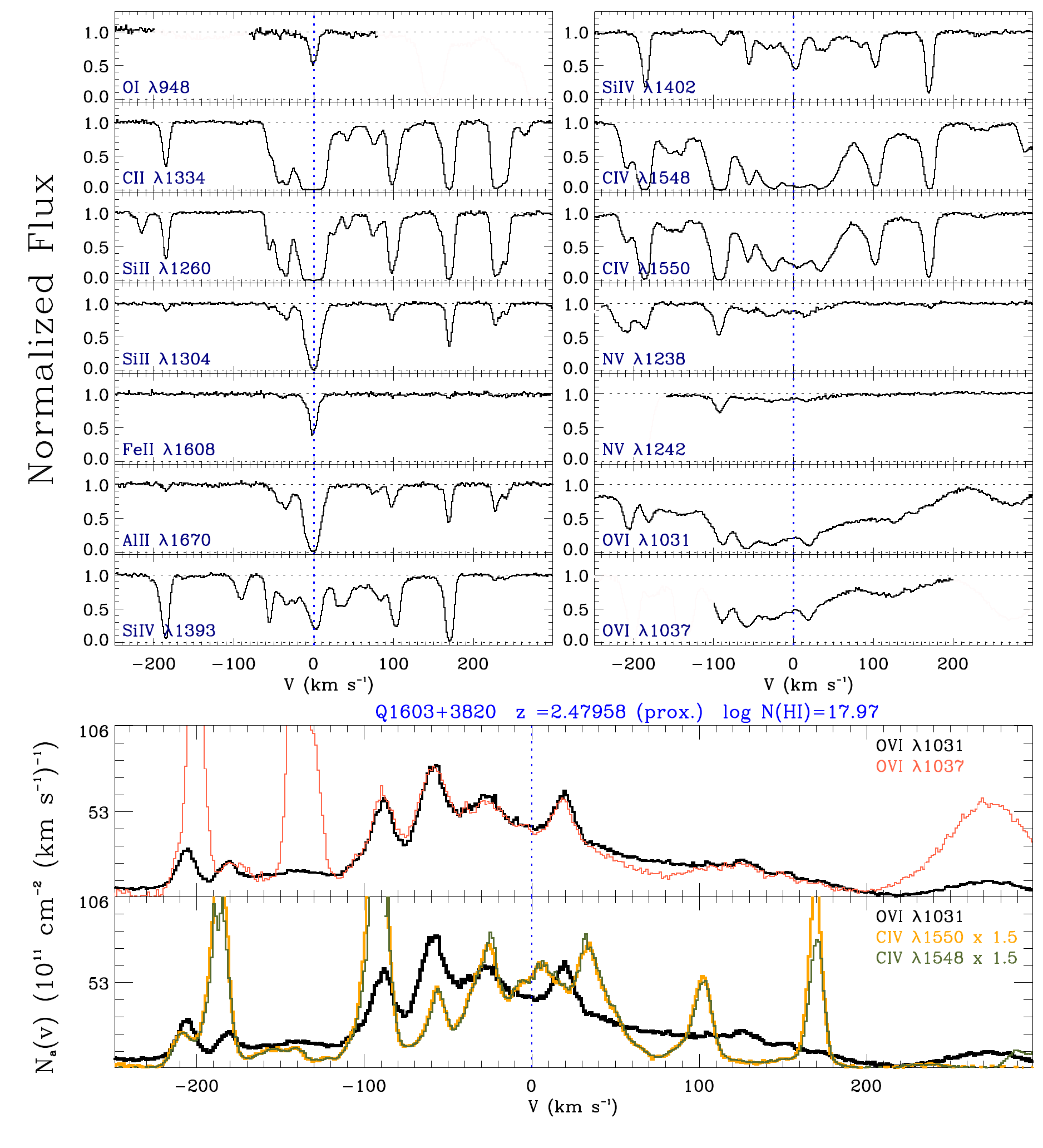}  
  \caption{Same as Fig.~\ref{f-q1009}, but for a different absorber.  \label{f-q1603}}
\end{figure*}

\begin{figure*}[tbp]
\epsscale{1} 
\plotone{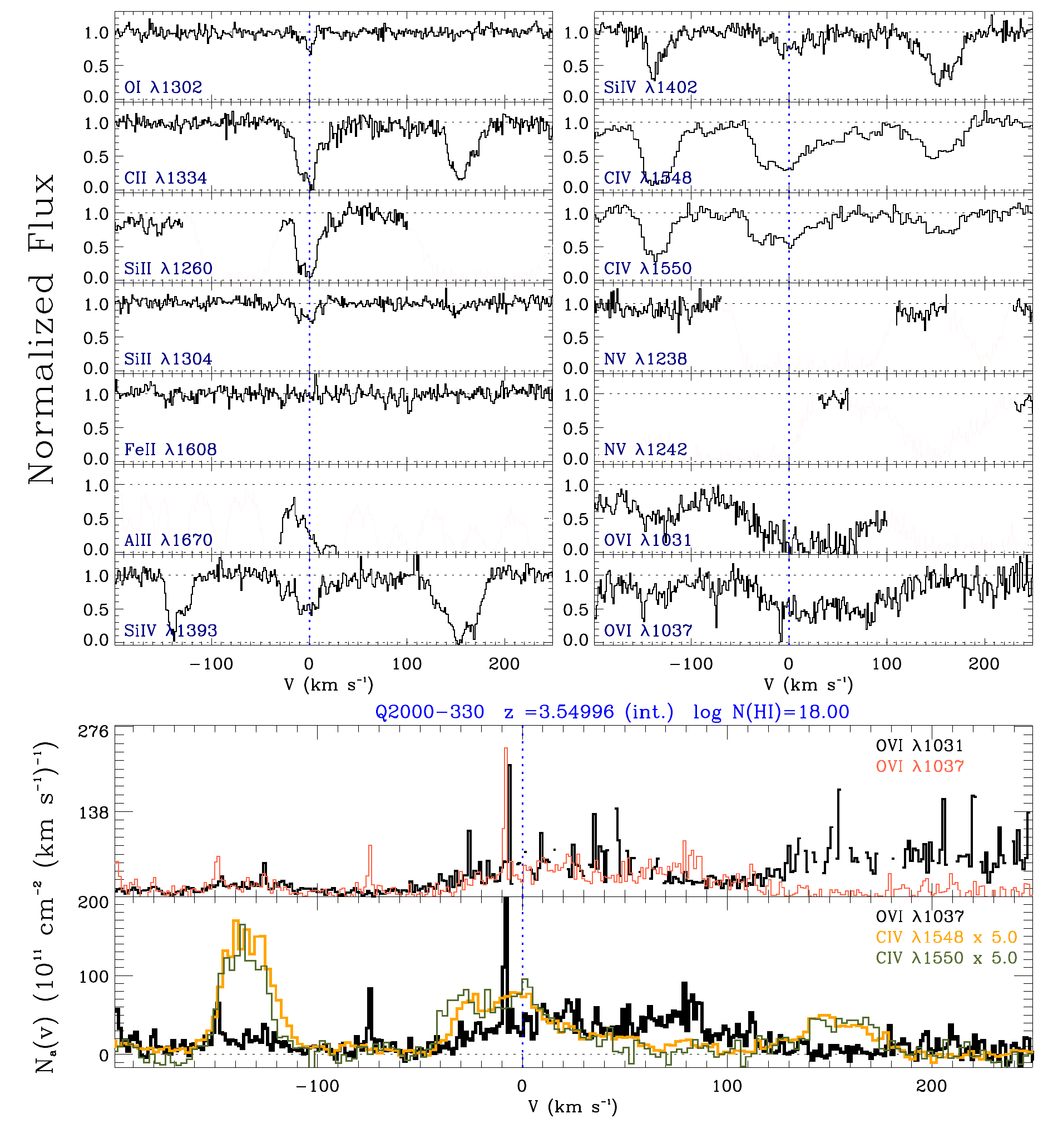}  
  \caption{Same as Fig.~\ref{f-q1009}, but for a different absorber.  \label{f-q2000}}
\end{figure*}

\begin{figure*}[tbp]
\epsscale{1} 
\plotone{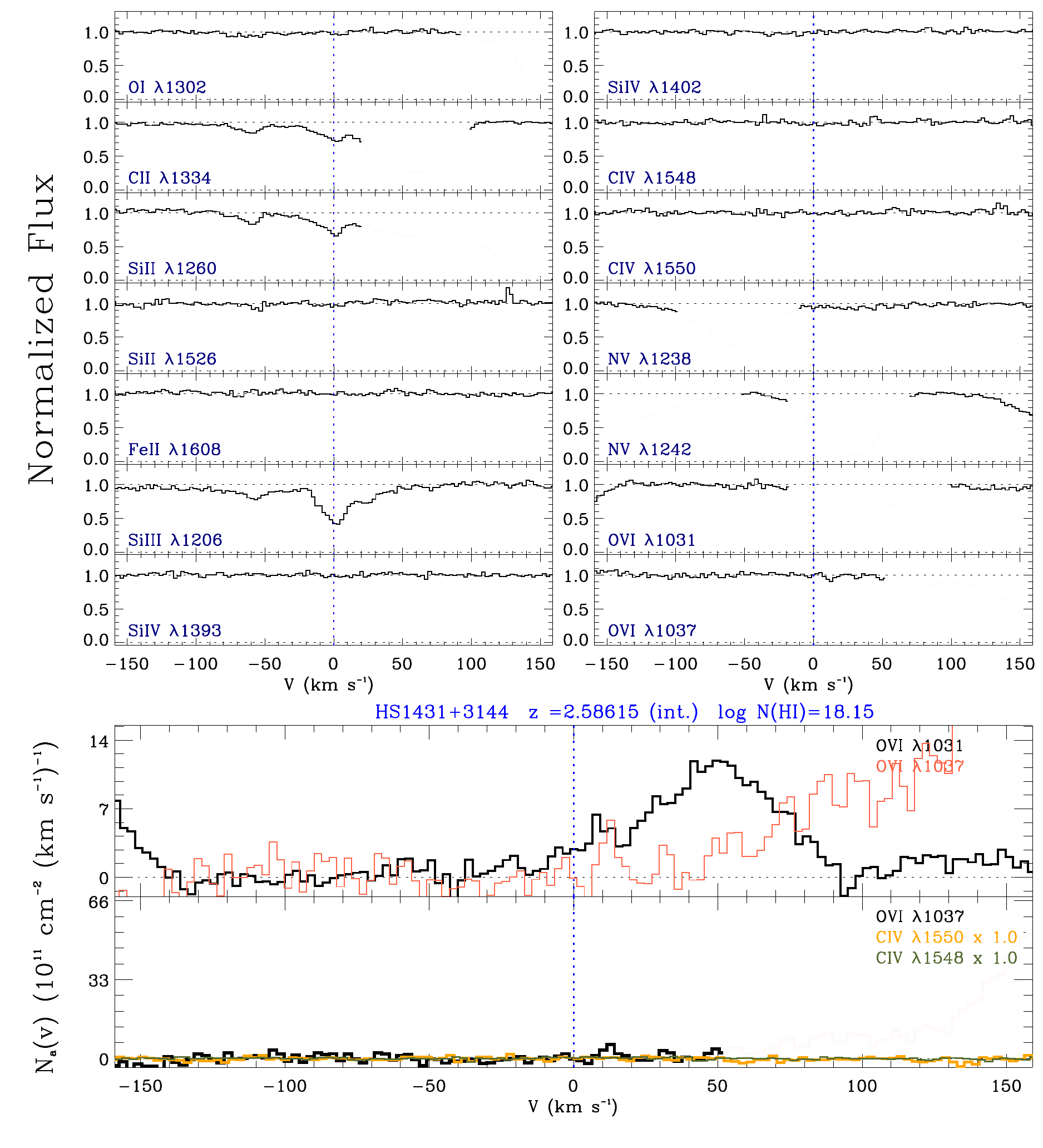}  
  \caption{Same as Fig.~\ref{f-q1009}, but for a different absorber.  \label{f-hs1431}}
\end{figure*}

\begin{figure*}[tbp]
\epsscale{1} 
\plotone{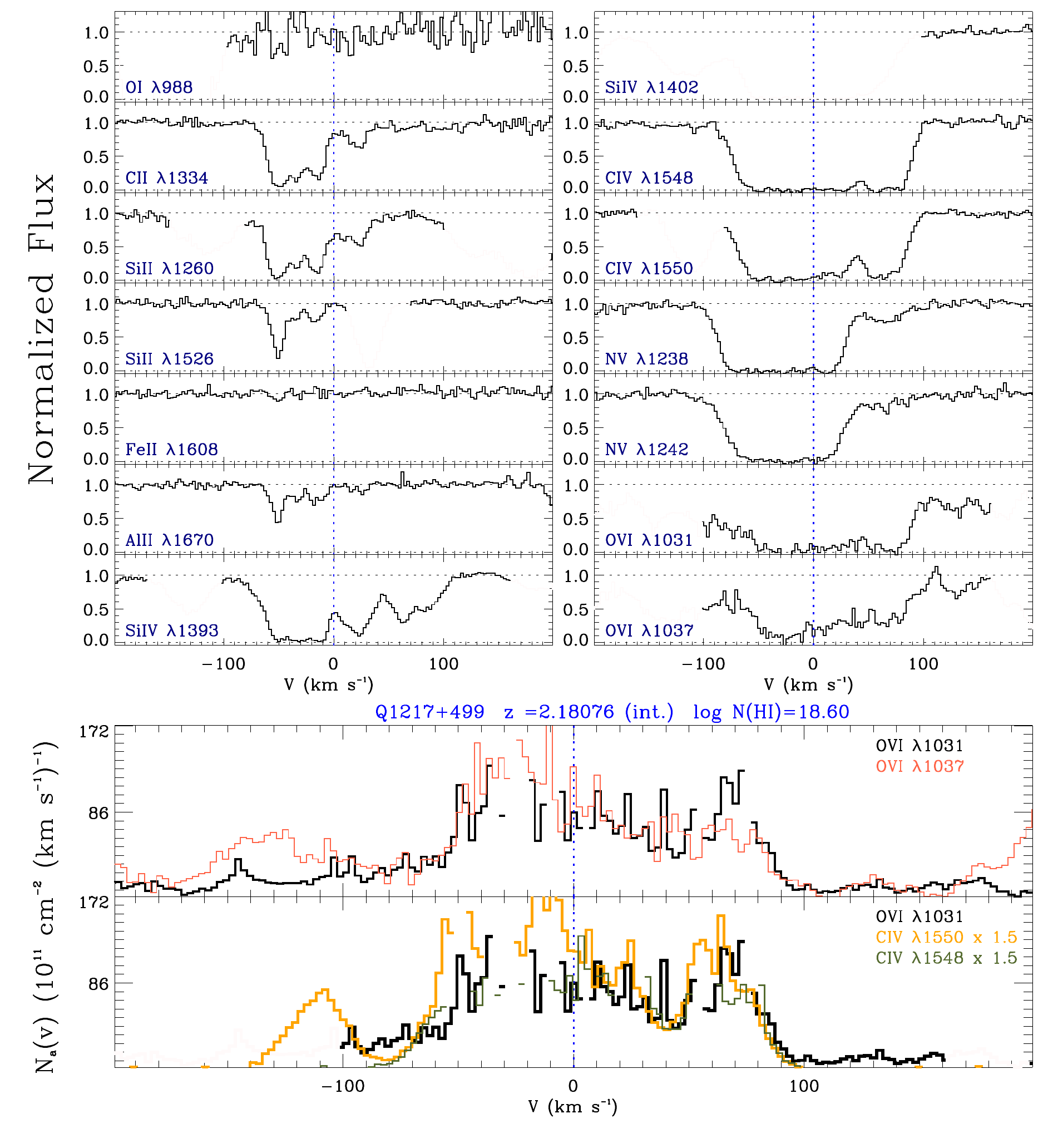}  
  \caption{Same as Fig.~\ref{f-q1009}, but for a different absorber.  \label{f-q1217}}
\end{figure*}

\clearpage

\begin{figure*}[tbp]
\epsscale{1} 
\plotone{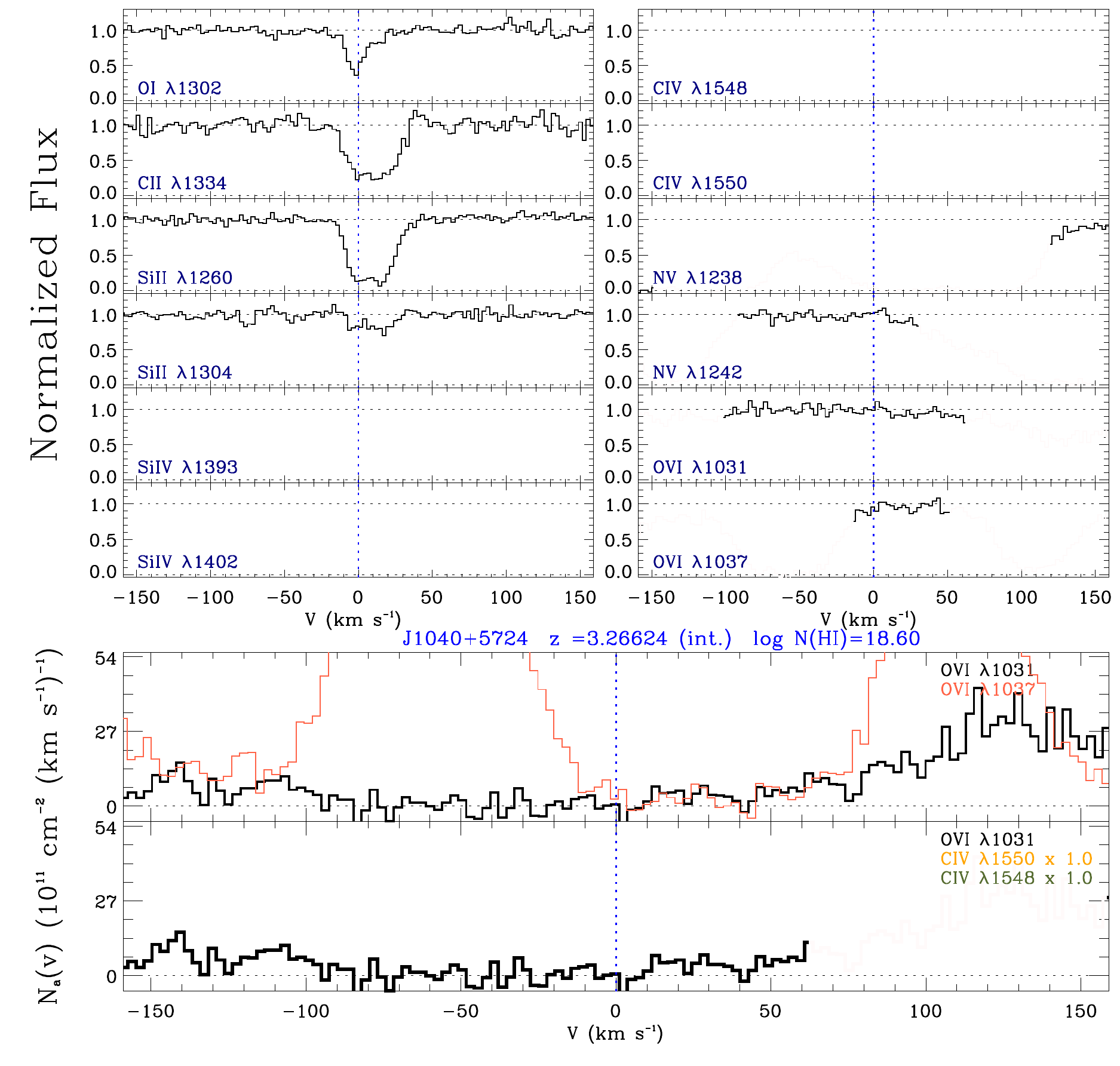}  
  \caption{Same as Fig.~\ref{f-q1009}, but for a different absorber.  \label{f-j1040}}
\end{figure*}

\begin{figure*}[tbp]
\epsscale{1} 
\plotone{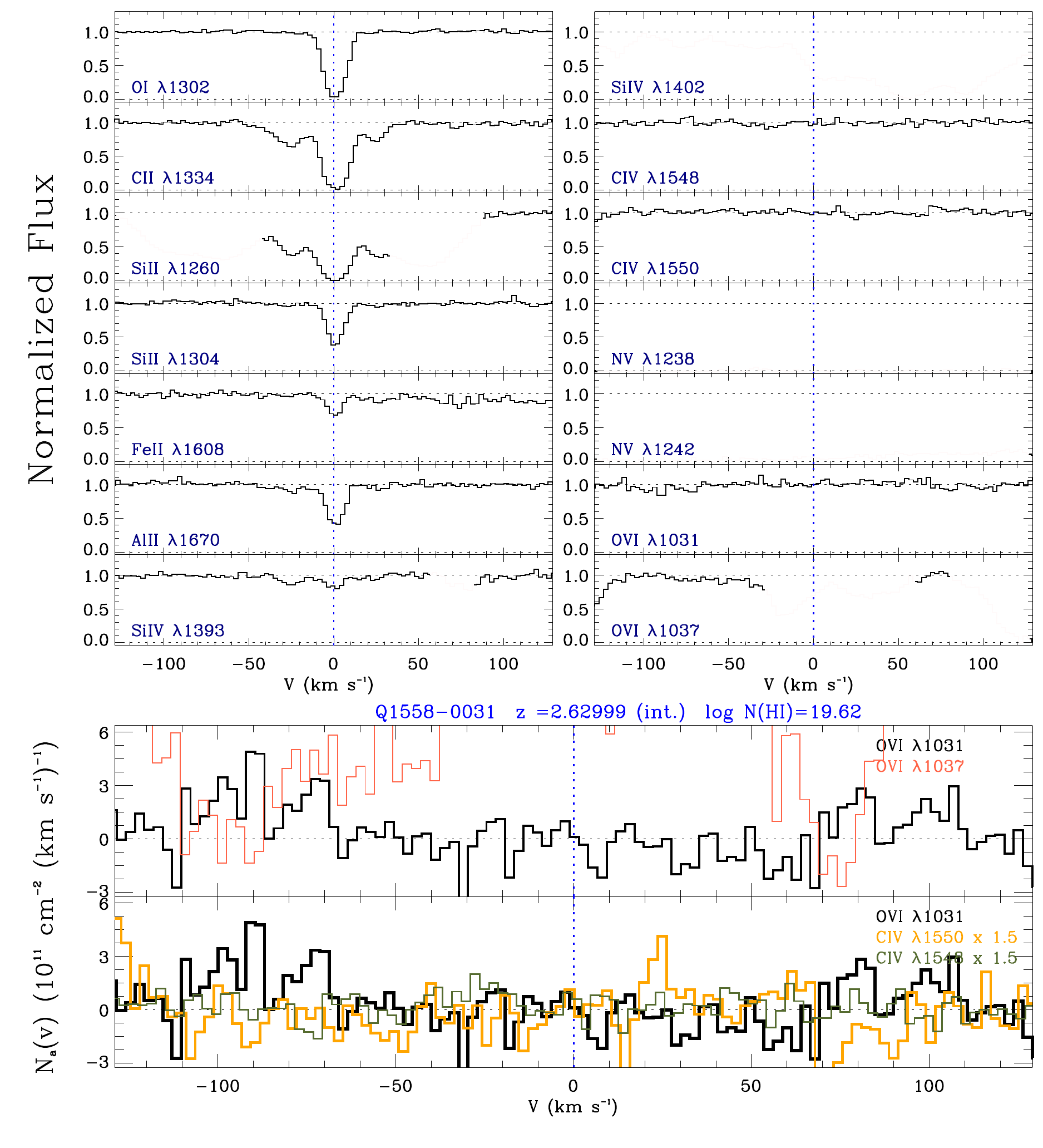}  
  \caption{Same as Fig.~\ref{f-q1009}, but for a different absorber.  \label{f-q1558}}
\end{figure*}

\begin{figure*}[tbp]
\epsscale{1} 
\plotone{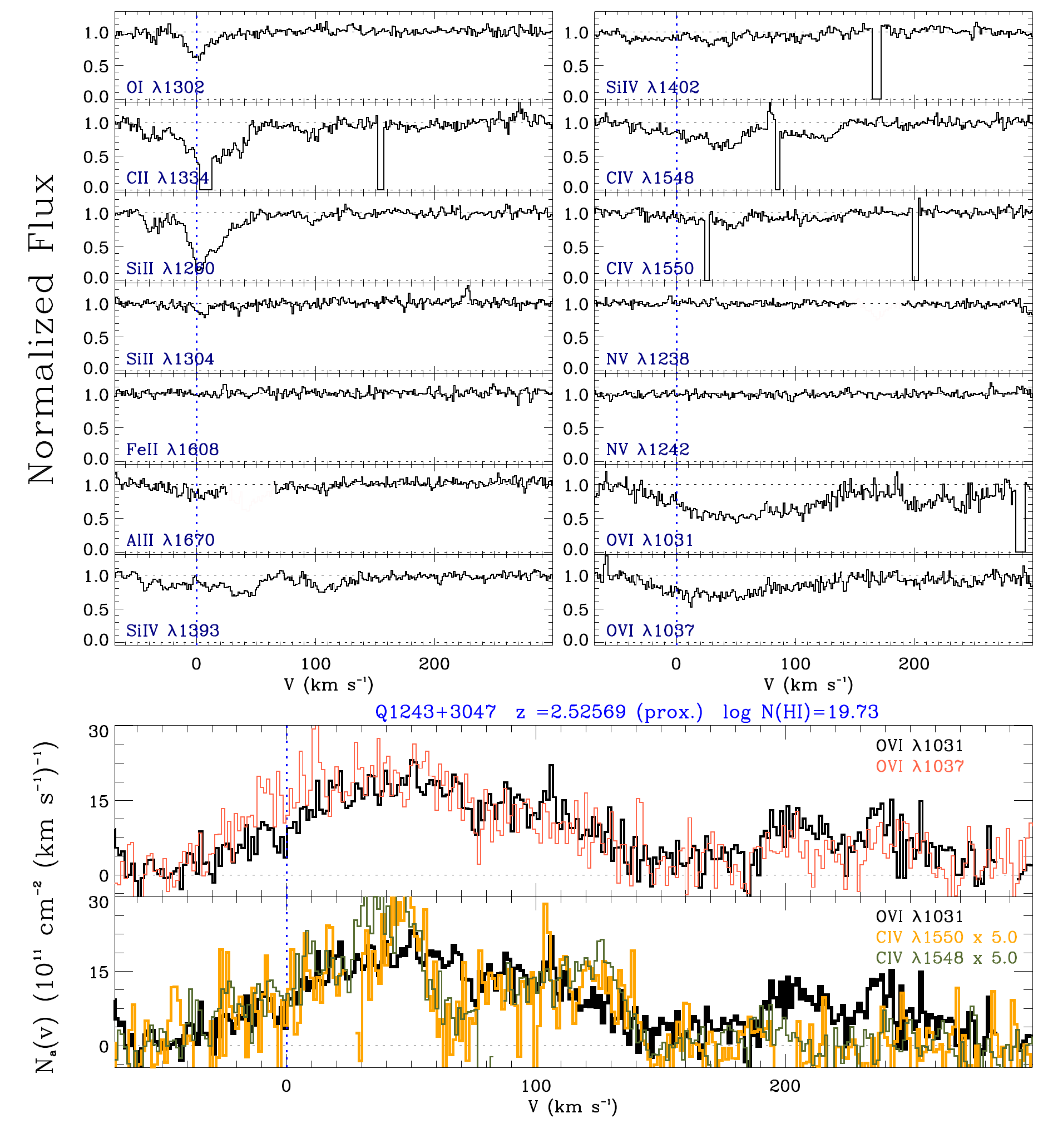}  
  \caption{Same as Fig.~\ref{f-q1009}, but for a different absorber.  \label{f-q1243}}
\end{figure*}

\begin{figure*}[tbp]
\epsscale{1} 
\plotone{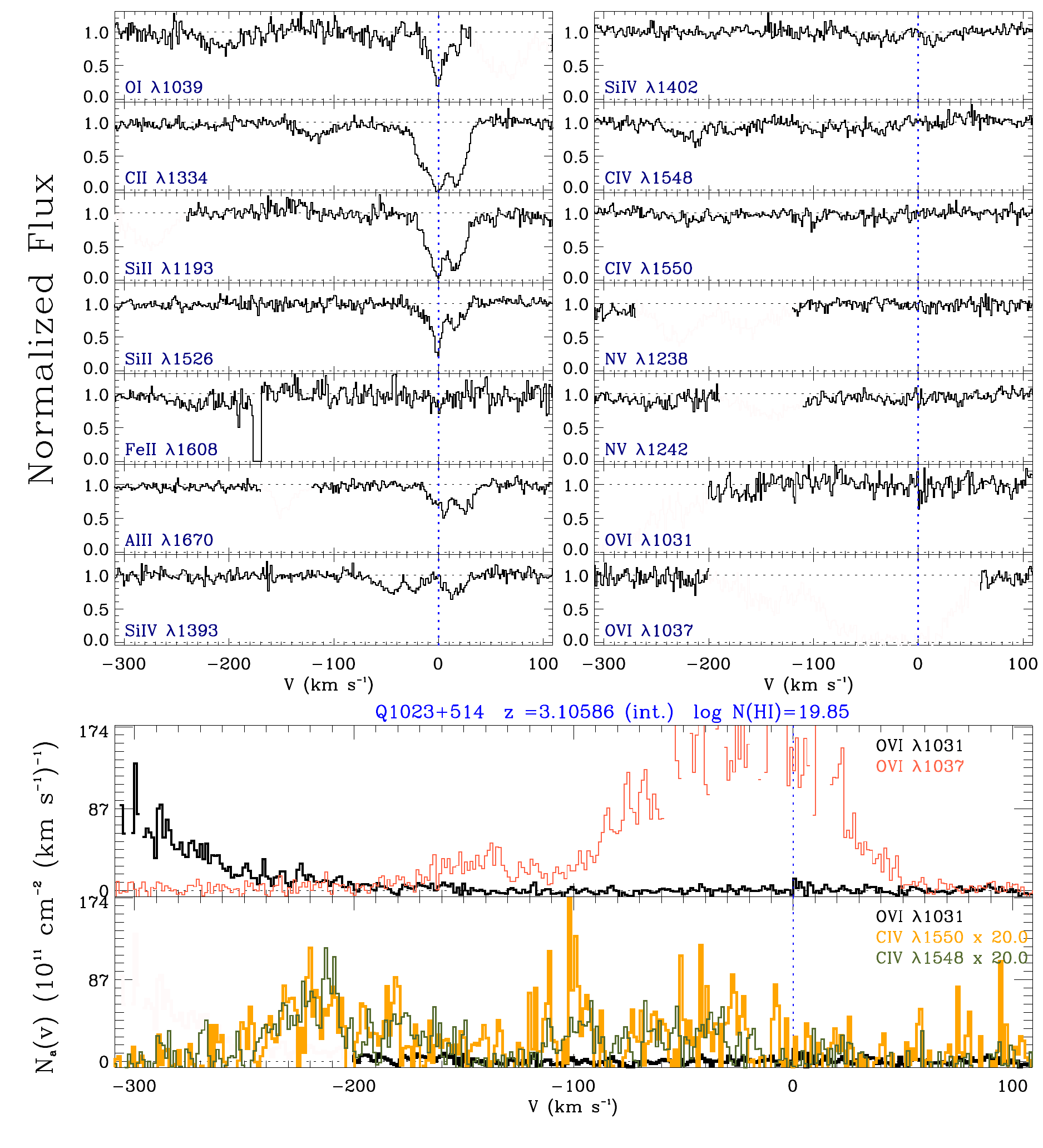}  
  \caption{Same as Fig.~\ref{f-q1009}, but for a different absorber.  \label{f-q1023}}
\end{figure*}

\begin{figure*}[tbp]
\epsscale{1} 
\plotone{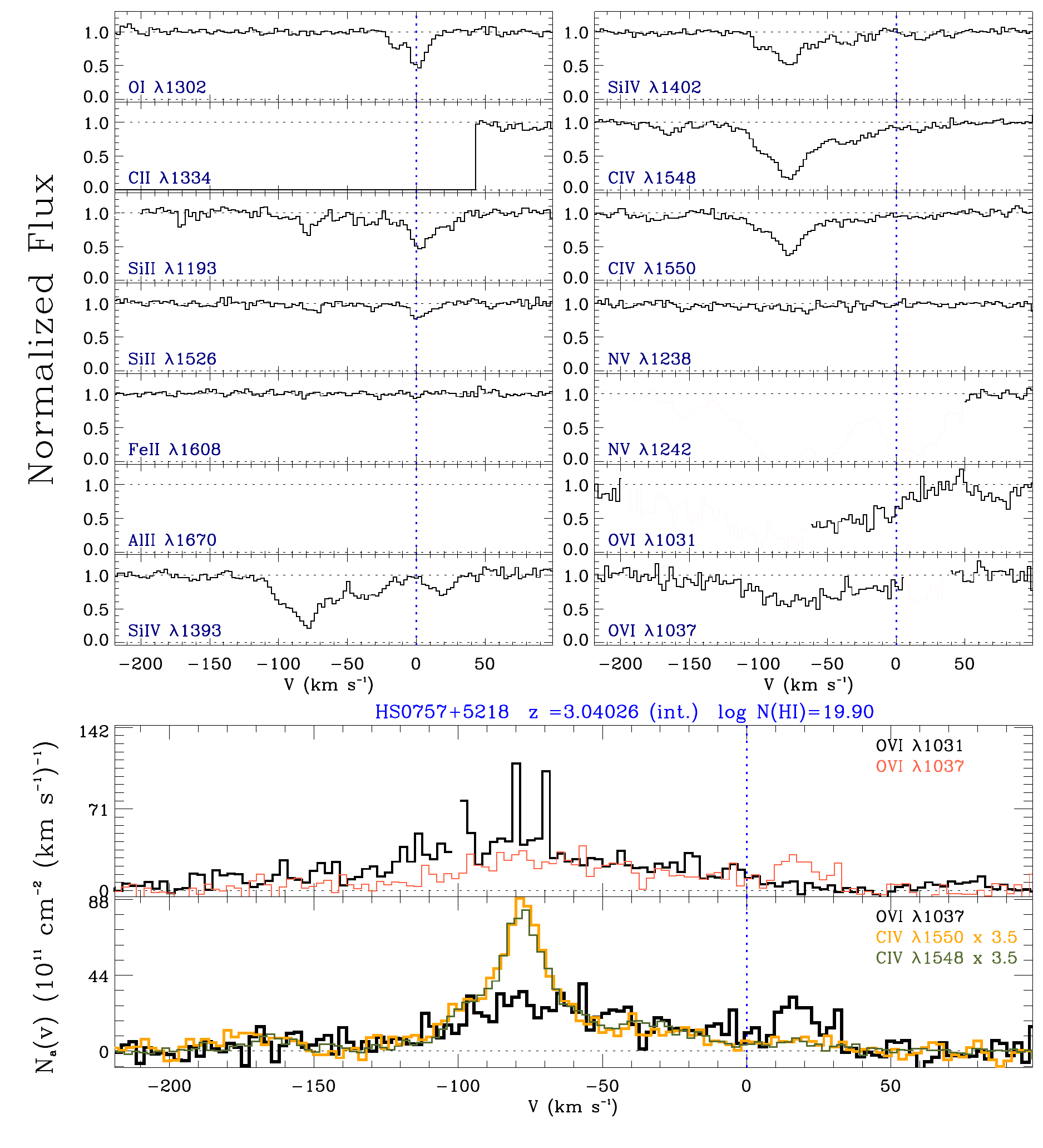}  
  \caption{Same as Fig.~\ref{f-q1009}, but for a different absorber.  \label{f-hs0757}}
\end{figure*}

\begin{figure*}[tbp]
\epsscale{1} 
\plotone{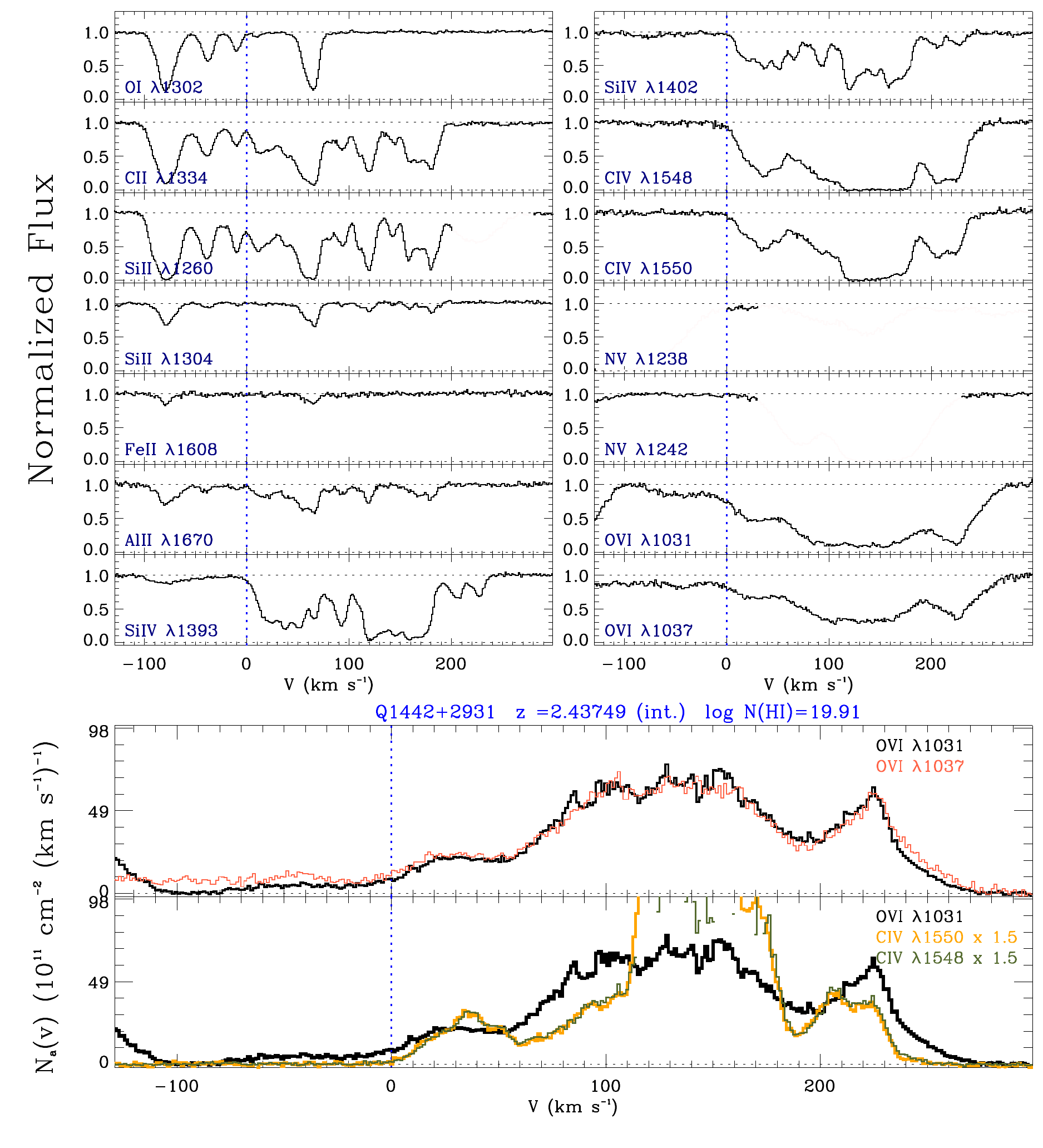}  
  \caption{Same as Fig.~\ref{f-q1009}, but for a different absorber.  \label{f-q1442}}
\end{figure*}

\begin{figure*}[tbp]
\epsscale{1} 
\plotone{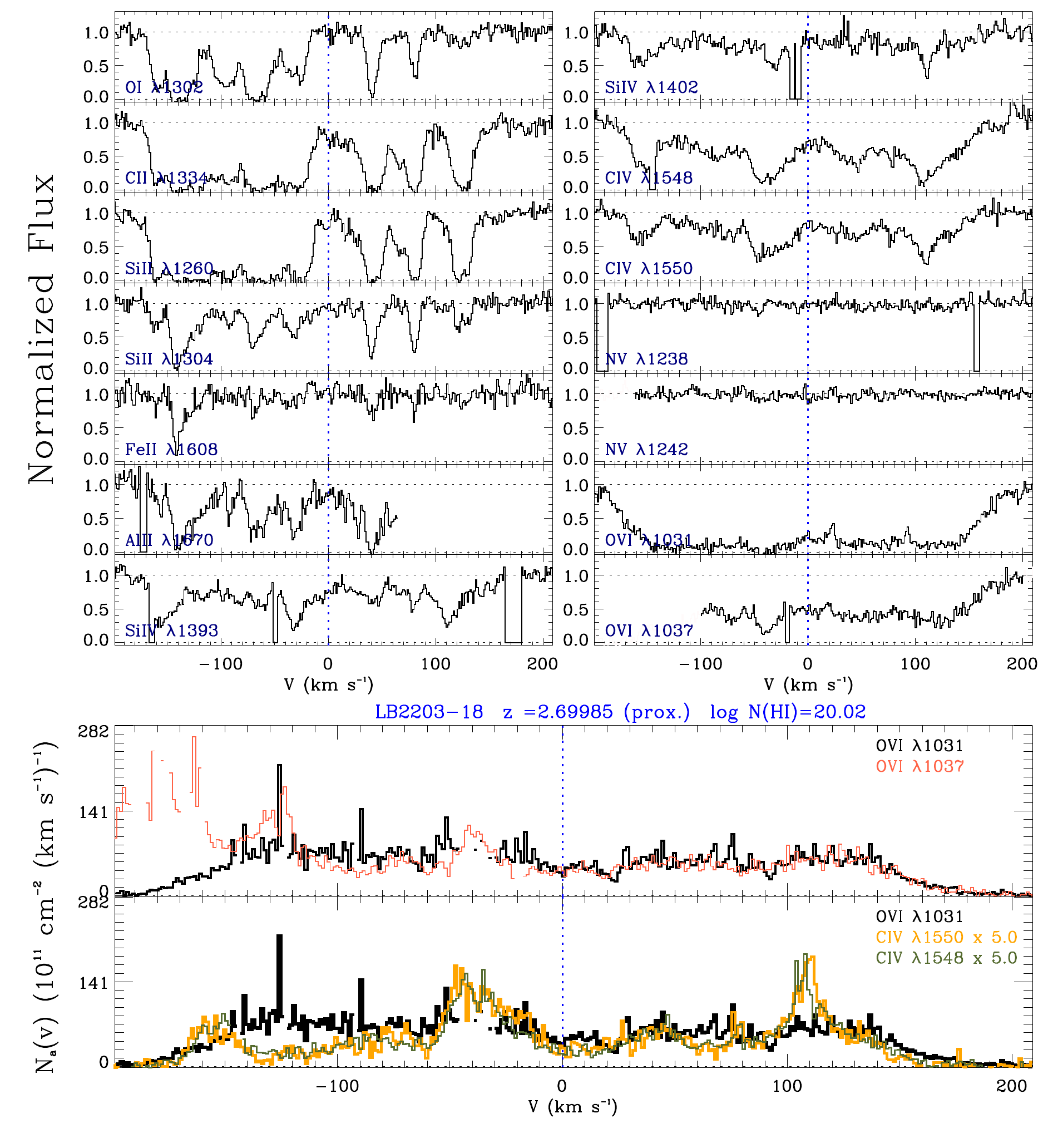}  
  \caption{Same as Fig.~\ref{f-q1009}, but for a different absorber.  \label{f-lb2203}}
\end{figure*}

\begin{figure*}[tbp]
\epsscale{1} 
\plotone{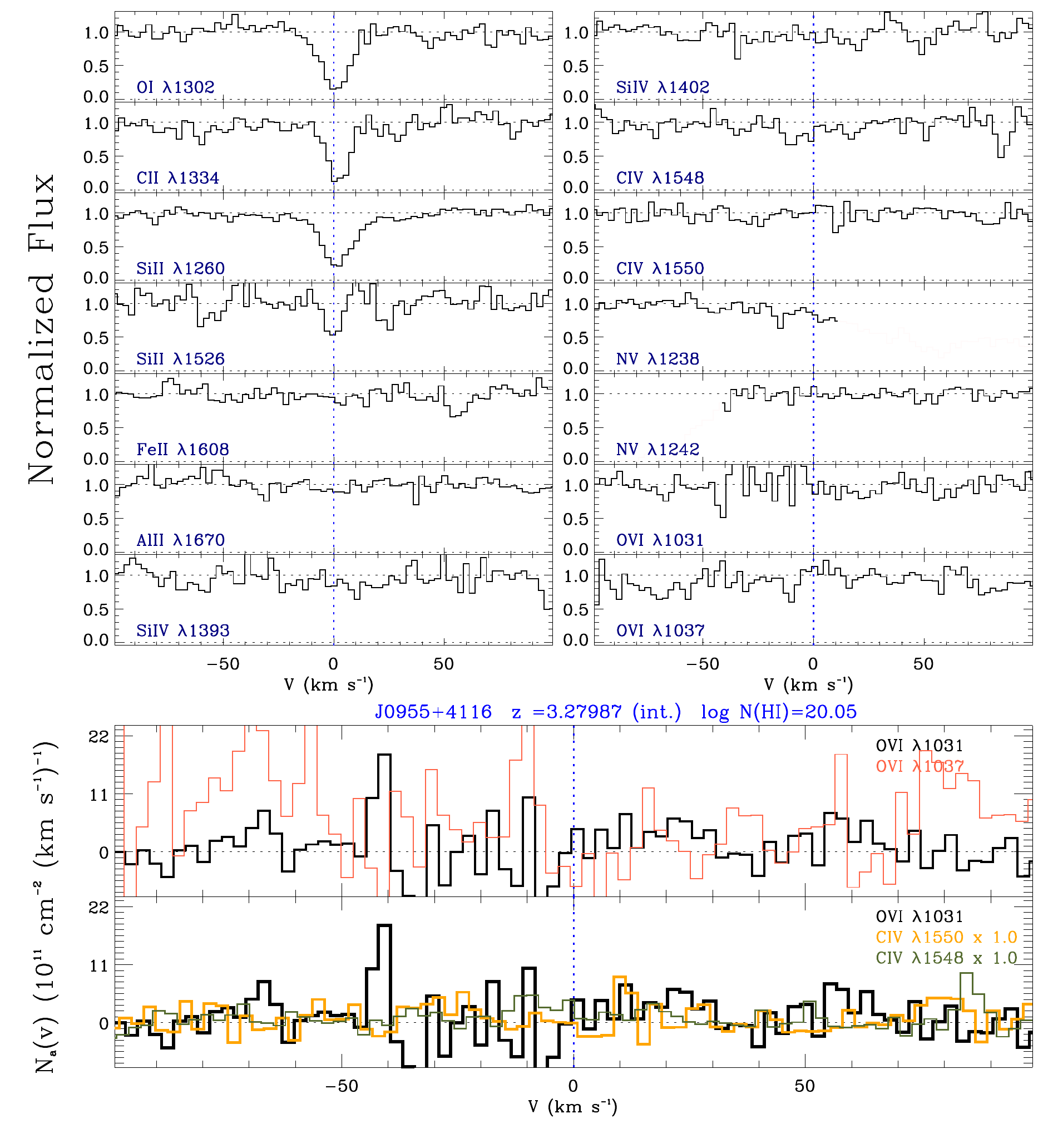}  
  \caption{Same as Fig.~\ref{f-q1009}, but for a different absorber.  \label{f-j0955}}
\end{figure*}

\begin{figure*}[tbp]
\epsscale{1} 
\plotone{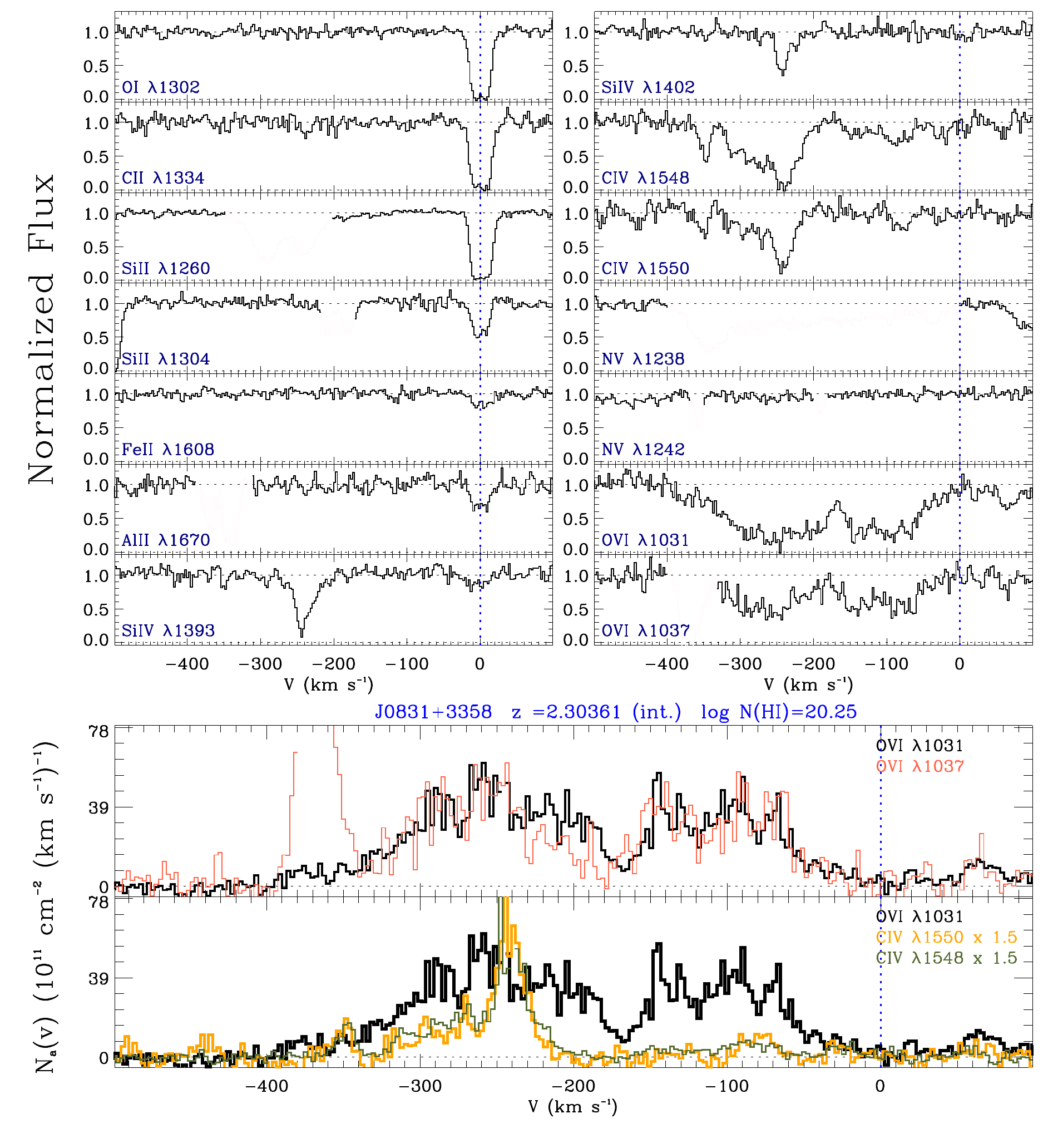}  
  \caption{Same as Fig.~\ref{f-q1009}, but for a different absorber.  \label{f-j0831}}
\end{figure*}

\begin{figure*}[tbp]
\epsscale{1} 
\plotone{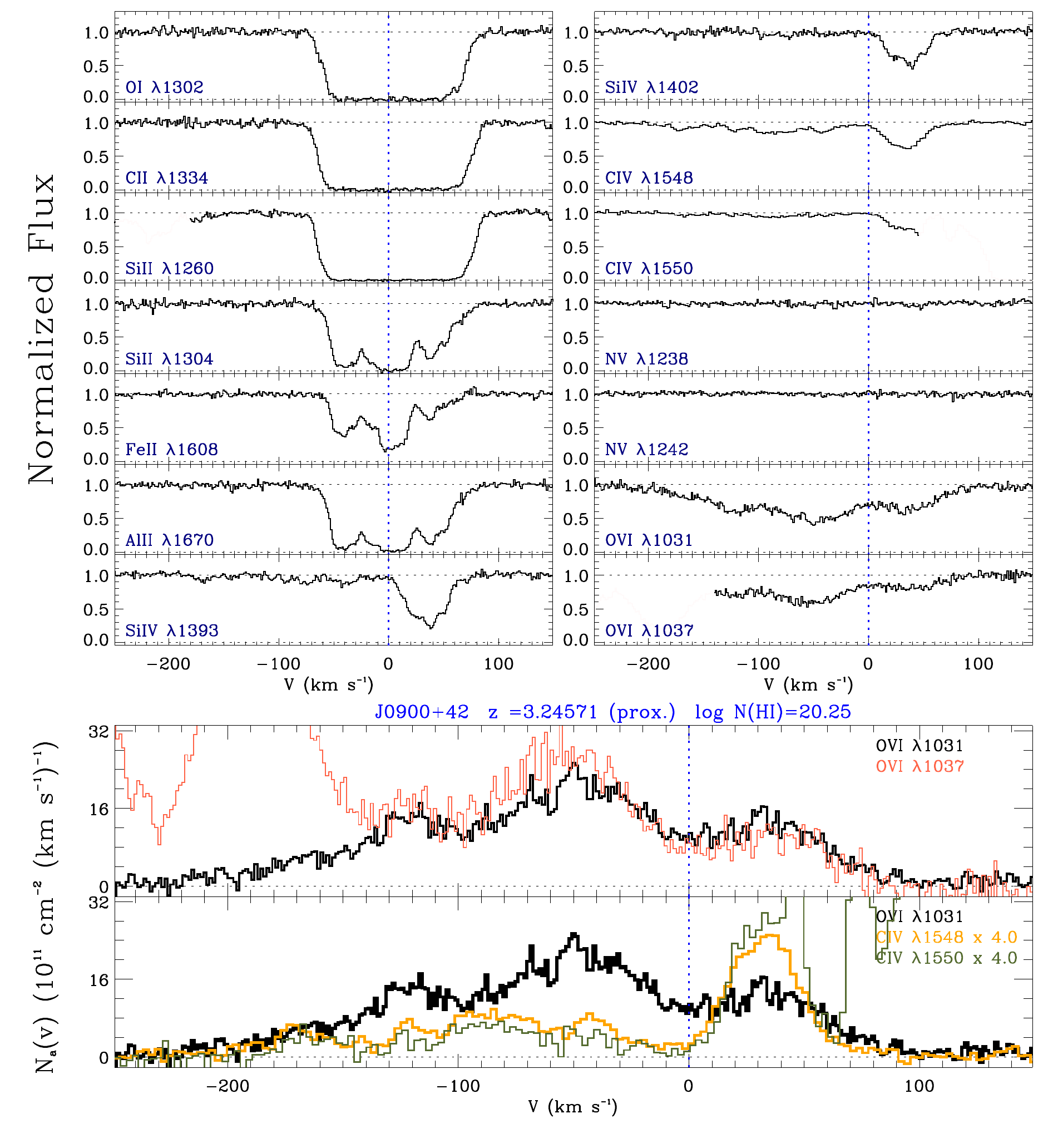}  
  \caption{Same as Fig.~\ref{f-q1009}, but for a different absorber.  \label{f-j0900} }
\end{figure*}

\begin{figure*}[tbp]
\epsscale{1} 
\plotone{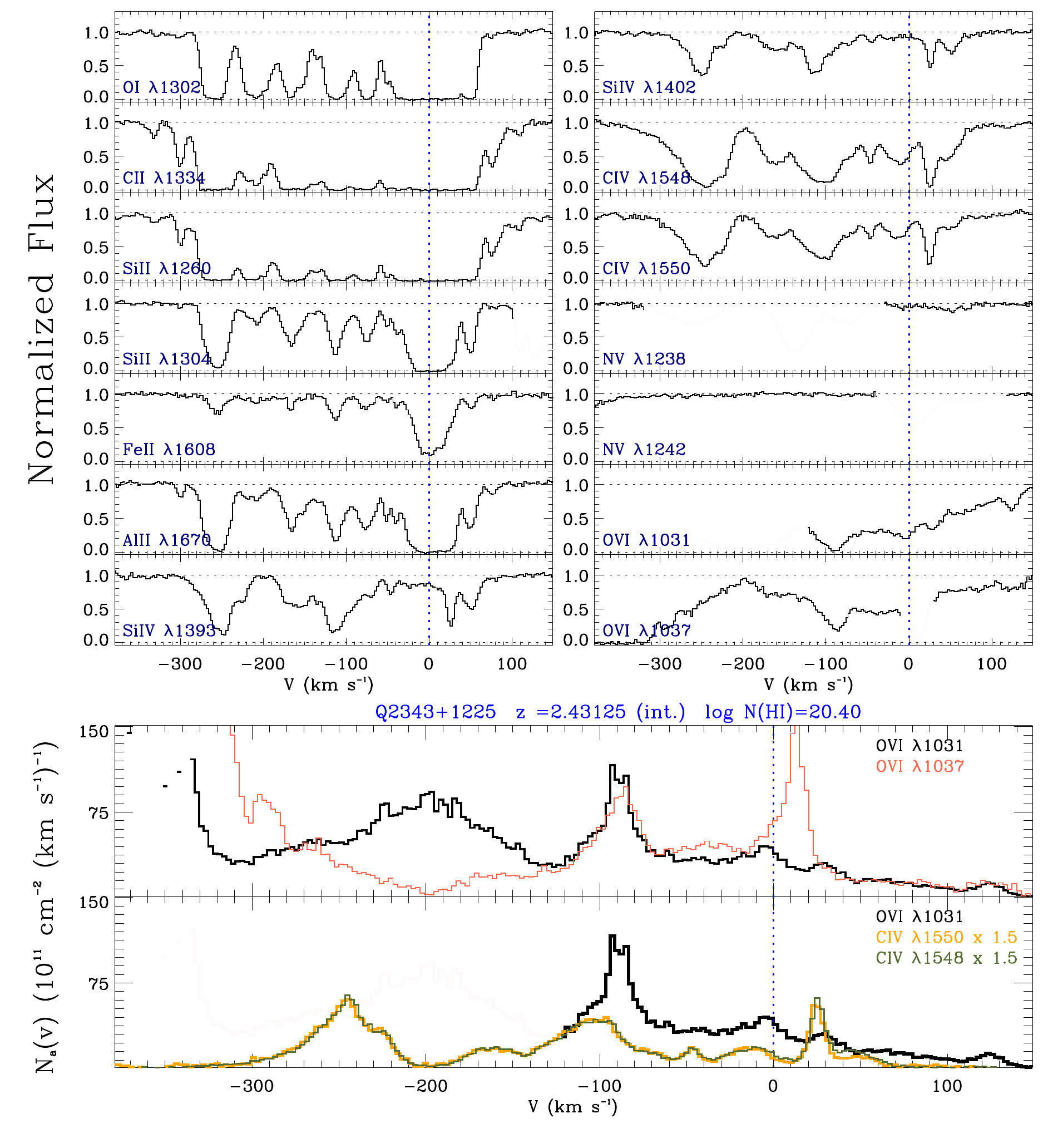}  
  \caption{Same as Fig.~\ref{f-q1009}, but for a different absorber.  \label{f-q2343}}
\end{figure*}

\begin{figure*}[tbp]
\epsscale{1} 
\plotone{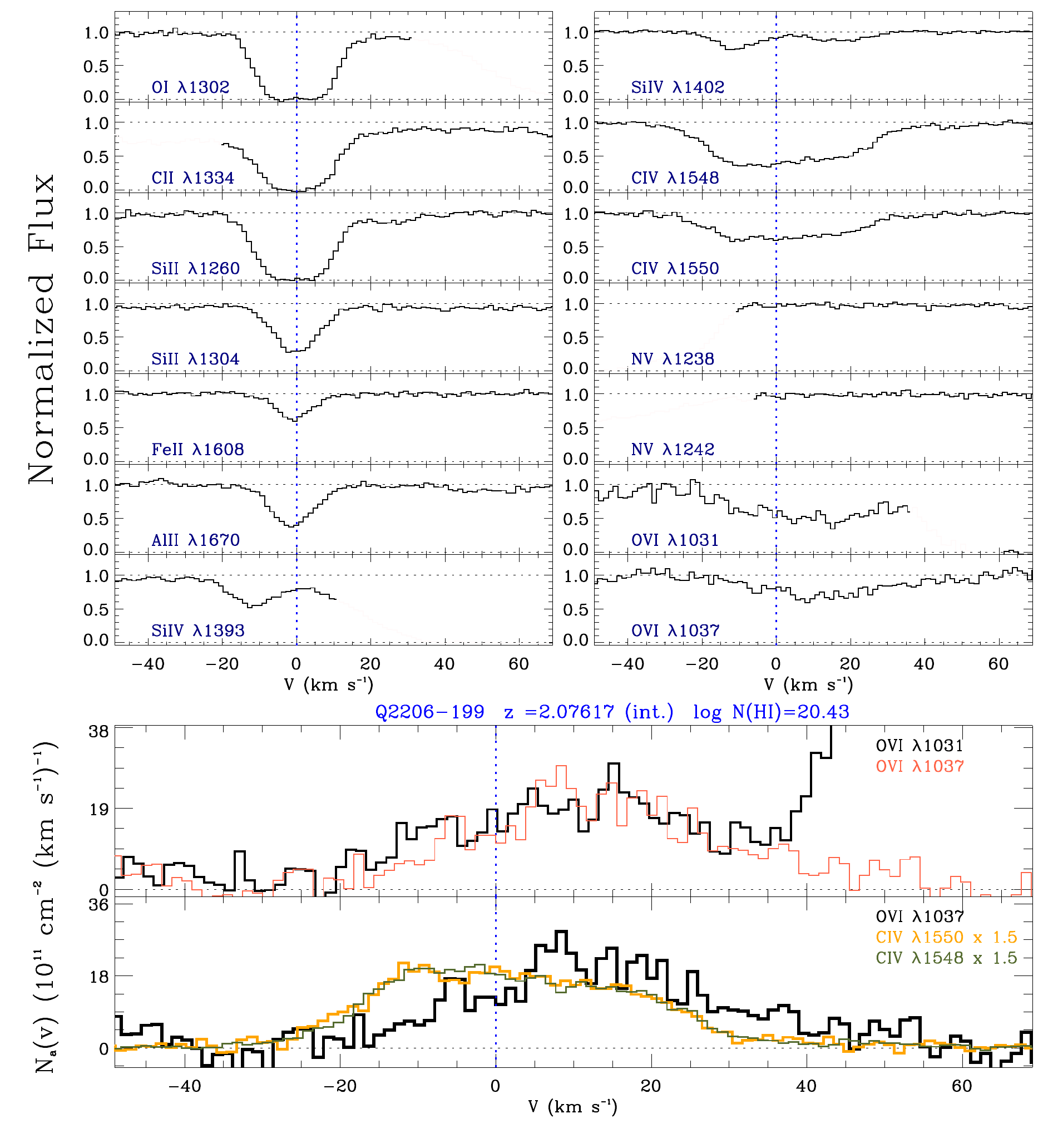}  
  \caption{Same as Fig.~\ref{f-q1009}, but for a different absorber.  \label{f-q2206}}
\end{figure*}

\begin{figure*}[tbp]
\epsscale{1} 
\plotone{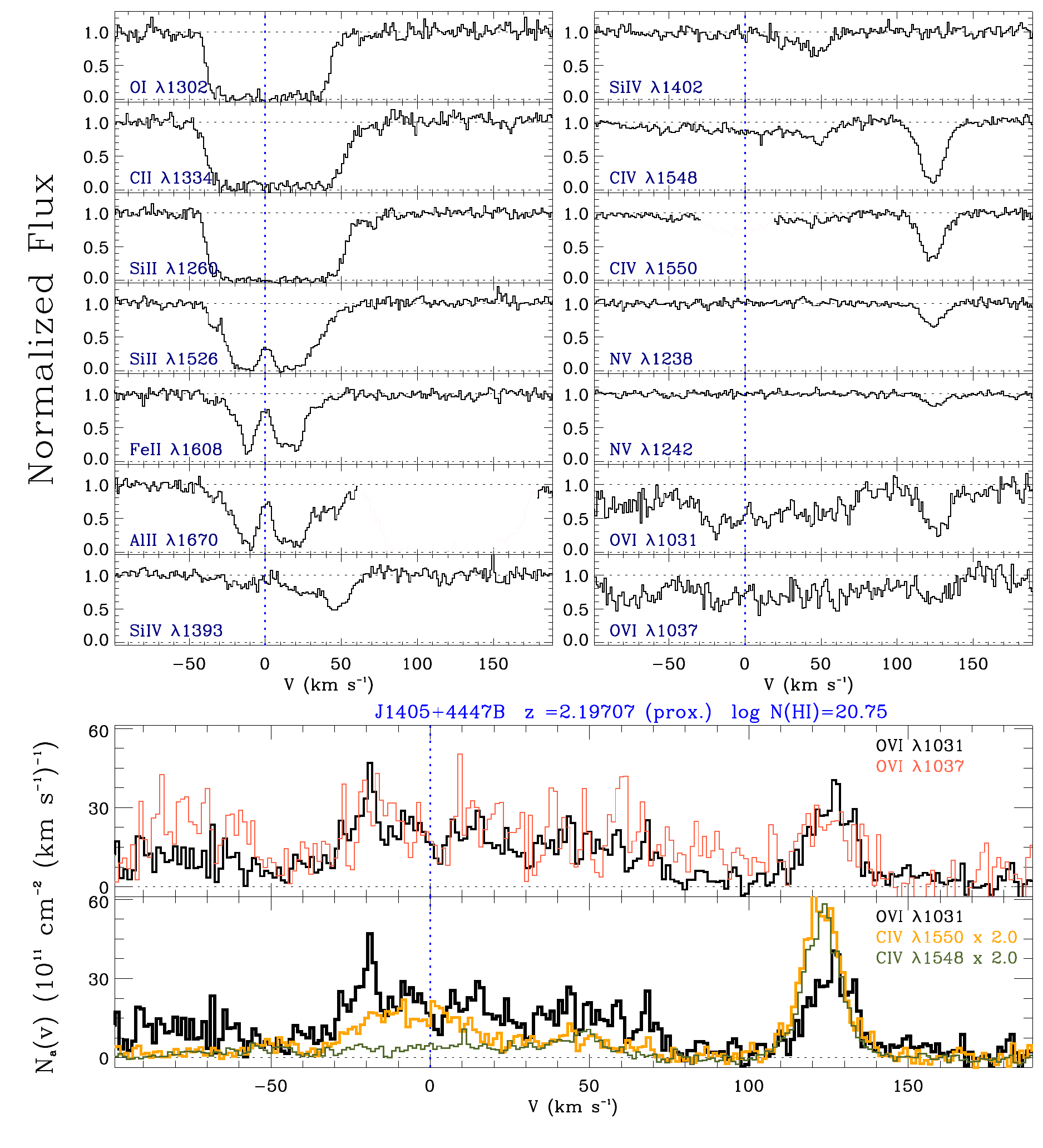}  
  \caption{Same as Fig.~\ref{f-q1009}, but for a different absorber.  \label{f-j1405}}
\end{figure*}

\begin{figure*}[tbp]
\epsscale{1} 
\plotone{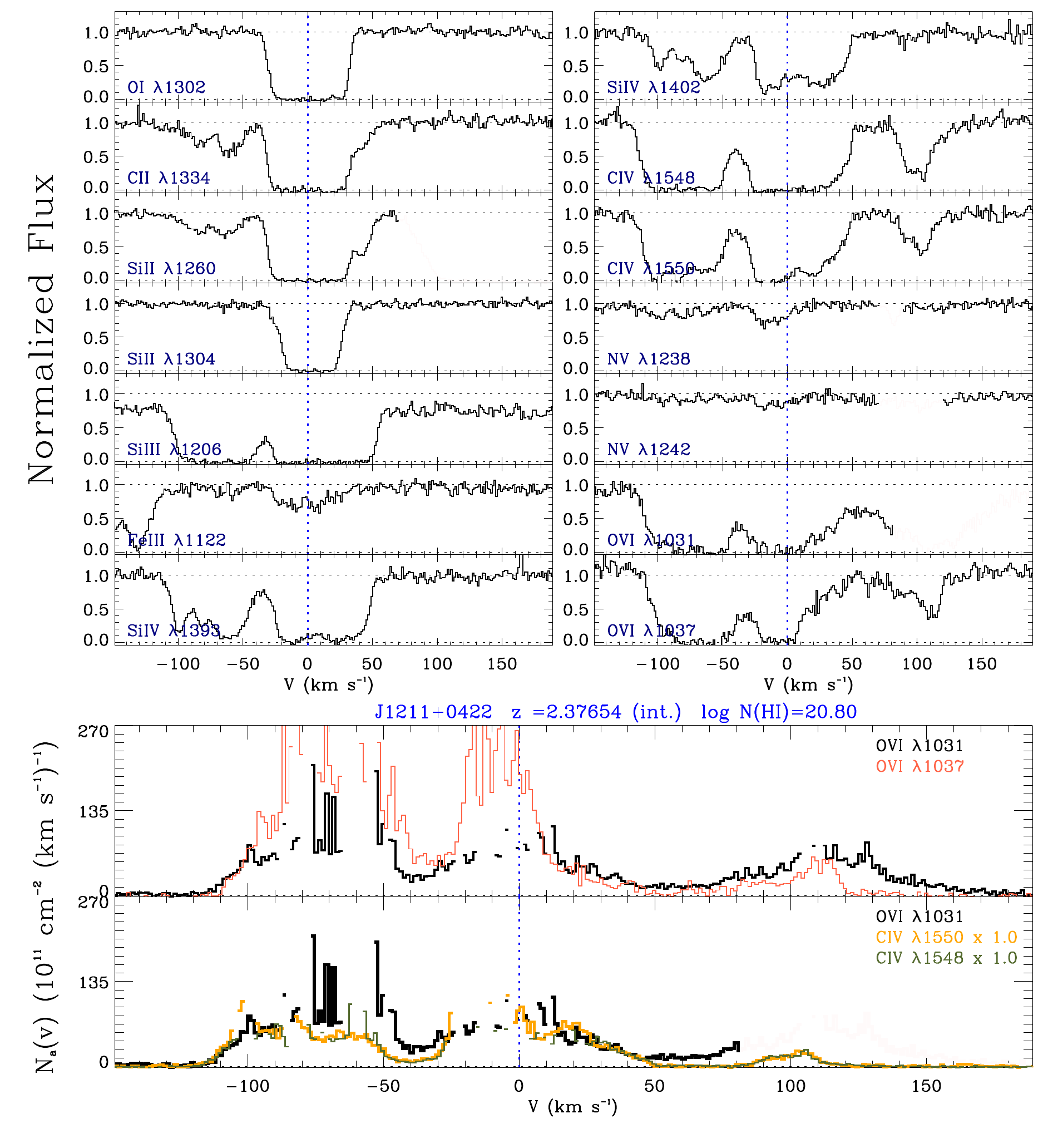}  
  \caption{Same as Fig.~\ref{f-q1009}, but for a different absorber.  \label{f-j1211}}
\end{figure*}

\clearpage
\input{tabA1.tex}

\clearpage

\begin{figure}[tbp]
\epsscale{1} 
\plottwo{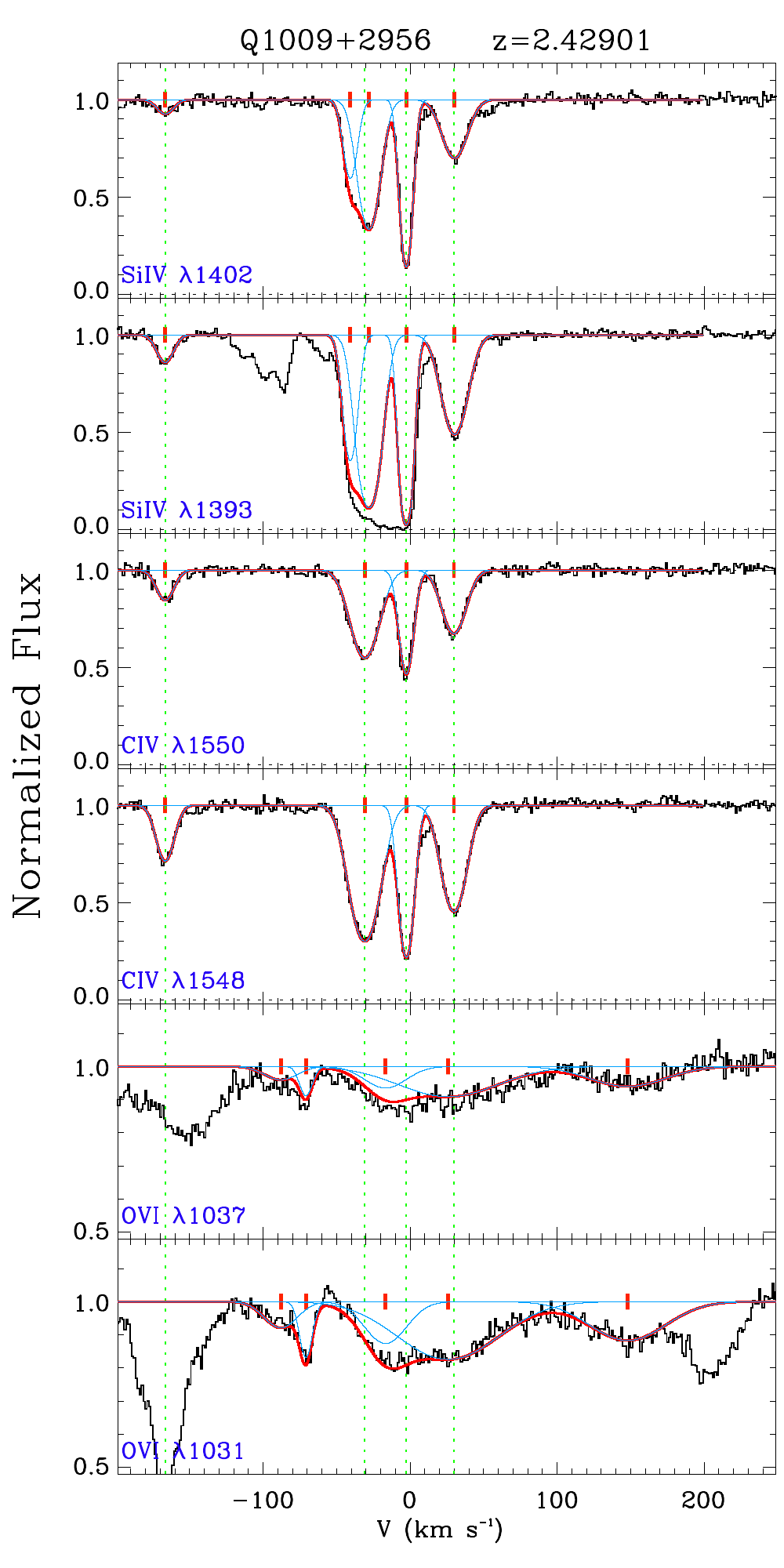}{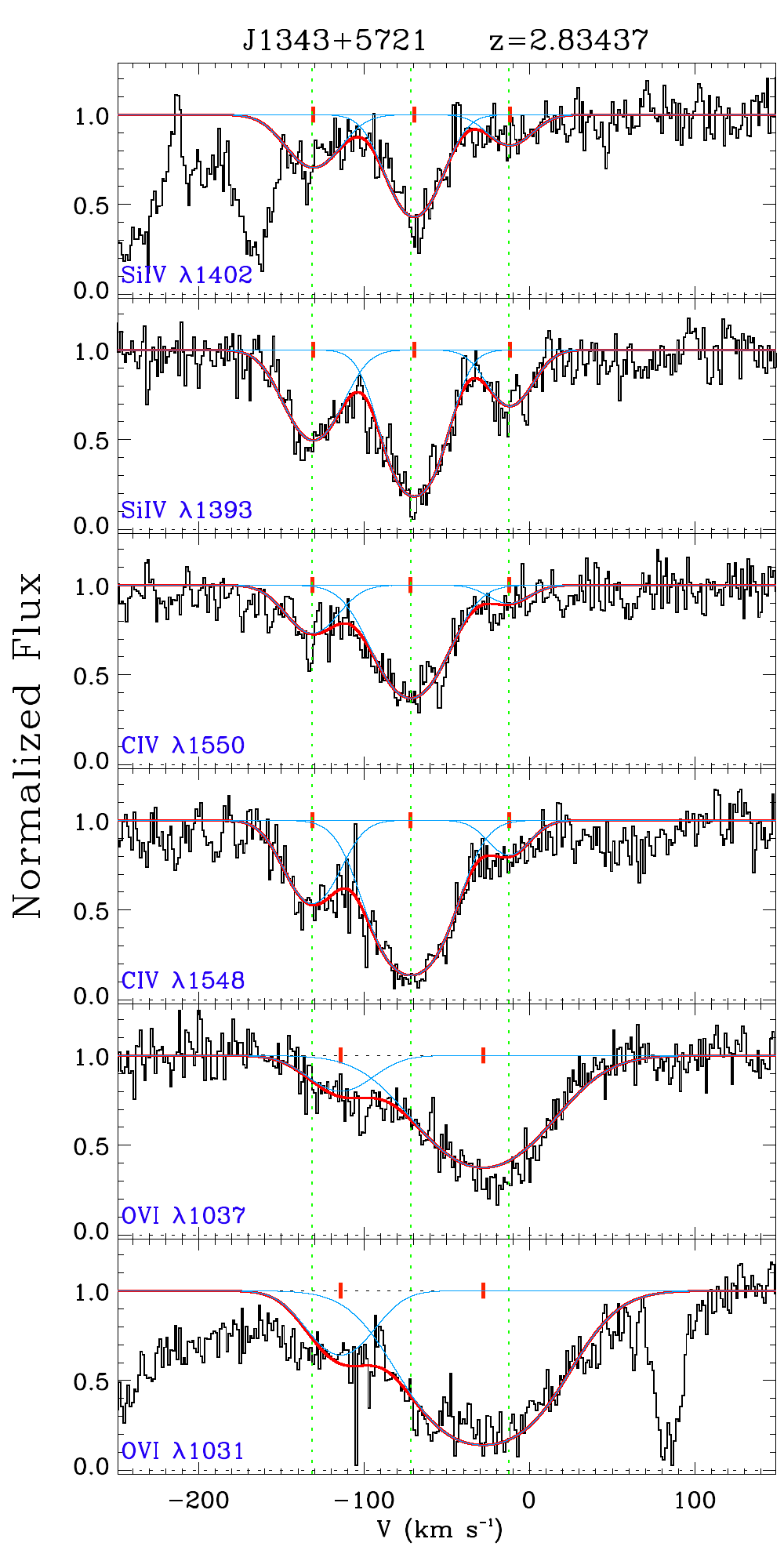}  
  \caption{Normalized profiles of the \siivt, \civt, and \ovit\ doublets. Each doublet was fitted simultaneously, but each species was fitted independently. The red lines indicate the global component model and the blue lines show the individual components. The vertical green dotted lines represent the velocity centroids for \civt\ while the red tick-marks show the velocity centroids for each species. Absorption that is not fitted corresponds to a contamination from an unrelated absorber. 
 \label{f-fitq1009}}
\end{figure}

\begin{figure}[tbp]
\epsscale{1} 
\plottwo{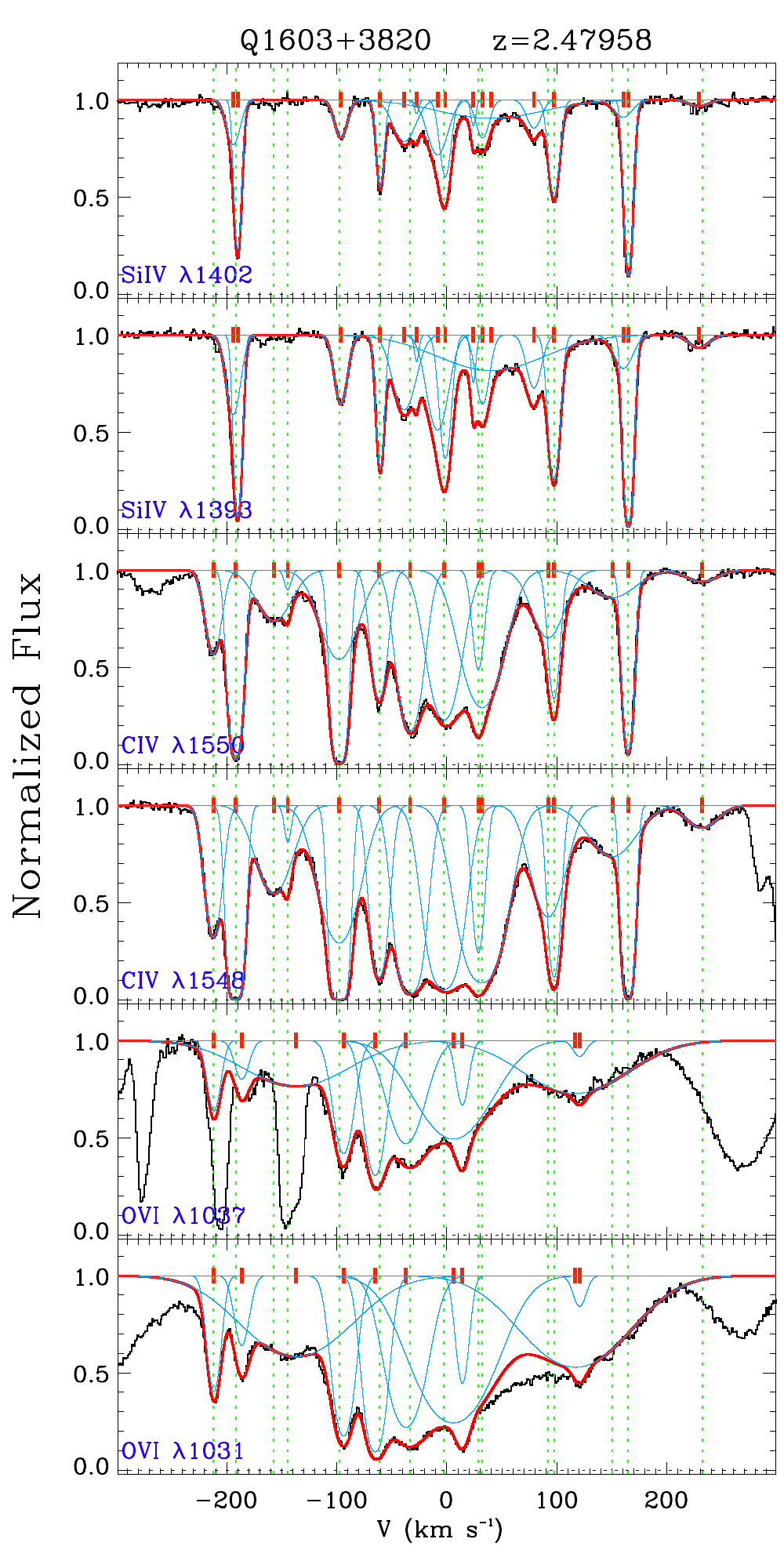}{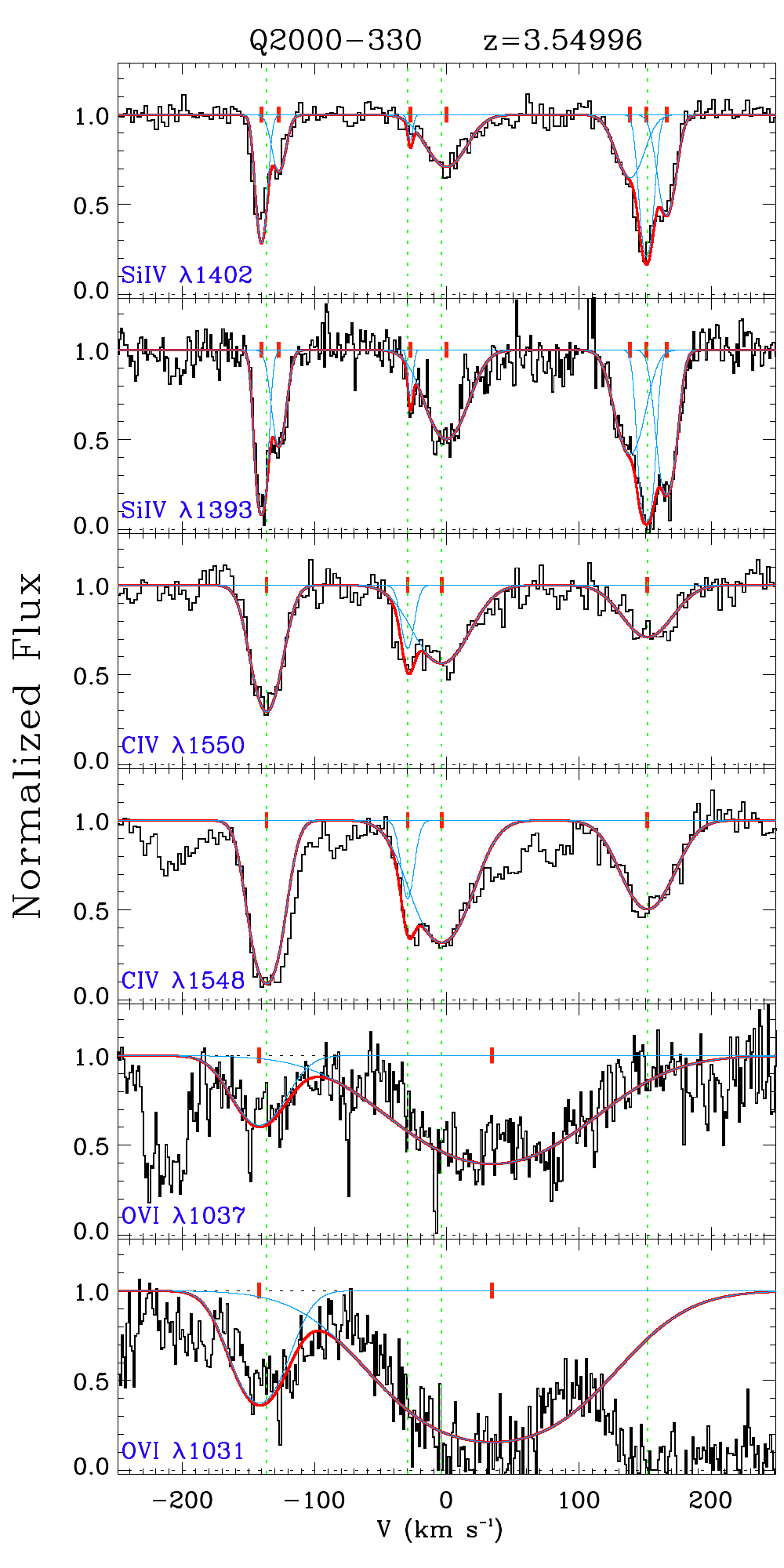}  
  \caption{Same as Fig.~\ref{f-fitq1009}, but for different absorbers. \label{f-fitq1603}}
\end{figure}

\begin{figure}[tbp]
\epsscale{1} 
\plottwo{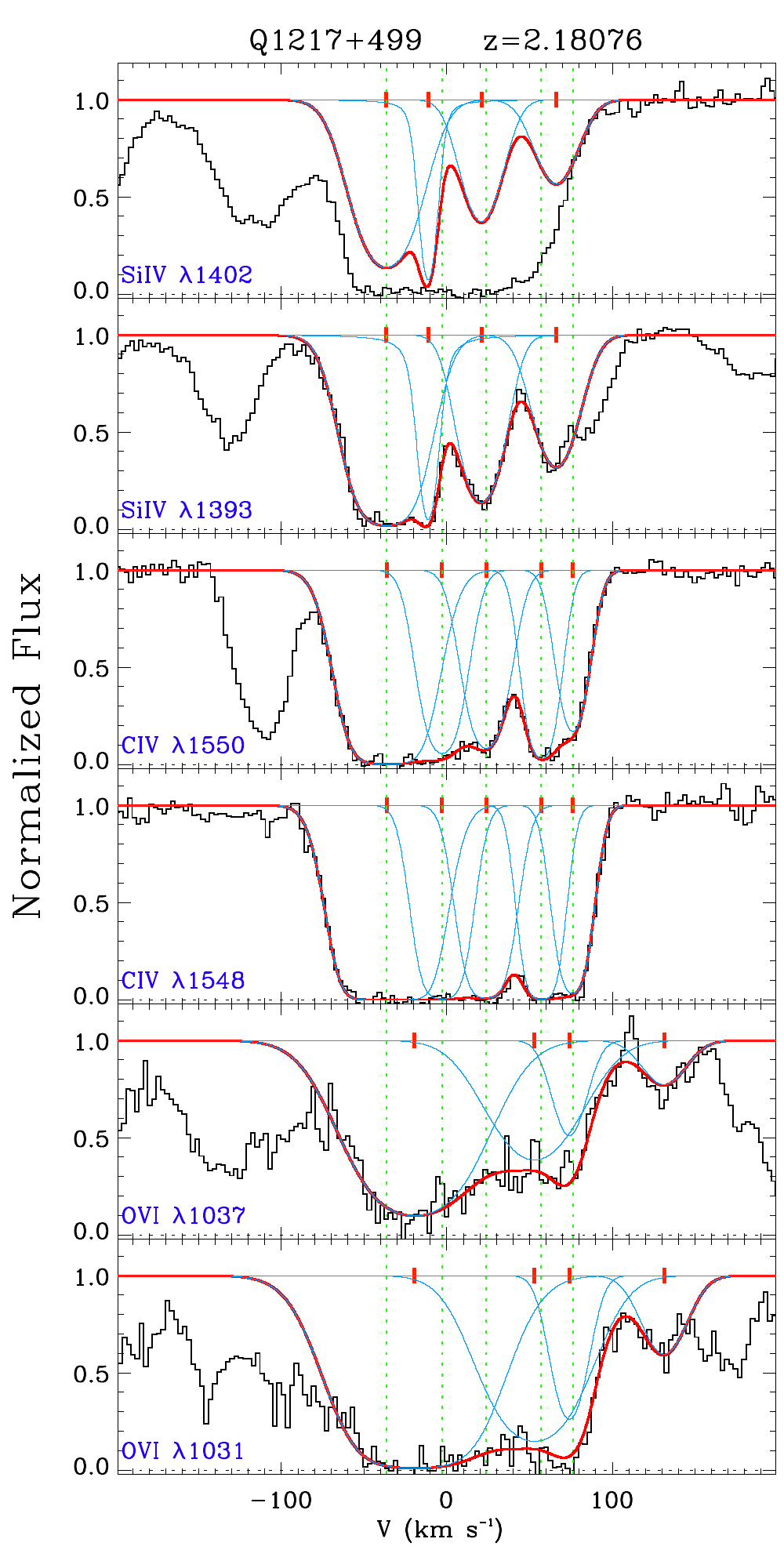}{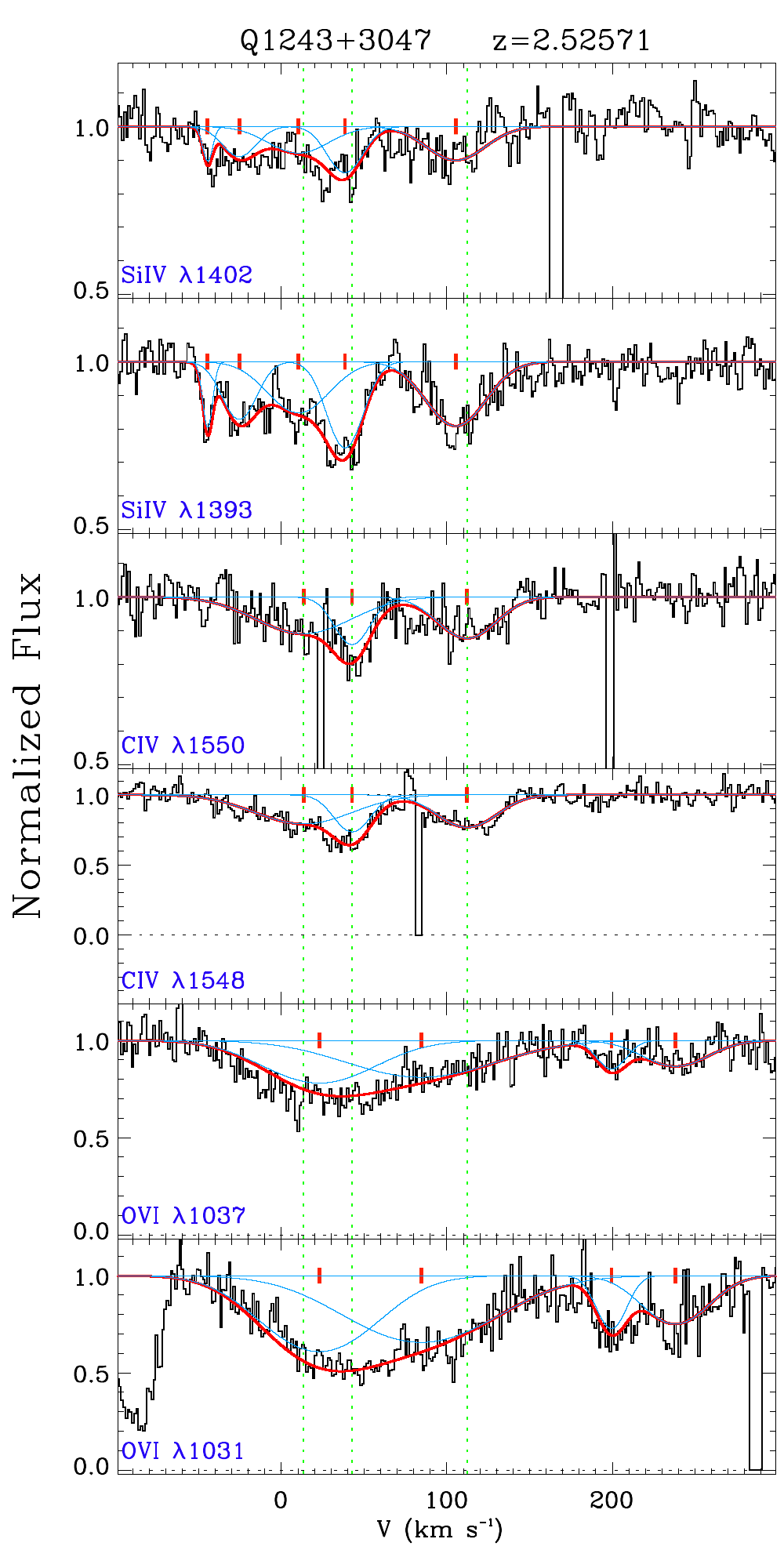}  
  \caption{Same as Fig.~\ref{f-fitq1009}, but for different absorbers.  \label{f-fitq1217}}
\end{figure}

\begin{figure}[tbp]
\epsscale{1} 
\plottwo{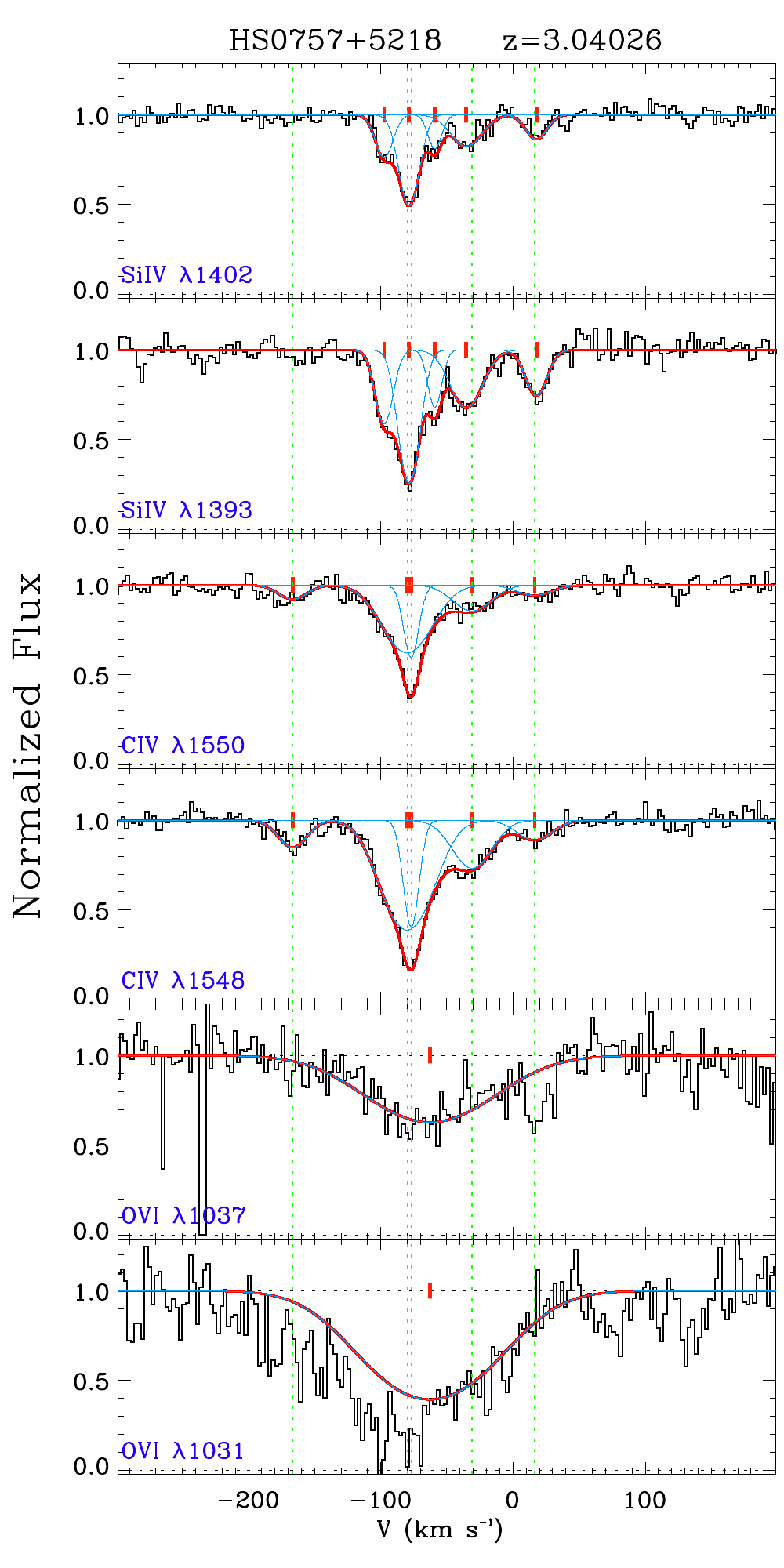}{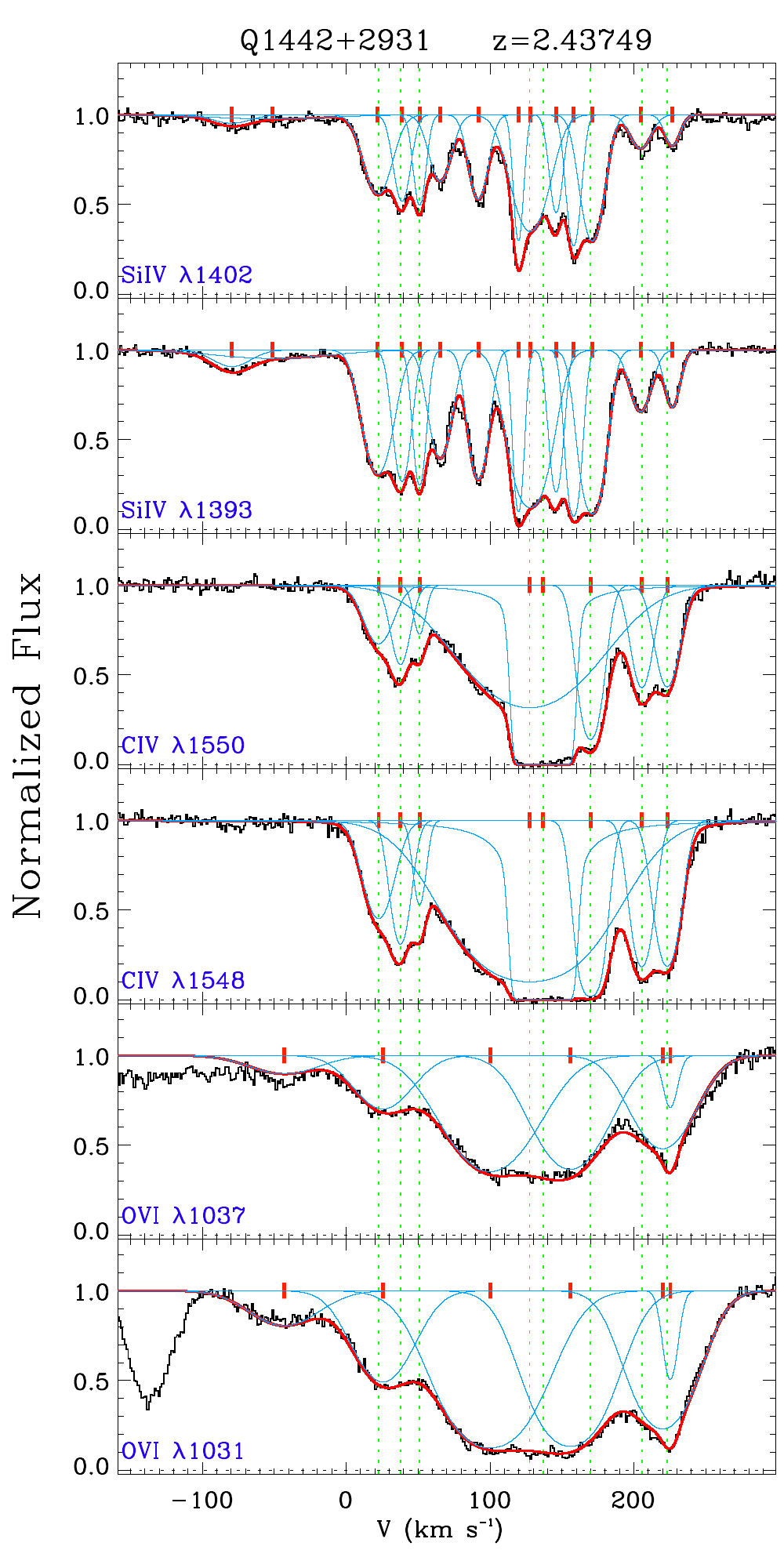}  
  \caption{Same as Fig.~\ref{f-fitq1009}, but for different absorbers.  \label{f-fiths0757}}
\end{figure}

\begin{figure}[tbp]
\epsscale{1} 
\plottwo{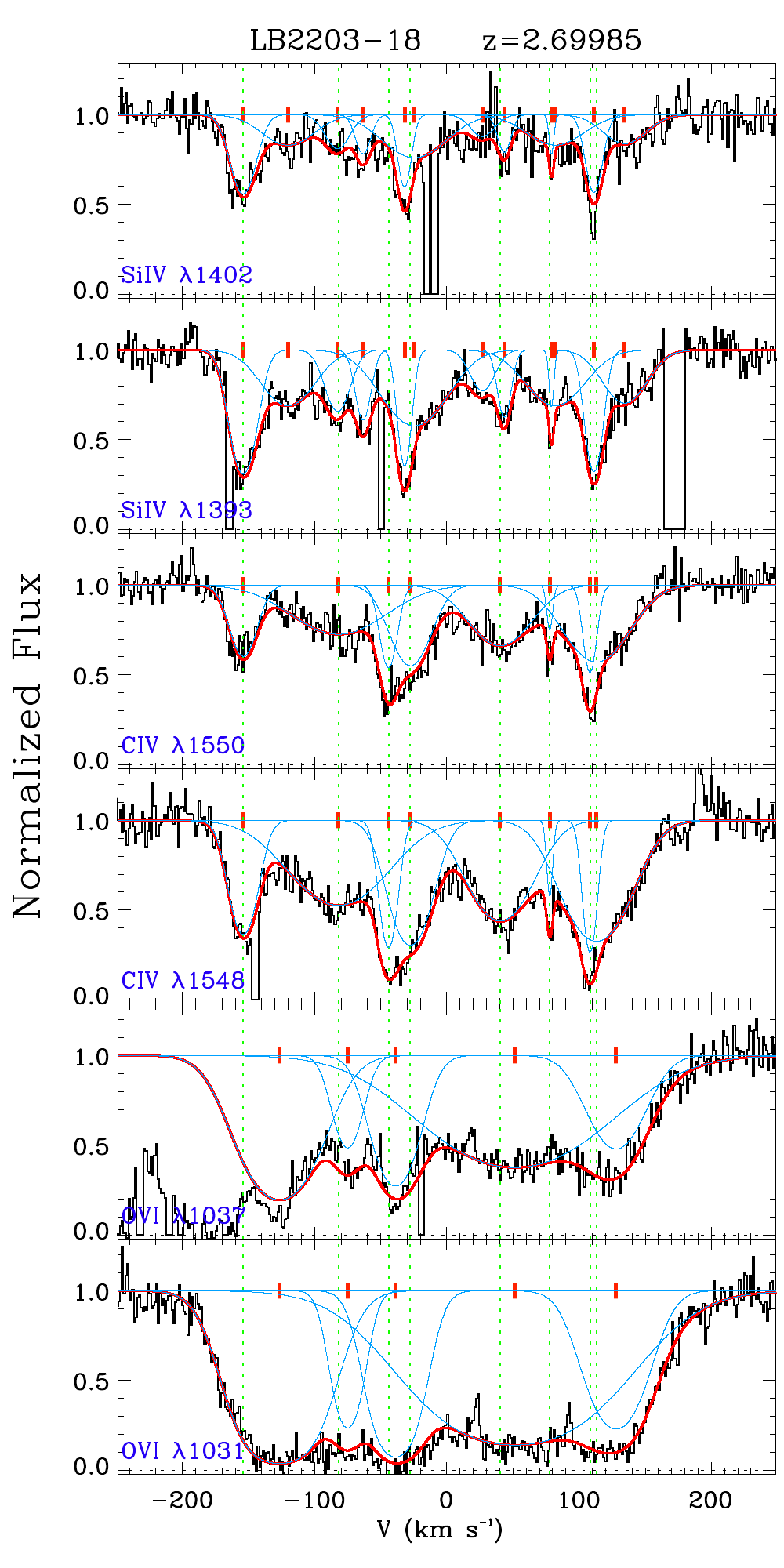}{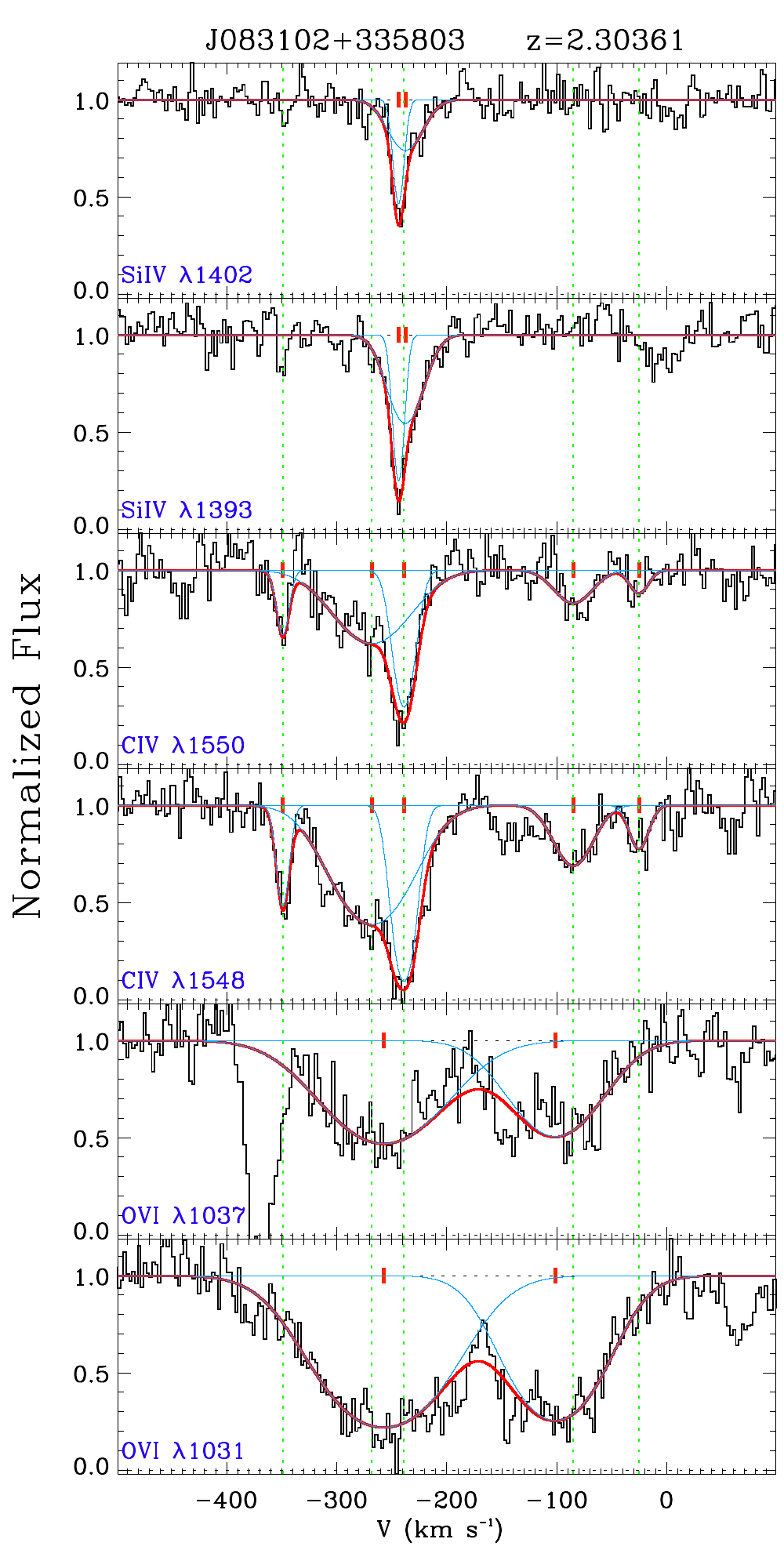}  
  \caption{Same as Fig.~\ref{f-fitq1009}, but for different absorbers.  \label{f-fitqlb2203}}
\end{figure}

\begin{figure}[tbp]
\epsscale{1} 
\plottwo{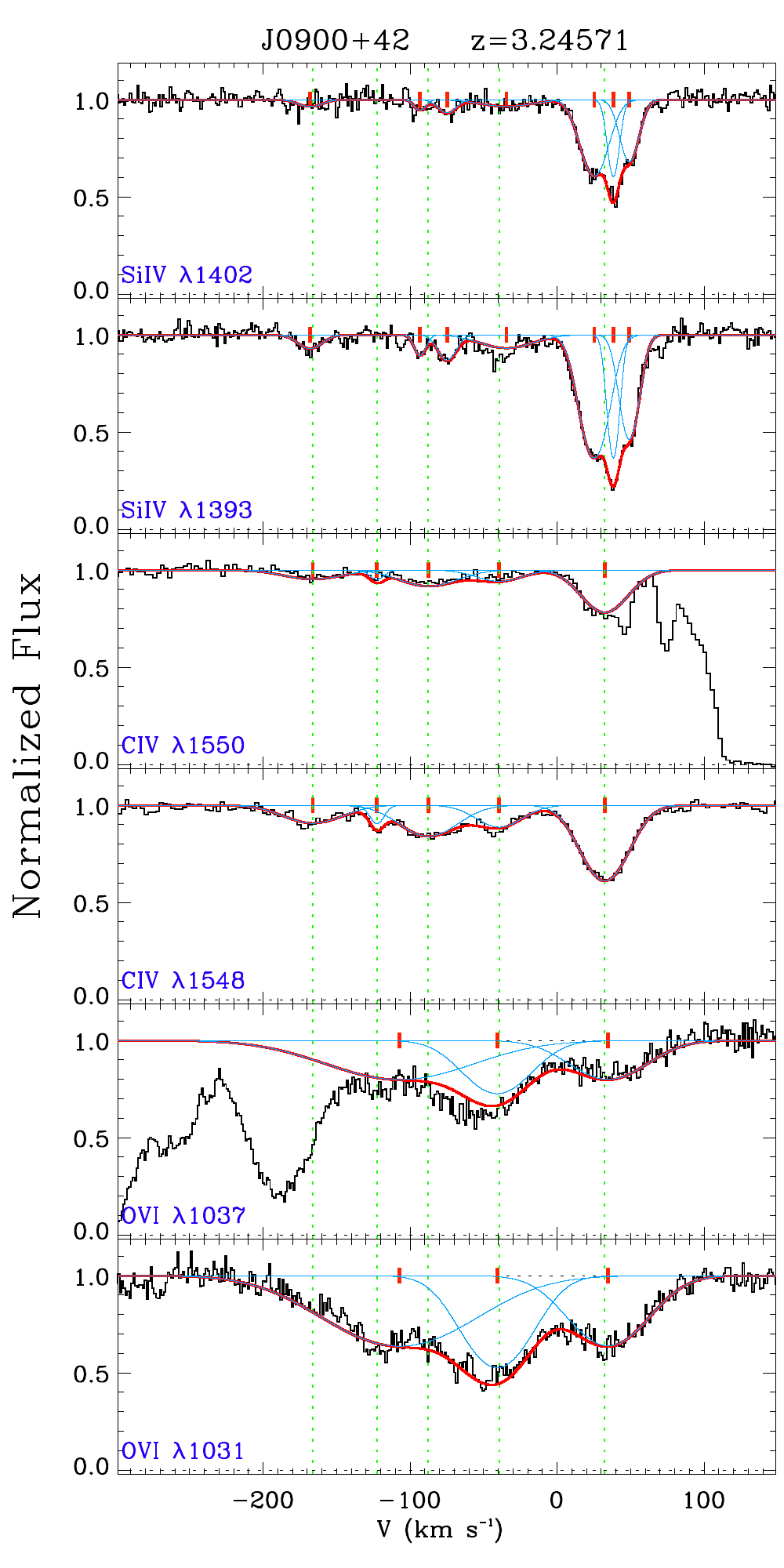}{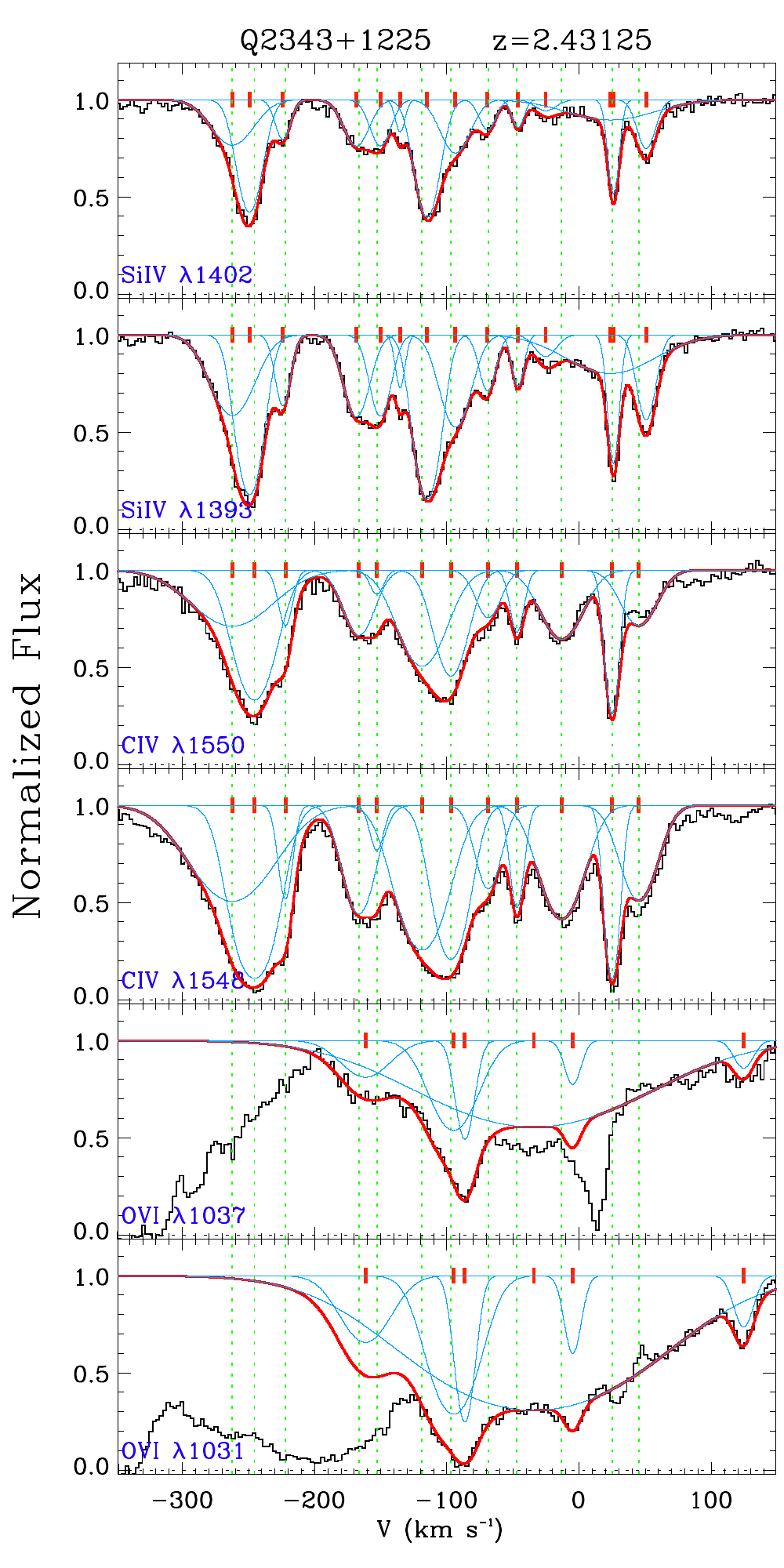}  
  \caption{Same as Fig.~\ref{f-fitq1009}, but for different absorbers.  \label{f-fitj0900}}
\end{figure}

\begin{figure}[tbp]
\epsscale{1} 
\plottwo{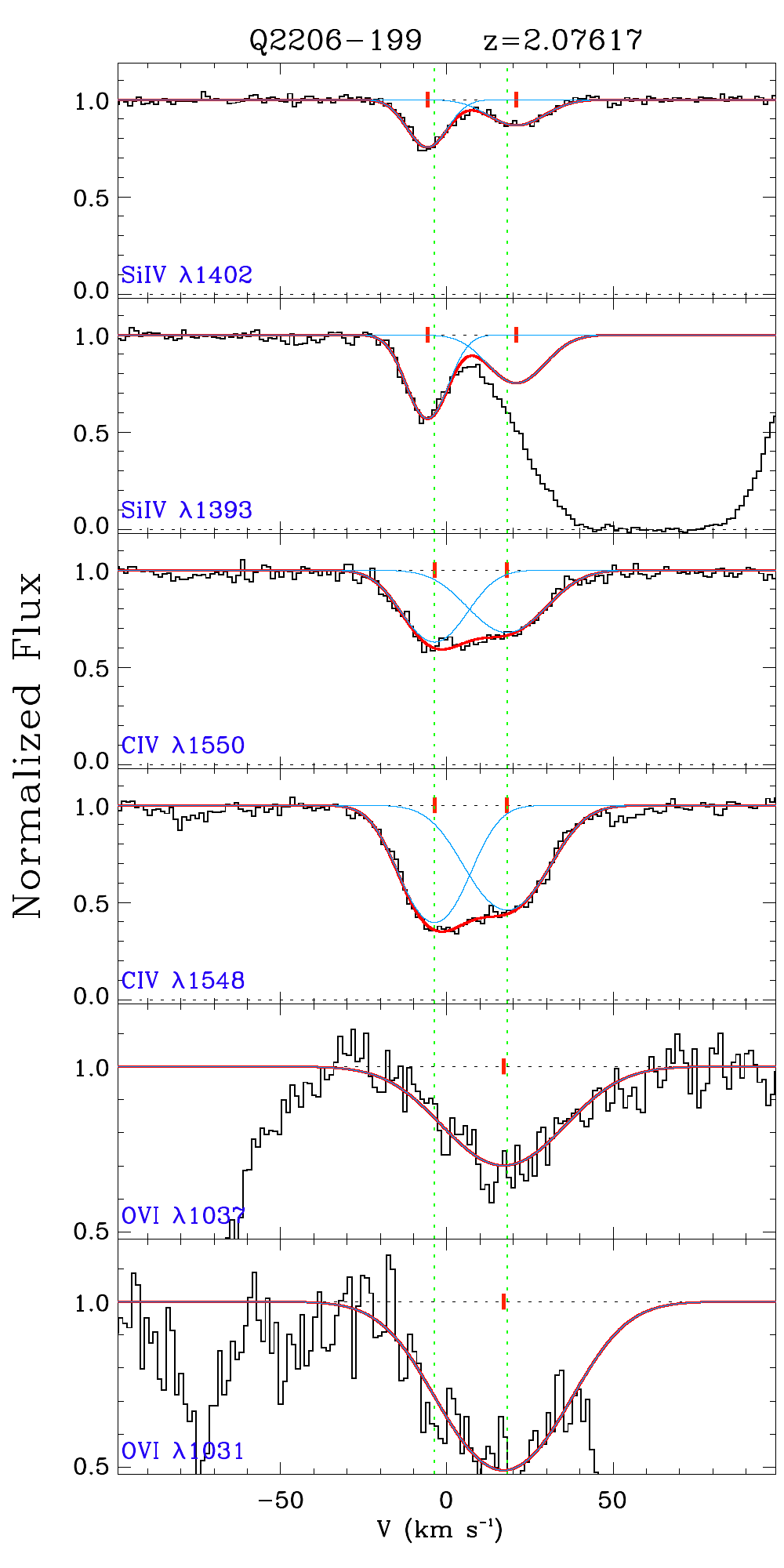}{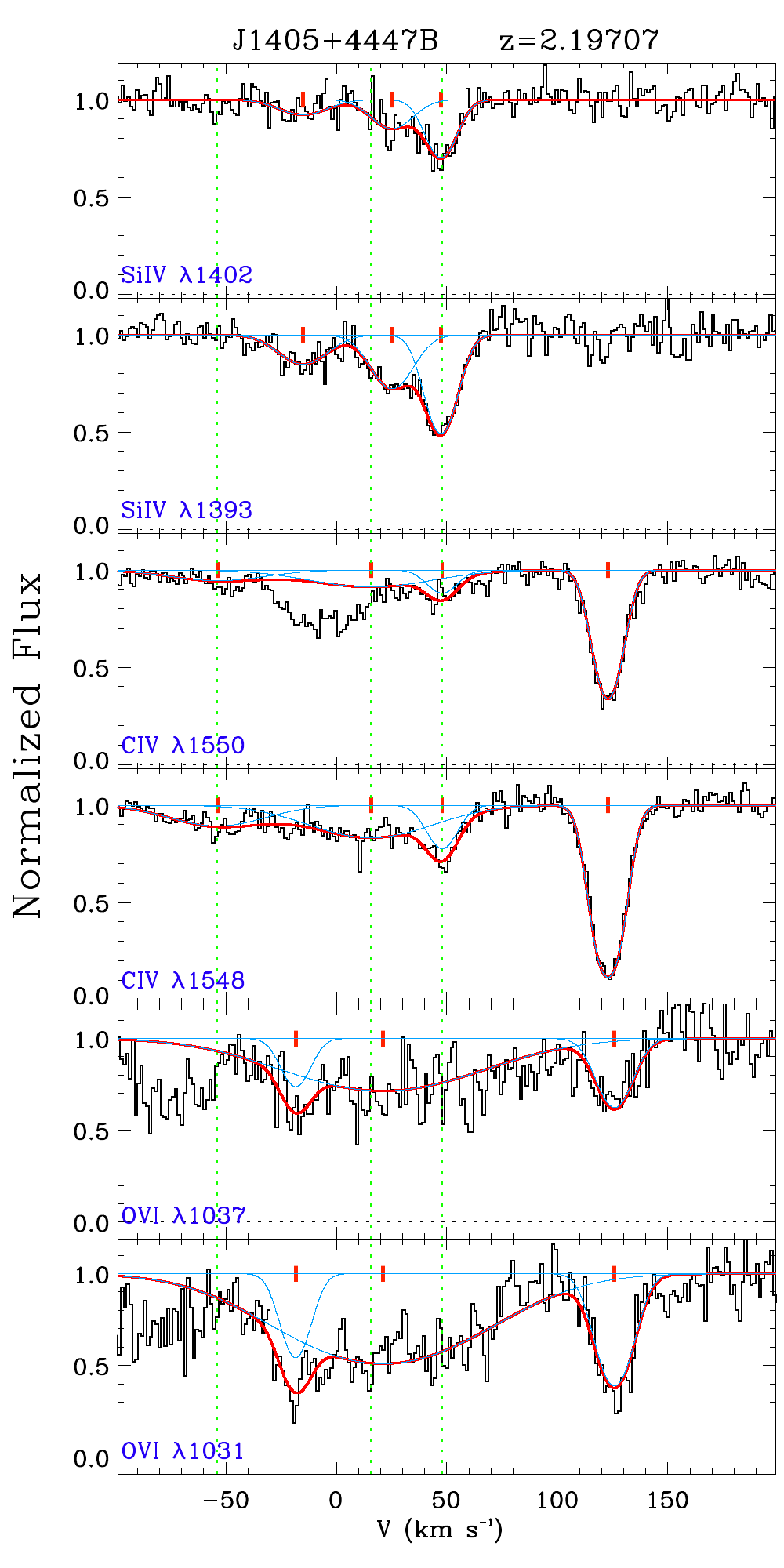} 
  \caption{Same as Fig.~\ref{f-fitq1009}, but for different absorbers.  \label{f-fitq2206}}
\end{figure}

\begin{figure}[tbp]
\epsscale{0.5} 
\plotone{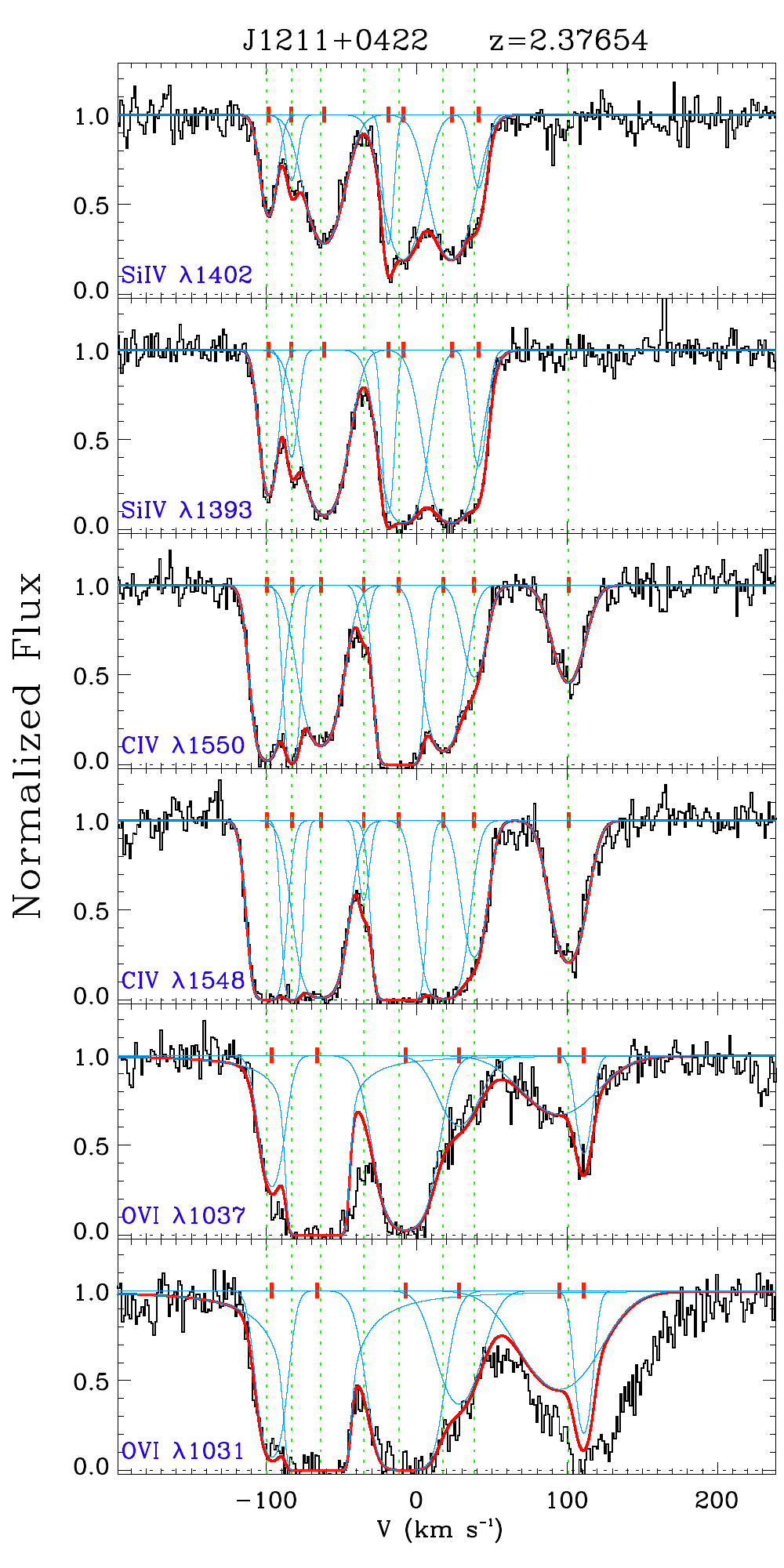}
  \caption{Same as Fig.~\ref{f-fitq1009}, but for a different absorber.  \label{f-fitj1211}}
\end{figure}

\end{appendix}

\end{document}

%% file: tab1.tex
\begin{deluxetable*}{lcclccccc}
\tablewidth{0pc}
\tablecaption{Data Sample Summary\label{t-data}}
\tabletypesize{\footnotesize}
\tablehead{\colhead{Quasar} &
\colhead{RA} &
\colhead{DEC} &
\colhead{PI} & 
\colhead{$R$} & 
\colhead{S/N\tablenotemark{a}} & 
\colhead{$z_{\rm{em}}$} & 
\colhead{$z_{\rm{abs}}$\tablenotemark{b}} &
\colhead{$\delta v$ \tablenotemark{c}} \\
\colhead{} & 
\colhead{(J2000)} &
\colhead{(J2000)} &
\colhead{} & 
\colhead({$\times 10^3$)} & 
\colhead{} & 
\colhead{} & 
\colhead{} &
\colhead{($10^3$ \km)} 
}
\startdata
Q1009+2956 	 & 10:11:55.60 & $+$29:41:41.6 &Sargent		     & 48   & 37   & 2.652 & 2.42901&  19.5	     \\
J1343+5721 	 & 13:43:29.00 & $+$57:21:47.9 & Prochaska	     & 48   & 10   & 3.034 & 2.83437&  15.6   \\
Q1603+3820 	 & 16:04:55.40 & $+$38:12:00.7 &Sargent		     & 48   & 54   & 2.510 & 2.47958&	2.6   \\
Q2000-330	 & 20:03:24.12 & $-$32:51:45.0 & Cowie  	     & 72\tablenotemark{d}   &  7   & 3.783 & 3.54996&  15.4   \\
HS1431+3144 	 & 14:33:16.10 & $+$31:31:26.3 &Prochaska	     & 36   & 29   & 2.940 & 2.58615&  29.6	     \\
Q1217-499	 & 12:19:30.88 & $+$49:40:54.3 & Pettini	     & 48   & 10   & 2.633 & 2.18076&  42.6   \\
J1040+5724	 & 10:40:18.52 & $+$57:24:48.1 & Prochaska	     & 36   & 13   & 3.426 & 3.26620&  11.2   \\
Q1558-0031	 & 15:58:10.16 & $-$00:31:20.0 & Prochaska	     & 36   & 20   & 2.827 & 2.62999&  16.3   \\
Q1243+3047	 & 12:46:10.90 & $+$30:31:31.0 & Prochaska	     & 48   &  9   & 2.560 & 2.52569&	2.9   \\
Q1023+5142	 & 10:23:25.25 & $+$51:42:50.8 & Hamann	     	     & 48   &  9   & 3.447 & 3.10586&	24.9 \\  
HS0757+5218	 & 08:01:17.80 & $+$04:22:23.9 & Cowie	     	     & 36   & 6   & 3.235 & 2.37654&	14.4 \\  
Q1442+2931 	 & 14:42:44.52 & $+$29:31:45.3 & Sargent	     & 48   & 34   & 2.638 & 2.43749&	17.4  \\
LB2203-18 	 & 22:06:39.61 & $-$18:18:46.3 & Prochaska	     & 48   & 14   & 2.728 & 2.69985&	2.4   \\
J0955+4116 	 & 09:55:41.94 & $+$41:16:53.9 & Pettini	     & 36   &  7   & 3.420 & 3.27987&	9.8   \\
J0831+3358 	 & 08:31:02.57 & $+$33:58:02.5 & Pettini	     & 36   & 10   & 2.429 & 2.30361&	11.4  \\
J0900+42 	 & 09:00:32.9 & $+$42:15:45.0 & Prochaska	     & 48   & 21   & 3.290 & 3.24571&	3.2  \\
Q2343+1225 	 & 23:46:28.32 & $+$12:48:59.1 & Steidel	     & 48   & 31   & 2.763 & 2.43125&	29.0	     \\
Q2206-199	 & 22:08:52.22 & $-$19:43:57.0 & Steidel	     & 36   & 11   & 2.559 & 2.07617&	47.1   \\
J1405+4447B 	 & 14:05:01.47 & $+$44:47:56.5 &Ellison, Martin      & 36   &  9   & 2.218 & 2.19707&	2.0   \\ 
J1211+0422	 & 12:11:17.65 & $+$52:10:34.6 & Prochaska	     & 48   & 12   & 2.526 & 2.37654&	43.8  
\enddata
\tablenotetext{a}{Signal-to-noise near \ovit\ $\lambda$1031; generally the S/N is higher at higher wavelengths.}
\tablenotetext{b}{The absorber redshift is determined by the centroid
  relative to the neutral (e.g., \hit\ or \oit) or low ionization (e.g., \ciit\ or \siiit) species, and
  will not generally match the exact redshift of the centroid to the
  observed \ovit.}
\tablenotetext{c}{$\delta v = c (z_{\rm em} - z_{\rm abs})/(1+z_{\rm abs})$}
\tablenotetext{d}{Also $R=36,000$ in the red (\civt, as well \siivt\ at high S/N).}
\end{deluxetable*}

%% file: tab2.tex
\begin{deluxetable*}{lccccccc}
\tablewidth{0pc}
\tablecaption{Total Column Densities \label{t-ions}}
\tabletypesize{\footnotesize}
\tablehead{\colhead{Quasar} &
\colhead{$z_{\rm{abs}}$} &
\colhead{$\log N_{\rm HI}$} &
\colhead{$\log N_{\rm SiII}$} &
\colhead{$\log N_{\rm OVI}$} &
\colhead{$\log N_{\rm NV}$} &
\colhead{$\log N_{\rm CIV}$} &
\colhead{$\log N_{\rm SiIV}$} }
\startdata
Q1009+2956  & 2.42901 & $   17.75 \pm	0.15	    $ & $12.72 \pm 0.03$ & $14.00 \pm 0.08$ & $<12.3$		  	& $13.92 \pm 0.01$      & $13.73 \pm 0.01$ \\
J1343+5721  & 2.83437 & $   17.78 \pm	0.20	    $ & $13.45 \pm 0.03$ & $14.77 \pm 0.03$ & $\le 14.06 \pm 0.02$	& $14.22 \pm 0.03$	& $13.80 \pm 0.04$ \\
Q1603+3820  & 2.47958 & $   17.97 \pm	0.20	    $ & $14.27 \pm 0.05$ & $15.10 \pm 0.05$ & $14.08 \pm 0.01$		& $14.98 \pm 0.01$      & $13.96 \pm 0.01$ \\
Q2000-330   & 3.54996 & $   18.00 \pm	0.25	    $ & $13.24 \pm 0.02$ & $14.99 \pm 0.03$ & \nodata			& $14.31 \pm 0.02$   	& $13.91 \pm 0.02$ \\
HS1431+3144 & 2.58615 & $   18.15 \pm   0.15	    $ & $12.19 \pm 0.13$ & $<13.0	  $ & \nodata			& $<12.2	 $   	& $<11.8	 $ \\
Q1217-499   & 2.18076 & $   18.60 \pm   0.15	    $ & $13.67 \pm 0.01$ & $> 15.15       $ & $> 15.18 		$	& $ > 15.07$  	     	& $14.18 \pm 0.04$ \\
J1040+5724  & 3.26620 & $   18.60 \pm   0.20	    $ & $13.32 \pm 0.05$ & $<12.9         $ & $<12.9   		$       & \nodata		& \nodata	\\
Q1558-0031  & 2.62999 & $   19.62 \pm	0.10	    $ & $13.53 \pm 0.02$ & $<12.7	  $ & \nodata			& $<12.6	     $	& $12.49 \pm 0.03$ \\
Q1243+3047  & 2.52569 & $   19.73 \pm	0.04	    $ & $13.10 \pm 0.05$ & $14.41 \pm 0.02$ & $<12.6		      $	& $13.64 \pm 0.02$      & $13.18 \pm 0.02$ \\
Q1023+5142  & 3.10586 & $   19.85 \pm   0.15 	    $ & $13.59 \pm 0.01$ & $<13.2	  $ & $<13.1		$	& $13.48 \pm 0.08$      & $12.97 \pm 0.02$ \\
HS0757+521  & 3.04026 & $   19.90 \pm   0.20        $ & $13.01 \pm 0.01$ & $14.38 \pm 0.05$ & $13.09 \pm 0.10$    	& $13.98 \pm 0.01$      & $13.46 \pm 0.02$ \\
Q1442+2931  & 2.43749 & $   19.91\,^{+0.10}_{-0.25} $ & $13.92 \pm 0.01$ & $15.04 \pm 0.01$ & \nodata			& $> 14.9  	  $  	& $14.15 \pm 0.02$ \\
LB2203-18   & 2.69985 & $   20.02 \pm	0.15	    $ & $14.58 \pm 0.04$ & $15.28 \pm 0.01$ & $<12.9		$	& $14.54 \pm 0.01$      & $13.98 \pm 0.01$ \\
J0955+4116  & 3.27987 & $   20.05 \pm   0.20        $ & $12.74 \pm 0.04$ & $<13.2         $ & $<13.0		$	& $<12.6	      $	& $<12.2$		\\
J0831+3358  & 2.30361 & $   20.25 \pm	0.15	    $ & $13.71 \pm 0.04$ & $15.00 \pm 0.02$ & $\le 13.59$	     	& $14.33 \pm 0.02$      & $13.42 \pm 0.02$ \\
J0900+42    & 3.24571 & $   20.30 \pm	0.10	    $ & $> 14.95       $ & $14.59 \pm 0.01$ & $12.59 \pm 0.14$	     	& $13.65 \pm 0.02$      & $13.44 \pm 0.01$ \\
Q2343+1225  & 2.43125 & $   20.40 \pm	0.20	    $ & $> 15.06       $ & $14.96 \pm 0.01$ & \nodata			& $14.70 \pm 0.01$      & $14.01 \pm 0.01$ \\
Q2206-199   & 2.07617 & $   20.43 \pm	0.06	    $ & $13.64 \pm 0.01$ & $13.93 \pm 0.05$ & $<12.2		      $	& $13.75 \pm 0.01$      & $12.87 \pm 0.02$ \\
J1405+4447B & 2.19707 & $   20.65 \pm	0.15	    $ & $14.85 \pm 0.05$ & $14.45 \pm 0.02$ & $13.21 \pm 0.08        $	& $13.83 \pm 0.03$      & $13.10 \pm 0.04$ \\
J1211+0422  & 2.37654 & $   20.80 \pm	0.10	    $ & $14.91 \pm 0.04$ & $> 15.33	  $ & $13.64 \pm 0.02        $	& $>14.92  	  $  	& $14.19 \pm 0.01$ 
\enddata
\tablecomments{Detailed descriptions and uncertainties (blending, continuum, etc.) on the column densities are discussed in \S\S\ref{s-ion} and \ref{s-nhi} and in the appendix. The ``$>$'' symbol means that the weakest available line is saturated. The ``$<$'' symbol implies that a given species has not been detected and the reported value is a $3\sigma$ estimate. The ``$\le$'' implies that the profile may be contaminated. Absence of value implies that the lines are contaminated for \nvt\ and not observed for \civt\ and \siivt.   }
\end{deluxetable*}

%% file: tab3.tex
\begin{deluxetable*}{lcccccc}
\tablewidth{0pc}
\tablecaption{Metallicity of the Cool Gas Phase \label{t-met}}
\tabletypesize{\footnotesize}
\tablehead{\colhead{Quasar} &
\colhead{$z_{\rm{abs}}$} &
\colhead{$\log N_{\rm HI}$} &
\colhead{$\log N_{X}$\tablenotemark{a}} &
\colhead{$\log ({\rm X/H})_\sun$\tablenotemark{a}} &
\colhead{$[{\rm X/H}]$}&
\colhead{$[{\rm X/H}]_{\rm adopt}$\tablenotemark{b}}
}
\startdata
Q1009+2956   & $   2.42901 $ & $   17.75 \pm   0.15        $ & $   < 12.02	    $ & $   -3.31$ & $    <-2.42		  $ & $    -2.10 \pm 0.20	   $ \\
J1343+5721   & $   2.83437 $ & $   17.78 \pm   0.20        $ & $   13.82 \pm   0.02 $ & $   -3.31$ & $    -0.65\,^{+0.14}_{-0.20} $ & $    -0.60 \pm 0.20	   $ \\
Q1603+3820   & $   2.47958 $ & $   17.97 \pm   0.20        $ & $   14.66 \pm   0.02 $ & $   -3.31$ & $     0.00\,^{+0.14}_{-0.20} $ & $    -0.05 \pm 0.20	   $ \\
 Q2000-330   & $   3.54996 $ & $   18.00 \pm   0.25        $ & $   13.31 \pm   0.04 $ & $   -3.31$ & $    -1.38\,^{+0.16}_{-0.26} $ & $    -1.35 \pm 0.25	   $\tablenotemark{c} \\
HS1431+3144  & $   2.58615 $ & $   18.15 \pm   0.15        $ & $   < 12.24	    $ & $   -3.31$ & $   <-2.60			  $ & $    <-2.60		   $\tablenotemark{d}  \\
 Q1217-499   & $   2.18076 $ & $   18.60 \pm   0.15        $ & $   < 13.79  	    $ & $   -3.31$ & $    <-1.50		  $ & $    -1.60:		   $\tablenotemark{e} \\
J1040+5724   & $   3.26620 $ & $   18.60 \pm   0.20        $ & $   13.92 \pm   0.03 $ & $   -3.31$ & $    -1.37\,^{+0.14}_{-0.21} $ & $    -1.37\,^{+0.14}_{-0.21} $ \\
Q1558-0031   & $   2.62999 $ & $   19.62 \pm   0.10        $ & $   14.83 \pm   0.04 $ & $   -3.31$ & $    -1.48\,^{+0.09}_{-0.11} $ & $    -1.48\,^{+0.09}_{-0.11} $ \\
Q1243+3047   & $   2.52569 $ & $   19.73 \pm   0.04        $ & $   13.80 \pm   0.03 $ & $   -3.31$ & $    -2.62 \pm 0.05	  $ & $    -2.62 \pm 0.05	   $ \\
Q1023+5142   & $   3.10586 $ & $   19.85 \pm   0.15 	   $ & $   14.92 \pm   0.03 $ & $   -3.31$ & $    -1.72\,^{+0.11}_{-0.21} $ & $    -1.62\,^{+0.11}_{-0.15} $ \\
HS0757+521   & $   3.04026 $ & $   19.90 \pm   0.20        $ & $   13.87 \pm   0.02 $ & $   -3.31$ & $    -2.72\,^{+0.14}_{-0.20} $ & $    -2.72\,^{+0.14}_{-0.20} $ \\
Q1442+2931   & $   2.43749 $ & $   19.91\,^{+0.10}_{-0.25} $ & $   14.61 \pm   0.01 $ & $   -3.31$ & $    -1.99\,^{+0.16}_{-0.25} $ & $    -1.99\,^{+0.16}_{-0.25} $ \\
 LB2203-18   & $   2.69985 $ & $   20.02 \pm   0.15        $ & $   14.89 \pm   0.02 $ & $   -4.86$ & $    -0.27\,^{+0.11}_{-0.15} $ & $    -0.27\,^{+0.11}_{-0.15} $ \\
J0955+4116   & $   3.27987 $ & $   20.05 \pm   0.20        $ & $   14.12 \pm   0.04 $ & $   -3.31$ & $    -2.62 \,^{+0.14}_{-0.21} $ & $   -2.62 \,^{+0.14}_{-0.21}$ \\
J0831+3358   & $   2.30361 $ & $   20.25 \pm   0.15        $ & $   14.82 \pm   0.05 $ & $   -3.31$ & $    -2.12\,^{+0.12}_{-0.16} $ & $    -2.12\,^{+0.12}_{-0.16} $ \\
J0900+42     & $   3.24571 $ & $   20.30 \pm   0.10        $ & $   14.64 \pm   0.02 $ & $   -4.86$ & $    -0.80\,^{+0.08}_{-0.10} $ & $    -0.80\,^{+0.08}_{-0.10} $ \\
Q2343+1225   & $   2.43125 $ & $   20.40 \pm   0.20        $ & $   14.91 \pm   0.03 $ & $   -4.86$ & $    -0.63\,^{+0.14}_{-0.20} $ & $    -0.63\,^{+0.14}_{-0.20} $ \\
 Q2206-199   & $   2.07617 $ & $   20.43 \pm   0.06        $ & $   13.64 \pm   0.01 $ & $   -4.49$ & $    -2.30\,^{+0.05}_{-0.06} $ & $    -2.30\,^{+0.05}_{-0.06} $ \\
J1405+4447   & $   2.19707 $ & $   20.65 \pm   0.15        $ & $   14.30 \pm   0.03 $ & $   -4.86$ & $    -1.49\,^{+0.11}_{-0.15} $ & $    -1.49\,^{+0.11}_{-0.15} $ \\
J1211+0422   & $   2.37654 $ & $   20.80 \pm   0.10        $ & $   14.53 \pm   0.04 $ & $   -4.86$ & $    -1.41\,^{+0.09}_{-0.11} $ & $    -1.41\,^{+0.09}_{-0.11} $ 
\enddata
\tablenotetext{a}{Solar abundances are from \citet{asplund09}. X is \oit\ when $[{\rm X/H}]_\sun = -3.31$, \siiit\ when $[{\rm X/H}]_\sun = -4.49$, \siit\ when $[{\rm X/H}]_\sun = -4.86$}
\tablenotetext{b}{For all the absorbers with $\log N_{\rm HI}\le 20$, we use Cloudy photoionization models to predict the column densities of typically \ciit, \oit, \siiit, \aliit, \siit, \siivt. For absorbers that required an ionization correction, the metallicity is updated in the last column.}
\tablenotetext{c}{This absorber has 3 main components at $-140, 0,+140$ \km\ seen in the low ions. The $+140$ \km\ component corresponds to the LLS. The \oit\ appears solely detected  at 0 \km\ where  $\log N_{\rm HI}=16.82$ according to  \citet{prochter10}. This absorber shows evidence of metallicity variation across the profiles: \citet{prochter10} show using Cloudy photoionization models that $[{\rm Si/H}] = -0.10 \pm 0.35, -0.61 \pm 0.20, -2.19 \pm 0.29$ for the $-140, 0,+140$ \km\ components, respectively. We list here the metallicity based on the total column density comparison, but we consider this metallicity variation when estimating \nhii\ (see Table~\ref{t-nh}). }
\tablenotetext{d}{We cannot constrain the metallicity better than the upper limit given by \oit\ because the column densities of \siiit\ and \siiiit\ remain uncertain. However, the non-detection of \siivt\ and low value of \siiit\ are consistent in the photoionization model with the upper limit derived from \oit.}
\tablenotetext{e}{This absorber has no detection of \oit, but shows a complex velocity structure for the singly ionized species with a strong absorption at $-30$ \km\ and a much weaker absorption at positive velocities. The complex ionization structure and uncertain \nhi\ in each component make uncertain the metallicity estimate. We emphasize that with the colon symbol. }
\end{deluxetable*}

%% file: tab4.tex
\begin{deluxetable}{lcccc}
\tablewidth{0pc}
\tablecaption{Mean and Median of the Doppler Parameters and Fraction of Broad Components \label{t-bval}}
\tabletypesize{\footnotesize}
\tablehead{
\colhead{Ion}  &\colhead{m} & \colhead{$\langle b \rangle$} & \colhead{$\bar{b}$} &\colhead{$f(b>b_c)$} \\
\colhead{} 	&\colhead{} & \colhead{(\km)} & \colhead{(\km)}  &\colhead{(\%)}}
\startdata
  \cutinhead{Entire Sample}
  \siivt\ & 105& $     12 \pm	   10  $ & $	     9 $ & $ 71 \pm 5  $   \\
   \civt\ & 91 & $     16 \pm	   11  $ & $	    13 $ & $ 63 \pm 5  $   \\
   \ovit\ & 60 & $     31 \pm	   23 $ & $	    26 $ & $ 80 \pm 5  $   \\
  \cutinhead{Robust Entire Sample}
  \siivt\ & 94 & $     10 \pm	    6  $ & $	     8 $ & $ 68 \pm 6  $   \\
   \civt\ & 78 & $     14 \pm	    8  $ & $	    13 $ & $ 60 \pm 6  $   \\
   \ovit\ & 40 & $     25 \pm	   17  $ & $	    23 $ & $ 75 \pm 8  $   \\
  \cutinhead{Intervening Sample}
  \siivt\ & 69 & $     11 \pm	    9  $ & $	     9 $ & $ 71 \pm 6  $   \\
   \civt\ & 70 & $     16 \pm	   10  $ & $	    13 $ & $ 69 \pm 6  $   \\
   \ovit\ & 39 & $     31 \pm	   24  $ & $	    26 $ & $ 83 \pm 6  $   \\
  \cutinhead{Robust Intervening Sample}
  \siivt\ & 63 & $     10 \pm	    6  $ & $	     9 $ & $ 71 \pm 6  $   \\
   \civt\ & 56 & $     15 \pm	    8  $ & $	    13 $ & $ 66 \pm 7  $   \\
   \ovit\ & 29 & $     27 \pm	   17  $ & $	    25 $ & $ 79 \pm 9  $  
\enddata 
\tablecomments{The $m$ value represents the size of each sample. $\langle b \rangle$ is the mean value of $b$ with its 1$\sigma$ dispersion for a given sample, while $\bar{b}$ is the median. $f(b>b_c)$ is the fraction of components with $b >b_c$, the $b$-value cutoff between broad and narrow components where the narrow components probe gas temperature of about $7 \times 10^4$ K in the absence of nonthermal broadening ( $b_c= 6.5$ \km\ for \siivt, 10 \km\ for \civt, 9 \km\ for \ovit). }
\end{deluxetable}

%% file: tab5.tex
\begin{deluxetable*}{lcccccc}
\tablewidth{0pc}
\tablecaption{Neutral and Ionized Gas Column Densities \label{t-nh}}
\tabletypesize{\footnotesize}
\tablehead{\colhead{Quasar} &
\colhead{$z_{\rm{abs}}$} &
\colhead{$\log N_{\rm HI}$} &
\colhead{$\log N^{\rm OVI}_{\rm HII}$}& 
\colhead{$\log N^{\rm SiIV}_{\rm HII}$}& 
\colhead{$N^{\rm OVI}_{\rm HII}/N_{\rm HI}$}&
\colhead{$N^{\rm SiIV}_{\rm HII}/N_{\rm HI}$}
}
\startdata
Q1009+2956   & $   2.42901 $ & $   17.75 \pm   0.15        $ & $ >      20.1	$ & $ >  20.7 	$ & $>  228.5	$ & $ > 928.8	$ \\
J1343+5721   & $   2.83437 $ & $   17.78 \pm   0.20        $ & $ >      19.8 	$ & $ >  19.3 	$ & $>  39.7    $ & $ >  32.2	$ \\
Q1603+3820   & $   2.47958 $ & $   17.97 \pm   0.20        $ & $ >      19.2 	$ & $ >  18.9 	$ & $>  15.5    $ & $ >   8.5	$ \\
Q2000-330\tablenotemark{a}   & $  3.54996 $ & $18.00 \pm 0.25$ & $ >    20.1	$ & $ >  19.8 	$ & $>  120.2	$ & $>   68.3 	$ \\
HS1431+314   & $   2.58615 $ & $   18.15 \pm   0.15        $ & $ <     19.6 $     & $ <    20.5 $ & $<     28.8$ & $ <      217.7 $ \\
Q1217-499    & $   2.18076 $ & $   18.60 \pm   0.15        $ & $ >      20.8:	$ & $ >  20.7:	$ & $>  144.2:	$ & $ > 116.9: 	$ \\
J1040+5724   & $   3.26620 $ & $   18.60 \pm   0.20        $ & $ <     18.2 $ 	  & \nodata	  & $< 	   0.4$ & \nodata \\
Q1558-0031   & $   2.62999 $ & $   19.62 \pm   0.10        $ & $ <      18.2	$ & $ >	 18.9  	$ & $<    0.04	$ & $ >   0.2  	$ \\
Q1243+3047   & $   2.52569 $ & $   19.73 \pm   0.04        $ & $ >      21.0	$ & $ >	 20.7  	$ & $>   20.4	$ & $ >   9.1  	$ \\
Q1023+5142   & $   3.10586 $ & $   19.85 \pm   0.15	   $ & $ <      18.8	$ & $ >	 19.5  	$ & $<    0.1	$ & $ >   0.4  	$ \\
HS0757+5218  & $   3.04026 $ & $   19.90 \pm   0.20	   $ & $ >      21.1	$ & $ >	 21.1  	$ & $>   16.2	$ & $ >  14.7	$ \\
Q1442+2931   & $   2.43749 $ & $   19.91\,^{+0.10}_{-0.25} $ & $ >      21.0	$ & $ >	 21.0  	$ & $>   13.5	$ & $ >  13.1  	$ \\
 LB2203-18   & $   2.69985 $ & $   20.02 \pm   0.15        $ & $ >      19.6	$ & $ >	 19.1	$ & $>    0.4	$ & $ >   0.1   $ \\
J0955+4116   & $   3.27987 $ & $   20.05 \pm   0.20       $ & $ <      19.9	$ & $ <	 19.7	$ & $<    0.6	$ & $ <   0.5   $ \\
J0831+3358   & $   2.30361 $ & $   20.25 \pm   0.15        $ & $ >      21.1	$ & $ >	 20.4  	$ & $>    7.6	$ & $ >   1.5  	$ \\
J0900+42     & $   3.24571 $ & $   20.30 \pm   0.10        $ & $ >      19.4	$ & $ >	 19.1	$ & $>     0.1	$ & $ >   0.1	$ \\
Q2343+1225   & $   2.43125 $ & $   20.40 \pm   0.20        $ & $ >      19.6	$ & $ >	 19.5  	$ & $>     0.2	$ & $ >   0.1  	$ \\
 Q2206-199   & $   2.07617 $ & $   20.43 \pm   0.06        $ & $ >      20.2	$ & $ >	 20.1	$ & $>     0.6	$ & $ >   0.4  	$ \\
J1405+4447   & $   2.19707 $ & $   20.65 \pm   0.15        $ & $ >      20.0	$ & $ >	 19.5 	$ & $>     0.2	$ & $ >   0.1 	$ \\
J1211+0422   & $   2.37654 $ & $   20.80 \pm   0.10        $ & $ >      20.8	$ & $ >	 20.5  	$ & $>     0.9	$ & $ >   0.5  	$ 
\enddata
\tablecomments{See \S\ref{s-nh} for details on how \nhii\ is derived from \novi\ and \nsiiv. For the cases where \ovit\ or \siivt\ are not detected, we have set the ionization fraction of \ovit\ and \siivt\ to 0.2 and 0.4, respectively. The colon indicates that the values are uncertain because the metallicity correction is uncertain.}
\tablenotetext{a}{From \citet{prochter10}, the metallicity is different between the main 3 components: $[{\rm Si/H}] = -0.10 \pm 0.35, -0.61 \pm 0.20, -2.19 \pm 0.29$ for the $-140, 0,+140$ \km\ components. We estimate  \novi\ and \nsiiv\ in the 3 components and corrected for the respective metallicity to estimate \nhii. The column densities in the $-140, 0,+140$ \km\ components are $\log N_{\rm OVI} =  14.24 \pm 0.02, 14.90 \pm 0.03, 13.73 \pm 0.10$ and $\log N_{\rm SiIV} =  13.23 \pm 0.01, 13.17 \pm 0.02, 13.63 \pm 0.02$, respectively. 
}
\end{deluxetable*}

%% file: tab6.tex
\begin{deluxetable}{lcccc}
\tablewidth{0pc}
\tablecaption{Cosmic Baryonic Census of $\tau_{\rm LL}>1$ absorbers at $z\sim 2$--$3$ \label{t-omegab}}
\tabletypesize{\footnotesize}
\tablehead{
\colhead{Absorbers}  &\colhead{$\Omega_{\rm g}^{\rm HI}$}& \colhead{$\Omega_{\rm g}^{\rm HII}({\rm SiIV})$}  & \colhead{$\Omega_{\rm g}^{\rm HII}({\rm OVI})$} & \colhead{$\Omega_{\rm g}/\Omega_{\rm b}$}\\
\colhead{} 	&\colhead{($10^{-3}$)} & \colhead{($10^{-3}$)} & \colhead{($10^{-3}$)}& \colhead{(\%)}}
\startdata
  LLS 	  &  0.0077   &   0.04--1.9   &   0.08--1.2     &  0.4--7   \\
  SLLS 	  &  0.17     &   0.04--2.5   &   0.2--4.4      &  0.9--16 \\
  DLA  	  &  0.79     &   0.02--0.2  &   0.03--0.4      &  2--3 	
\enddata 
\tablecomments{We define $\Omega_{\rm g} = \Omega_{\rm g}^{\rm HI} + \Omega_{\rm g}^{\rm HII}({\rm SiIV}) +\Omega_{\rm g}^{\rm HII}({\rm OVI})$ and use $\Omega_b = 0.044$ as the ratio of the total baryon density to the critical density \citep[e.g,][]{spergel03,omeara01}. The calculation of $\Omega_{\rm g}^{\rm HII}$ is inversely proportional to the ionization fraction of \siivt\ and \ovit\ and the metallicity; we use values from Table~\ref{t-nh} to estimate $\Omega_{\rm g}^{\rm HII}$ (see \S\ref{s-barbud} for more details). Note that the average metallicity for the SLLSs is a factor 4 smaller than for the DLAs and LLSs in the KODIAQ sample, which drive the high value of $\Omega_{\rm g}$ for the SLLSs. }
\end{deluxetable}

%% file: tab7.tex
\begin{deluxetable}{lccc}
\tablewidth{0pc}
\tablecaption{Cosmic Metal Census of $\tau_{\rm LL}>1$ absorbers at $z\sim 2$--$3$ \label{t-omega}}
\tabletypesize{\footnotesize}
\tablehead{
\colhead{Absorbers}  &\colhead{$\Omega_{\rm Z}^{\rm OVI}$} & \colhead{$\Omega_{\rm Z}^{\rm SiIV}$} & \colhead{$\Omega_{\rm Z}^{\rm h}$} \\
\colhead{} 	&\colhead{($10^{-7}$)} & \colhead{($10^{-7}$)} & \colhead{($10^{-7}$)}}
\startdata
  LLS 	  &  4--14 &   4--17 & 8--31   \\
  SLLS 	  &  2--8  &   3--11 & 5--19  \\
  DLA  	  &  1--5  &   1--5  & 2--9	 
\enddata 
\tablecomments{$\Omega_{\rm Z}^{\rm h} =\Omega_{\rm Z}^{\rm OVI} + \Omega_{\rm Z}^{\rm SiIV} $. The calculation of $\Omega_{\rm Z}$ is inversely proportional to the ionization fraction of \siivt\ and \ovit; we adopted here $f_{\rm OVI}=f_{\rm SiIV}=0.1$;  $\Omega_{\rm Z}$ is, however, independent of the metallicity of the gas (see \S\ref{s-metbud} for more details).}
\end{deluxetable}

%% file: tabA1.tex
\begin{deluxetable}{ccccc}
\tablewidth{0pc}
\tablecaption{Results from the Profile Fits \label{t-fit}}
\tabletypesize{\footnotesize}
\tablehead{\colhead{No.} & \colhead{$v$} & \colhead{$b$} &\colhead{$\log N$} &\colhead{Flag} \\
\colhead{} & \colhead{(\km)} & \colhead{(\km)} &\colhead{[cm$^{-2}$]} &\colhead{}}
\startdata
  \cutinhead{  Q1009+2956   -- $ z = 2.42903 $ (int.)}
  \cutinhead{-- \siivt\ --}
   1 & $       -166.6  \pm          0.3  $ & $          7.4  \pm          0.4 $ & $       12.05  \pm         0.02  $ & 0  \\
   2 & $        -40.6  \pm          0.3  $ & $          5.8  \pm          0.3 $ & $       12.79  \pm         0.04  $ & 0  \\
   3 & $        -28.0  \pm          0.3  $ & $          9.2  \pm          0.3 $ & $       13.29  \pm         0.01  $ & 0  \\
   4 & $         -2.6  \pm          0.1  $ & $          4.6  \pm          0.1 $ & $       13.30  \pm         0.01  $ & 0  \\
   5 & $         30.3  \pm          0.1  $ & $         11.1  \pm          0.1 $ & $       12.88  \pm         0.01  $ & 0  \\
  \cutinhead{-- \civt\ --}
   1 & $       -166.6  \pm          0.1  $ & $          7.9  \pm          0.2 $ & $       12.80  \pm         0.01  $ & 0  \\
   2 & $        -30.8  \pm          0.1  $ & $         12.5  \pm          0.1 $ & $       13.54  \pm         0.01  $ & 0  \\
   3 & $         -2.5  \pm          0.1  $ & $          5.9  \pm          0.1 $ & $       13.35  \pm         0.01  $ & 0  \\
\enddata
\tablecomments{Each doublet of a given ion was simultaneously fitted, but each species was independently fitted. Flag: 0: result is robust; 1: uncertain owing to the complexity of the absorption profile or on the necessity to rely on only one transition of the doublet; 2: uncertain owing to the relative low S/N; 3: uncertain results because the component is saturated in both transitions of the fitted doublet.
}
\end{deluxetable}